\documentclass[twocolumn,secnumarabic,amssymb, nobibnotes, aps, prd]{revtex4-2}

\usepackage{subfig}
\usepackage{graphicx}
\usepackage{amsmath}
\usepackage{array} 
\newcolumntype{L}[1]{>{\raggedright\let\newline\\\arraybackslash\hspace{0pt}}m{#1}}
\newcolumntype{C}[1]{>{\centering\let\newline\\\arraybackslash\hspace{0pt}}m{#1}}
\newcolumntype{R}[1]{>{\raggedleft\let\newline\\\arraybackslash\hspace{0pt}}m{#1}}
\newcolumntype{N}{@{}m{0pt}@{}}
\usepackage{multirow}
\usepackage{fixltx2e}
\usepackage{color}
\begin{document}

\title{Modeling and Simulation of Transitional Taylor-Green Vortex Flow with PANS}%

\author{F.S. Pereira\textsuperscript{1,2}}\email[F.S. Pereira: ]{fspereira@lanl.gov}
\author{F.F. Grinstein\textsuperscript{2}}
\author{D.M. Israel\textsuperscript{2}}
\author{R. Rauenzahn\textsuperscript{2}}
\author{S.S. Girimaji\textsuperscript{3}}
\affiliation{\textsuperscript{1}Los Alamos National Laboratory, T-Division, Los Alamos, New Mexico, USA}
\affiliation{\textsuperscript{2}Los Alamos National Laboratory, X-Division, Los Alamos, New Mexico, USA}
\affiliation{\textsuperscript{3}Texas A\&M Univeristy, Ocean Engineering, College Station, Texas, USA}


\begin{abstract}
{\color{blue}The Partially-Averaged Navier-Stokes (PANS) equations model is used to predict the transitional Taylor-Green vortex (TGV) flow at Reynolds number $3000$. A new form of the closure is proposed, in which the PANS procedure is applied to a variant of the BHR turbulence model. The TGV is} a benchmark case for transitional flows in which the onset of turbulence is driven by vortex-stretching and reconnection mechanisms. Since these physical phenomena are observed in numerous flows of variable-density (e.g., oceanography and material mixing), this study constitutes the first step toward extending the PANS method to such a class of problems. We start by deriving the governing equations of the model and analyze the selection of the parameters controlling its physical resolution, $f_\varepsilon$ and $f_k$, through \textit{a-priori} testing. Afterward, we conduct PANS computations at different constant physical resolutions to evaluate the model's accuracy and cost predicting the TGV flow. This is performed through simple verification and validation exercises, and the physical and modeling interpretation of the numerical predictions. The results confirm that PANS can efficiently (accuracy vs. cost) predict the present flow problem. Yet, this is closely dependent on the physical resolution of the model. Whereas high-physical resolution ($f_k<0.50$) computations are in good agreement with the reference DNS studies, low-physical resolution ($f_k\ge 0.50$) simulations lead to large discrepancies with the reference data. The physical and modeling interpretation of the results demonstrates that the origin of these distinct behaviors lies in the model's ability to {\color{blue}resolve the phenomena driving the onset of turbulence not amenable to modeling by the closure}. The comparison of the computations' cost indicates that high-physical resolution PANS achieves the accuracy of DNS ($f_k=0.00$) at a fraction of the cost. We observe a cost reduction of one order of magnitude at the current Re, which is expected to grow with Re. 
\end{abstract}

\maketitle

\section{Introduction}	
\label{sec:1}

The onset of turbulence is a complex phenomenon governing the physics of numerous transient flows of practical interest - e.g., material mixing \cite{DIMOTAKIS_ARFM_2005,CABOT_NP_2006,ORLICZ_POF_2015}, combustion \cite{MARBLE_AIAA23_1987,YANG_AIAA_1993,YANG_AME_2014}, astrophysics \cite{ARNETT_ARAA_1989,HAMMER_TAJ_2010,THOMAS_PR_2012}, and bluff-bodies in cross-flow \cite{WILLIAMSON_ARFM_1996,ZDRAVKOVICH_BOOK_1997,PEREIRA_JCP_2018,PEREIRA_IJHFF_2019}. Transition is characterized by its transient dynamics, sensitivity to the flow's geometry and initial conditions, triggering mechanisms \cite{MORKOVING_NASAREP_1988,KIDA_ARFM_1994,ZAKI_FTC_2013,ECKHADT_PA_2018}, and dependence on hydrodynamic instabilities and coherent structures \cite{HELMHOLTZ_PMJS_1868,KELVIN_PMJS_1871,RAYLEIGHT_PLMS_1882,TAYLOR_PRSA_1950,RICHTMEYER_CPAM_1960,MESHKOV_SFD_1969,CROW_AIAA_1970,LAPORTE_AIAA_2002,YAO_JFMR_2020,LEWEKE_JFM_1998,LEWEKE_ARFM_2016}. This set of features poses unique challenges to modeling and simulation of transitional flows.

Accurate computations of this class of problems are closely dependent on the mathematical model's ability to capture the fundamental physics of the onset, development, and decay of turbulence. Direct numerical simulation (DNS) and high-resolution large-eddy simulation (LES) \cite{SMAGORINSKY_MWR_1963,BORIS_FDR_1992,GRINSTEIN_BOOK_2010} are the optimal techniques to predict transitional flows since these methods resolve all (DNS) or most (LES) flow scales. Yet, the caveat of such scale-resolving simulation (SRS) strategies is the inherent computational cost \cite{SPALART_FORSR_1997,SPALART_IJHFF_2000,MENTER_TFMCAAIA_2011} and the difficulty of selecting proper initial and boundary flow conditions. The Reynolds-averaged Navier-Stokes (RANS) equations, on the other hand, reduce the cost burden of simulating transitional flows through its fully statistical description of turbulence, in which a closure models the entire spectrum of turbulence scales \cite{WILCOX_BOOK_2010,PIQUET_BOOK_1999,POPE_BOOK_2000,HANJALIC_BOOK_2011}. However, RANS closures have not been designed for transitional flows and, as such, often lead to inaccurate predictions \cite{WILCOX_BOOK_2010,PEREIRA_IJHFF_2019}.
A possible alternative to these modeling strategies is the utilization of bridging methods \cite{SPEZIALLI_ICNMFD_1996,FASEL_JFE_2002,GIRIMAJI_JAM_2005,SCHIESTEL_TCFD_2005,CHAOUAT_PF_2005}. This class of SRS models can operate at any degree of physical resolution (from DNS to RANS) and has been expressly designed to efficiently (accuracy vs. cost) predict complex flows by resolving only the flow scales not amenable to modeling \cite{PEREIRA_JCP_2018,PEREIRA_IJHFF_2019}. The dynamics of the remaining scales can be accurately represented through an appropriate turbulence closure. In this manner, bridging models embed the concept of accuracy-on-demand, enabling significant enhancement of the efficiency of SRS predictions.

In recent years, the importance of predicting transitional flows has grown due to industrial needs and advances in computational capabilities. These provided the computational resources to allow scientists and engineers to study complex flows involving separation and transition. This has been driving intensive research in transition modeling and simulation, and diverse validation initiatives \cite{NASA_TRAN_2017,AIAA_TRAN_2020} using benchmark problems. The Taylor-Green vortex (TGV) \cite{TAYLOR_PRSA_1937} is a test-case designed to study transition to turbulence driven by vortex-stretching and reconnection mechanisms \cite{BRACHET_JFM_1983,KIDA_ARFM_1994,YANG_JFM_2011}. This problem initially features multiple well-defined laminar vortices (figure \ref{fig:3_1}), which interact and evolve in time. Their spatio-temporal development comprises five key stages:
\begin{itemize}
 \item[$i)$] vortex stretching mechanisms generate vortex sheets which gradually get closer;
 \item[$ii)$] vortex sheets roll-up and reconnect \cite{KIDA_ARFM_1994,YANG_JFM_2011};
 \item[$iii)$] onset of turbulence and subsequent intensification of vorticity;
 \item[$iv)$] coherent structures breakdown and fully-developed turbulence;
 \item[$v)$] turbulence decay.
\end{itemize}
There is a consensus that step \textit{iii)} can involve multiple instabilities \cite{BRACHET_JFM_1983,CROW_AIAA_1970,LAPORTE_AIAA_2002,YAO_JFMR_2020,LEWEKE_JFM_1998,LEWEKE_ARFM_2016}. In addition to the complex physics, this benchmark flow possesses well-characterized initial conditions which provide the consistency for rigorous verification and validation exercises. Brachet et al. \cite{BRACHET_JFM_1983} and Brachet \cite{BRACHET_FDR_1991} performed direct numerical simulations of the TGV flow at multiple Reynolds numbers to investigate the flow physics. These pioneering studies have been used as a reference in various numerical studies \cite{SHU_JSC_2005,DRIKAKIS_JOT_2007,YANG_JFM_2010,YANG_JFM_2011,CHAPELIER_AIAA42_2012,DEBONIS_NASAREP_2013,SHIROKOV_JOT_2014,BULL_AIAA_2015,DAIRAY_JCP_2017,MOURA_JCP_2017,PENG_PRF_2018,SHARMA_POF_2019,GRINSTEIN_CMA_2019}.

This work investigates the ability of a bridging partially-averaged Navier-Stokes (PANS) equations method to predict the transitional TGV flow at Reynolds number $3000$ \cite{BRACHET_JFM_1983}. Since our primary interest is variable-density flows and the present study constitutes the first step toward extending PANS to such class of flows, we use a simplified variable-density PANS closure based on Besnard-Harlow-Rauenzahn (BHR) multi-equation RANS framework \cite{BESNARD_TREP_1992} in the {\color{blue}BHR-LEVM (linear eddy viscosity model) closure version \cite{BANERJEE_PRE_2010,ZARLING_TREP_2011}}. The model's accuracy and cost is evaluated through simulation of the TGV flow at distinct constant degrees of physical resolution. This strategy prevents commutation errors \cite{HAMBA_PF_2011} and enables robust verification and validation studies. The results are compared against the DNS data of Brachet et al. \cite{BRACHET_JFM_1983} and Drikakis et al. \cite{DRIKAKIS_JOT_2007}, and interpreted through the assessment of the coherent and turbulent flow fields.

The structure of this manuscript is as follows. Section \ref{sec:2} presents the derivation of the governing equations of PANS {\color{blue}BHR-LEVM}. This step comprises the evaluation of the parameters defining the physical resolution of the model. Section \ref{sec:3} introduces the flow problem and the numerical settings of the computations. Section \ref{sec:4} discusses the results of the study, and Section \ref{sec:5} concludes the paper with the summary of the main findings. 
%
%
%
\section{Governing equations}
\label{sec:2}
The PANS equations are based upon the scale-invariance of the Navier-Stokes equations. This property has been initially demonstrated by Germano \cite{GERMANO_JFM_1992} for incompressible flow, and later extended to compressible flow by Suman and Girimaji \cite{SUMAN_FTC_2010}. {\color{blue}In this work, we propose a simplified PANS version of the BHR-LEVM closure, in which the model equations can include some variable-density effects. This constitutes the first step toward extending PANS to {\color{blue}variable-density (compressible and/or multi-material)} flow.}

To derive the governing equations of PANS {\color{blue}BHR-LEVM}, let us start by considering a general linear and constant preserving filtering operator, $\langle\ \cdot\ \rangle$. This filter commutes with spatial and temporal differentiation, and decomposes any instantaneous flow quantity, $\Phi$, into a resolved/filtered, $\langle \Phi \rangle$, and unresolved/modeled, $\phi$, component,
\begin{equation}
\label{eq:2_1}
\Phi \equiv \langle \Phi \rangle + \phi \; .
\end{equation}
From the concept of Favre-averaging \cite{FAVRE_CRAS_1958,FAVRE_JM_1965,FAVRE_JM2_1965,FAVRE_CRASP_1971}, the former decomposition can be extended to variable-density flow, 
\begin{equation}
\label{eq:2_2}
\Phi \equiv \{ \Phi \} + \phi^* \; .
\end{equation}
where  $\{ \Phi \}$ and $\phi^*$ stand for the density-weighted resolved and unresolved fluctuating fraction of $\Phi$. In the limit of all turbulent scales being modeled, relations \ref{eq:2_1} and \ref{eq:2_2} are equivalent to the Reynolds- \cite{REYNOLDS_PTRSL_1985} and Favre-averaged \cite{FAVRE_CRAS_1958,FAVRE_JM_1965,FAVRE_JM2_1965,FAVRE_CRASP_1971} decompositions,
\begin{equation}
\label{eq:2_3}
\Phi \equiv \overline{\Phi} + \phi' \; ,
\end{equation}
\begin{equation}
\label{eq:2_4}
\Phi \equiv \tilde{\Phi} + \phi'' \; ,
\end{equation}
where $\overline{\Phi}$ represents a Reynolds-averaged quantity, $\phi'$ the turbulent fluctuations, $\tilde{\Phi}$ a density-weighted quantity, and $\phi''$ the fluctuations uncorrelated with the fluid's density ($\overline{\rho \phi''}=0$). 

The application of the filtering operators \ref{eq:2_1} and \ref{eq:2_2} to the Navier-Stokes equations for compressible flow leads to their filtered or partially-averaged form \cite{GERMANO_JFM_1992,SUMAN_FTC_2010},
\begin{equation}
\label{eq:2_5}
\frac{\partial \langle \rho \rangle}{\partial t} + \frac{\partial \left( \langle \rho \rangle \{V_i \} \right)}{\partial x_i}=0\; ,
\end{equation}
\begin{equation}
\label{eq:2_6}
\begin{split}
\frac{\partial \left( \langle \rho \rangle\{V_i \} \right)}{\partial t} + \frac{ \partial \left(\langle \rho \rangle \{V_i \} \{V_j \} \right)}{\partial x_j}=&-
\frac{\partial\langle P \rangle}{\partial x_i} + \frac{\partial \langle \sigma_{ij} \rangle}{\partial x_j} \\
&+ \frac{\partial\left(\langle \rho \rangle \tau^1(V_i,V_j)\right)   }{\partial x_j}
\end{split}
\; ,
\end{equation}
\begin{equation}
\label{eq:2_7}
\begin{split}
\frac{\partial \left( \langle \rho \rangle\{E \} \right) }{\partial t} &+ \frac{ \partial \left( \langle \rho \rangle \{E \} \{V_j \} \right) }{\partial x_j}=
 -\frac{\partial\left( \langle \rho \rangle \tau^1(V_j,E)  \right)}{\partial x_j}  \\
& -\frac{\partial \left( \{ V_j \}\langle P \rangle \right)}{\partial x_j} -\frac{\partial \tau^2(V_j,P) }{\partial x_j}  \\
& +\frac{\partial \left( \{ V_i \}\langle \sigma_{ij} \rangle \right) }{\partial x_j} +\frac{\partial  \tau^2(V_i,\sigma_{ij}) }{\partial x_j}\\ 
& -\frac{\partial \langle q_j^c \rangle}{\partial x_j}  
\end{split}
\; ,
\end{equation}
where the unresolved turbulent stresses $\tau^1(\Phi_i,\Phi_j)$ and $\tau^2(\Phi_i,\Phi_j)$ are expressed in terms of generalized central second-moment tensors to guarantee scale-invariance \cite{GERMANO_JFM_1992}. These tensors account for the effect of the unresolved turbulent scales on the resolved field and are formally defined as \cite{GERMANO_JFM_1992,SUMAN_FTC_2010},
\begin{equation}
\label{eq:2_9}
\tau^1(\Phi_i,\Phi_j) \equiv \{ \Phi_i, \Phi_j \} - \{ \Phi_i \} \{ \Phi_j \}\; ,
\end{equation}
\begin{equation}
\label{eq:2_10}
\tau^2(\Phi_i,\Phi_j) \equiv \langle \Phi_i, \Phi_j \rangle - \{ \Phi_i \} \{ \Phi_j \}\; .
\end{equation}
In equations \ref{eq:2_5} to \ref{eq:2_7}, $x_i$ are the coordinates of a Cartesian system, $\rho$ is the fluid's density, $V_i$ are the Cartesian velocity components, $P$ is the pressure, $\sigma_{ij}$ is the viscous-stress tensor assuming Newtonian fluid,
\begin{equation}
\label{eq:2_11}
{\color{blue}
\langle \sigma_{ij} \rangle = 2 \mu \left( \{ S_{ij} \} - \frac{2}{3}\frac{\partial \{ V_k\}}{\partial x_k} \delta_{ij} \right) \; ,
}
\end{equation}
with $\{ S_{ij} \}$ being the resolved strain-rate tensor,
\begin{equation}
\label{eq:2_12}
\{ S_{ij} \} = \frac{1}{2} \left( \frac{\partial \{ V_i\}}{\partial x_j} + \frac{\partial \{ V_j\}}{\partial x_i} \right)\; ,
\end{equation}
$\mu$ is the dynamic viscosity, $E=\frac{1}{2}V_i^2+e$ is the fluid's total energy, and $e$ is the internal energy. The conductive heat flux $q^c$ is neglected because the present TGV flow is characterized by negligible thermal effects. The pressure is calculated assuming a thermally perfect gas ($P=\rho R T$) so that its resolved component is written as follows \cite{SUMAN_FTC_2010},
\begin{equation}
\label{eq:2_13}
\langle P \rangle = (\gamma - 1) \langle \rho \rangle \left( \{E\} - \frac{\{V_k\} \{V_k\} }{2} - k_u \right) \; ,
\end{equation}
where $T$ is the fluid's temperature, $\gamma$ is the ratio between specific heats, and $k_u$ is the unresolved or modeled turbulence kinetic energy. The generalized central second-moments create new variables that require modeling to close the system equations. This is accomplished through constitutive relationships.
%
%
\subsection{Constitutive relationships}
\label{sec:2.1}
%
The generalized central second-moment tensor $\tau^1(V_i,V_j)$ represents the effect of the unresolved turbulent scales on the resolved velocity field. In {\color{blue}BHR-LEVM closure \cite{BANERJEE_PRE_2010,ZARLING_TREP_2011}}, this tensor is modeled through the Boussinesq approximation \cite{BOUSSINESQ_MPDSAS_1877}, which establishes an analogy between the momentum transfer resultant from molecular gas and turbulent motion. Hence, $\tau^1(V_i,V_j)$ is assumed proportional to $\{S_{ij}\}$,
\begin{equation}
\label{eq:2.1_1}
 \tau^1(V_i,V_j) =2 \nu_u \{S_{ij}\} - \frac{2}{3} k_u \delta_{ij} \; .
\end{equation}
In equation \ref{eq:2.1_1}, $\nu_u$ is the turbulent kinematic viscosity of the unresolved or modeled scales. Throughout this work, the subscript \textit{u} denotes the unresolved fraction of a given (PANS) turbulence quantity, whereas \textit{t} indicates a total (RANS) turbulence variable. In the limit of all turbulence flow scales being modeled, the subscripts \textit{u} and \textit{t} are equivalent.

The remaining generalized central second-moment tensors invoke distinct constitutive relationships which derivation can be found in Besnard et al. \cite{BESNARD_TREP_1992}, Suman and Girimaji \cite{SUMAN_FTC_2010}, Stalsberg-Zarling et al. \cite{ZARLING_TREP_2011}, and Schwarzkopf et al. \cite{SCHWARZKOPF_JOT_2011,SCHWARZKOPF_FTC_2016}. In PANS form, these tensors lead to the following final form of the energy equation,
\begin{equation}
\label{eq:2.1_2}
\begin{split}
\frac{\partial \left( \langle \rho \rangle\{E \} \right) }{\partial t} &+ \frac{ \partial \left( \langle \rho \rangle \{E \} \{V_j \} \right) }{\partial x_j}= - \frac{\partial \left( \left\{ V_j \right\} \langle P \rangle \right) }{\partial x_j} \\
& +\frac{\partial \left( \{ V_i \}\langle \sigma_{ij} \rangle \right)}{\partial x_j} +\frac{\partial \left( \langle \rho \rangle \{V_i \} \tau^1(V_i,V_j) \right)}{\partial x_j}  \\
& +\frac{\partial}{\partial x_j}\left[ \left(  \mu + \frac{\mu_u}{\sigma_k} \right) \frac{\partial k_u}{\partial x_j} \right] 
\end{split}
\; ,
\end{equation}
where $\sigma_k$ is a coefficient. The former relations introduce two additional variables, $k_u$ and $\nu_u$, which need modeling to close the governing equations of PANS.
%
%
\subsection{PANS {\color{blue}BHR-LEVM} closure}
\label{sec:2.2}
%
In this work, the dynamics of the unresolved turbulent scales is represented through a variable-density PANS closure based on Besnard-Harlow-Rauenzahn (BHR) multi-equation RANS model \cite{BESNARD_TREP_1992} in the {\color{blue}BHR-LEVM closure version \cite{BANERJEE_PRE_2010,ZARLING_TREP_2011}}. Such RANS closure is featured by six dependent variables: turbulence kinetic energy, $k_t$, turbulence dissipation length-scale, $S_t$, velocity mass flux, $a_{i_t}$, and density-specific volume correlation, $b_t$. Since the last two quantities, $a_{i_t}$ and $b_t$, are not relevant for the present TGV computations, their equations will be presented in future work. 

The {\color{blue}BHR-LEVM} RANS closure calculates the total kinematic turbulent viscosity, $\nu_t$, as
\begin{equation}
\label{eq:2.2_1}
\nu_t= \frac{\mu_t}{\overline{\rho}} = c_\mu S_t \sqrt{k_t} \; ,
\end{equation}
where $c_\mu$ is a coefficient given in table \ref{tab:2_1}, $S_t$ is the turbulence dissipation length-scale,
\begin{equation}
\label{eq:2.2_2}
S_t=\frac{k_t^{3/2}}{\varepsilon_t} \; ,
\end{equation}
and $\varepsilon_t$ is the dissipation of $k_t$. {\color{blue}BHR-LEVM} models the former turbulence quantities, $k_t$ and $S_t$, through the following evolution equations,
\begin{equation}
\label{eq:2.2_3}
\frac{\partial k_t}{\partial t} + \tilde{V}_j\frac{\partial k_t}{\partial x_j} = {\cal{P}}_{b_t} + {\cal{P}}_{s_t} - \frac{k_t^{3/2}}{S_t} + \frac{1}{\overline{\rho}}\frac{\partial}{\partial x_j}\left( \frac{\overline{\rho}\nu_t}{\sigma_k} \frac{\partial k_t}{\partial x_j}\right)\; ,
\end{equation}
\begin{equation}
\label{eq:2.2_4}
\begin{split}
\frac{\partial S_t}{\partial t} + \tilde{V_j}\frac{\partial S_t}{\partial x_j} &= \frac{S_t}{k_t} \left( c_4 {\cal{P}}_{b_t} + c_1 {\cal{P}}_{s_t} \right) - c_2 \sqrt{k_t} \\
 &+ \frac{1}{\overline{\rho}}\frac{\partial}{\partial x_j}\left( \frac{\overline{\rho}\nu_t}{\sigma_S} \frac{\partial S_t}{\partial x_j} \right)
\end{split}
\; ,
\end{equation}
where ${\cal{P}}_{b_t}$ and ${\cal{P}}_{s_t}$ are the total production of turbulence kinetic energy by buoyancy and shear effects. The first production mechanism of $k_t$ is modeled as,
\begin{equation}
\label{eq:2.2_5}
{\cal{P}}_{s_t} = - R^1(V_i,V_j) \frac{\partial \tilde{V}_i}{\partial x_j}  \; ,
\end{equation}
whereas the second is neglected in this work owing to the fact that the TGV is a single-fluid and (nearly) incompressible flow problem. In equation \ref{eq:2.2_5}, $R^1(V_i,V_j)$ stands for the specific total turbulent or Reynolds-stress tensor - $R^1(V_i,V_j)=\tau^1(V_i,V_j)$ for $f_k=f_\varepsilon=1.00$. The coefficients $c_\mu$, $c_1$, $c_2$, $\sigma_k$, $\sigma_S$ are given in table \ref{tab:2_1} \cite{ZARLING_TREP_2011}.

The former set of governing equations has been developed to operate with RANS variables: $\tilde{\Phi}$,  $\overline{\Phi}$, or $\Phi_t$. We now derive their PANS counterpart \cite{GIRIMAJI_JAM_2005}. Toward this end, the parameters defining the ratios of modeled-to-total turbulence kinetic energy and dissipation length-scale,
\begin{equation}
\label{eq:2.2_8}
f_k\equiv \frac{k_u}{k_t}\;, \hspace{1cm}  f_S \equiv \frac{S_u}{S_t} \; ,
\end{equation}
need to be included in equations \ref{eq:2.2_3} and \ref{eq:2.2_4}. These enable the {\color{blue}BHR-LEVM} {\color{blue}to become a scale-dependent closure and operate at any degree of physical resolution}, i.e., from RANS to DNS. The ratio of modeled-to-total turbulence dissipation length-scale, $f_S$, can also be defined as follows,
\begin{equation}
\label{eq:2.2_9}
f_S\equiv \frac{S_u}{S_t} = \left(\frac{k_u^{3/2}}{\varepsilon_u}\right) \left( \frac{\varepsilon_t}{k_t^{3/2}}\right)=\frac{f_k^{3/2}}{f_\varepsilon}  \; ,
\end{equation}
where $f_\varepsilon$ is the ratio of modeled-to-total turbulence dissipation. Since $f_\varepsilon$ is physically more intuitive than $f_S$, the evolution equations for $k_u$ and $S_u$ are derived in terms of $f_k$ and $f_\varepsilon$.
\begin{table}[b!]
\centering
\setlength\extrarowheight{3pt}
\caption{Coefficients of {\color{blue}BHR-LEVM} closure.}
\label{tab:2_1}    
\begin{tabular}{C{1.0cm}C{1.0cm}C{1.0cm}C{1.0cm}C{1.0cm}}
\hline 
$c_1$ & $c_2$ &   $c_\mu$  & $\sigma_k
$ & $\sigma_S$  \\[3pt]
\hline 
0.06 & 0.42 & 0.28  & 1.00 & 0.10 \\ [3pt]
\hline
\end{tabular}
\end{table}
%
%
\subsubsection{$k_u$ evolution equation}
\label{sec:2.2.1}
%
It has been demonstrated by Girimaji \cite{GIRIMAJI_JAM_2005} and Suman and Girimaji \cite{SUMAN_FTC_2010} that the scale-invariant form of the evolution equation for $k_u$ can be written as follows,
\begin{equation}
\label{eq:2.2.1_1}
\frac{\partial k_u}{\partial t}  +  \{V_j\} \frac{\partial k_u}{\partial x_j} = {\cal{P}}_{s_u} - \varepsilon_u +  {\cal{T}}_u\; ,
\end{equation}
where ${\cal{P}}_{s_u}$, $\varepsilon_u$ and ${\cal{T}}_u$ represent the production by shear mechanisms, dissipation, and transport of unresolved turbulence kinetic energy, $k_u$. For constant $f_k$, differentiation commutes in time and space and so it is possible to establish a relationship between the evolution equation for total (RANS) and partial (PANS) turbulence kinetic energy,
\begin{equation}
\label{eq:2.2.1_2}
\frac{\partial k_u}{\partial t} + \tilde{V_j}\frac{\partial k_u}{\partial x_j} =  f_k \left[ \frac{\partial k_t}{\partial t} + \tilde{V_j}\frac{\partial k_t}{\partial x_j}  \right] \; .
\end{equation}
Considering that PANS calculates filtered or partial dependent variables, the left-hand side of the former equation can be rewritten as,
\begin{equation}
\label{eq:2.2.1_3}
\frac{\partial k_u}{\partial t} + \{V_j\}\frac{\partial k_u}{\partial x_j} =  f_k \left[ \frac{\partial k_t}{\partial t} + \tilde{V_j}\frac{\partial k_t}{\partial x_j}  \right] + \left(\{ V_j \} - \tilde{V}_j\right)\frac{\partial k_u}{\partial x_j} \; .
\end{equation}
Replacing the first term of the right-hand side by that of equation \ref{eq:2.2_3} (${\cal{P}}_{b_t}= 0$),
\begin{equation}
\label{eq:2.2.1_4}
\begin{split}
\frac{\partial k_u}{\partial t} + \{V_j\}\frac{\partial k_u}{\partial x_j} &=  f_k \left[ {\cal{P}}_{s_t} - \frac{k_t^{3/2}}{S_t} + \frac{1}{\overline{\rho}}\frac{\partial}{\partial x_j}\left( \frac{\overline{\rho} \nu_t}{\sigma_k} \frac{\partial k_t}{\partial x_j}\right) \right] \\
&+ \left(\{ V_j \} - \tilde{V}_j\right)\frac{\partial k_u}{\partial x_j}  
\end{split}
\; ,
\end{equation}
and applying a similar procedure to the left-hand side, the following relation is obtained,
\begin{equation}
\label{eq:2.2.1_5}
\begin{split}
{\cal{P}}_{s_u} - \varepsilon_u +   {\cal{T}}_u  &=  f_k \left[{\cal{P}}_{s_t }- \frac{k_t^{3/2}}{S_t} + \frac{1}{\overline{\rho}}\frac{\partial}{\partial x_j}\left( \frac{\overline{\rho} \nu_t}{\sigma_k} \frac{\partial k_t}{\partial x_j}\right) \right] \\
& + \left(\{ V_j \} - \tilde{V}_j\right)\frac{\partial k_u}{\partial x_j}
\end{split}
 \; .
\end{equation}
This equation illustrates the formal similarity between PANS (left-hand side) and RANS (right-hand side) production, dissipation, and transport terms. It is therefore possible to relate the source and sink terms of equation \ref{eq:2.2.1_5},
\begin{equation}
\label{eq:2.2.1_6}
{\cal{P}}_{s_u} - \frac{k_u^{3/2}}{S_u} =  f_k \left[ {\cal{P}}_{s_t} - \frac{k_t^{3/2}}{S_t} \right] =  f_k \left[ {\cal{P}}_{s_t} - \frac{1}{f_\varepsilon} \frac{k_u^{3/2}}{ S_u} \right] \; ,
\end{equation}
to obtain an expression for ${\cal{P}}_{s_t}$ in terms of unresolved quantities, $f_k$, and $f_\varepsilon$,
\begin{equation}
\label{eq:2.2.1_7}
{\cal{P}}_{s_t} =  \frac{1}{f_\varepsilon}\left( \frac{k_u^{3/2}}{S_u} \right) + \frac{1}{f_k}\left({\cal{P}}_{s_u}  - \frac{k_u^{3/2}}{S_u}\right)  \; .
\end{equation}
This relation is used in the derivation of the evolution equation for $S_u$. On the other hand, the transport terms ${\cal{T}}_u$ and ${\cal{T}}_t$ can be related as follows,
\begin{equation}
\begin{split}
\label{eq:2.2.1_8}
{\cal{T}}_u&= f_k\left[ \frac{1}{\overline{\rho}} \frac{\partial}{\partial x_j}\left(\frac{ \overline{\rho}\nu_t}{\sigma_k} \frac{\partial k_t}{\partial x_j}\right) \right] + \left(\{ V_j \} - \tilde{V}_j\right)\frac{\partial k_u}{\partial x_j} \\
&= \frac{1}{\langle \rho \rangle}\frac{\partial}{\partial x_j}\left( \frac{\langle \rho \rangle \nu_u}{\sigma_k}\frac{f_\varepsilon}{f_k^2} \frac{\partial k_u}{\partial x_j}\right) + \left(\{ V_j \} - \tilde{V}_j\right)\frac{\partial k_u}{\partial x_j} 
\end{split}
\; .
\end{equation}
Using scaling arguments, Girimaji \cite{GIRIMAJI_JAM_2005} showed that
\begin{equation}
\label{eq:2.2.1_9}
\left(\{ V_j \} - \tilde{V}_j\right)\frac{\partial k_u}{\partial x_j} \approx 0 \; ,
\end{equation}
leading to the so-called zero-transport model (ZTM). The accuracy of this model has been confirmed in the recent work of Tazraei and Girimaji \cite{TAZRAEI_PRF_2019}. Also, it is important to highlight that the velocity difference term tends to zero in the limit of $f_k=0.00$ and $1.00$ since
\begin{equation}
\label{eq:2.2.1_9.1}
{\color{blue}
\frac{\partial k_u}{\partial x_j}=0  \ \text{at}\   f_k=0.0 \; , \hspace{0.7cm} \{ V_j \} - \tilde{V}_j=0 \ \text{at}\  f_k=1.0 \; .
}
\end{equation}
This further reaffirms the validity of the ZTM. The derivation of the evolution equation for $k_u$ concludes by combining equations \ref{eq:2.2.1_4}, \ref{eq:2.2.1_8} and \ref{eq:2.2.1_9}, this leading to  its final form,
\begin{equation}
\label{eq:2.2.1_10}
\frac{\partial k_u}{\partial t} + \{V_j\}\frac{\partial k_u}{\partial x_j} =  {\cal{P}}_{s_u} - \frac{k_u^{3/2}}{S_u} +\frac{1}{\langle \rho \rangle}\frac{\partial}{\partial x_j}\left( \frac{\langle \rho \rangle \nu_u}{\sigma_k} \frac{f_\varepsilon}{f_k^2} \frac{\partial k_u}{\partial x_j}\right)   \; ,
\end{equation}
where
\begin{equation}
\label{eq:2.2.1_11}
{\cal{P}}_{s_u} = - \tau^1(V_i,V_j) \frac{\partial \{V_i\}}{\partial x_j}  \; .
\end{equation}
We recall that the derivation of equation \ref{eq:2.2.1_10} relies on the assumption that the filtering operator commutes with spatial and temporal differentiation. If this property does not hold, the model's derivation needs to consider additional terms \cite{GIRIMAJI_JOT_2013} and the modeled-to-total ratio of density, $f_\rho$. Despite being commonly neglected, this requirement holds for any bridging or hybrid formulation.
%
%
\subsubsection{$S_u$ evolution equation}
\label{sec:2.2.2}
%
Similarly to $k_u$, the derivation of the evolution equation for $S_u$ starts by establishing a relationship between the equations for $S_t$ (RANS) and $S_u$  (PANS),
\begin{equation}
\label{eq:2.2.2_1}
\frac{\partial S_u}{\partial t} + \tilde{V_j}\frac{\partial S_u}{\partial x_j} =  f_S \left[ \frac{\partial S_t}{\partial t} + \tilde{V_j}\frac{\partial S_t}{\partial x_j}  \right] \; .
\end{equation}
This equation can be rewritten as,
\begin{equation}
\label{eq:2.2.2_2}
\frac{\partial S_u}{\partial t} + \{V_j\}\frac{\partial S_u}{\partial x_j} =  f_S \left[ \frac{\partial S_t}{\partial t} + \tilde{V_j}\frac{\partial S_t}{\partial x_j}  \right] + \left(\{ V_j \} - \tilde{V}_j\right)\frac{\partial S_u}{\partial x_j} \; ,
\end{equation}
so that  the left-hand side contains only PANS variables. Then, applying the zero transport model \cite{GIRIMAJI_JAM_2005} and replacing the material derivative of $S_t$ by the right-hand side of equation \ref{eq:2.2_4} (${\cal{P}}_{b_t}= 0$) leads to,
\begin{equation}
\label{eq:2.2.2_5}
\begin{split}
\frac{\partial S_u}{\partial t} + \{V_j\}\frac{\partial S_u}{\partial x_j} &=  f_S \left[ \frac{S_t}{k_t} c_1 {\cal{P}}_{s_t} - c_2 \sqrt{k_t} \right] \\
&+ f_S \left[ \frac{1}{\overline{\rho}}\frac{\partial}{\partial x_j}\left( \frac{\overline{\rho} \nu_t}{\sigma_S} \frac{\partial S_t}{\partial x_j} \right)   \right]  \\
\end{split}\; .
\end{equation}
Finally, using equations \ref{eq:2.2_8}, \ref{eq:2.2_9} and \ref{eq:2.2.1_7}, the evolution equation for $S_u$ reads as,
\begin{equation}
\label{eq:2.2.2_6}
\begin{split}
\frac{\partial S_u}{\partial t} + \{V_j\}\frac{\partial S_u}{\partial x_j} &= c_1\frac{S_u}{k_u}  {\cal{P}}_{s_u} - c_2^*\sqrt{k_u}  \\
&+  \frac{1}{\langle \rho \rangle}\frac{\partial}{\partial x_j}\left( \frac{\langle \rho \rangle \nu_u}{\sigma_S} \frac{f_\varepsilon}{f_k^2} \frac{\partial S_u}{\partial x_j} \right)  
\end{split}\; ,
\end{equation}
where
\begin{equation}
\label{eq:2.2.2_7}
c_2^*= c_1 \left(1-\frac{f_k}{f_\varepsilon}\right)  + c_2 \frac{f_k}{f_\varepsilon}    \; .
\end{equation}
%
%
\subsection{Filter control parameter}
\label{sec:2.3}
%
The success of bridging methods is closely dependent on the physical resolution, i.e., the range of resolved flow scales. In PANS, this property is controlled by $f_\phi$ parameters (relations \ref{eq:2.2_8} and \ref{eq:2.2_9}). As the physical resolution increases, a wider range of flow scales is resolved and the numerical resolution necessary to accurately capture these physically resolved scales grows. Whereas excessively large physical resolutions lead to inefficient (accuracy vs. cost) calculations, reduced values of this parameter may reveal insufficient to resolve the flow phenomena not amenable to modeling \cite{PEREIRA_JCP_2018,PEREIRA_IJHFF_2019,PEREIRA_OE_2019,KAMBLE_POF_2020}. Since in practical problems the numerical resources are fixed, the selection of the physical resolution determines the efficiency and, consequently, the success of bridging computations.

The efficient determination of the physical resolution for given bridging model, flow problem, and quantities of interest considers three main factors:
\begin{itemize}
\item[$i)$] the length and time scales of the flow structures that need to be resolved;
\item[$ii)$] the smallest flow scales that a given spatio-temporal grid resolution (and numerical scheme) can accurately predict;
\item[$iii)$] the impact of the physical resolution on the closure's turbulence dependent quantities.
\end{itemize}
The first aspect has been recently addressed by Pereira et al. \cite{PEREIRA_IJHFF_2018,PEREIRA_JCP_2018,PEREIRA_IJHFF_2019,PEREIRA_OE_2019} for flows past circular cylinders in the sub-critical regime. {\color{blue}Supported by experimental evidence, these studies show that the simulations' accuracy is determined by the ability of the turbulence model (RANS or SRS) to predict the spatial development of the instabilities and coherent structures governing the flow dynamics: the vortex-shedding and Kelvin-Helmholtz.  Since these phenomena are usually not amenable to modeling by one-point closures, they need to be resolved and dictate the necessary physical resolution. Hence, the studies \cite{PEREIRA_IJHFF_2018,PEREIRA_JCP_2018,PEREIRA_IJHFF_2019,PEREIRA_OE_2019} provide clear examples of the importance of properly estimate the physical resolution of the model to efficiently simulate a given flow problem. Whereas insufficient physical resolution leads to inaccurate simulations, excessive physical resolution increases the cost of the calculations unnecessarily, reducing the efficiency of the computations.
}

\begin{figure}
\centering
\includegraphics[scale=0.105,trim=0 0 0 0,clip]{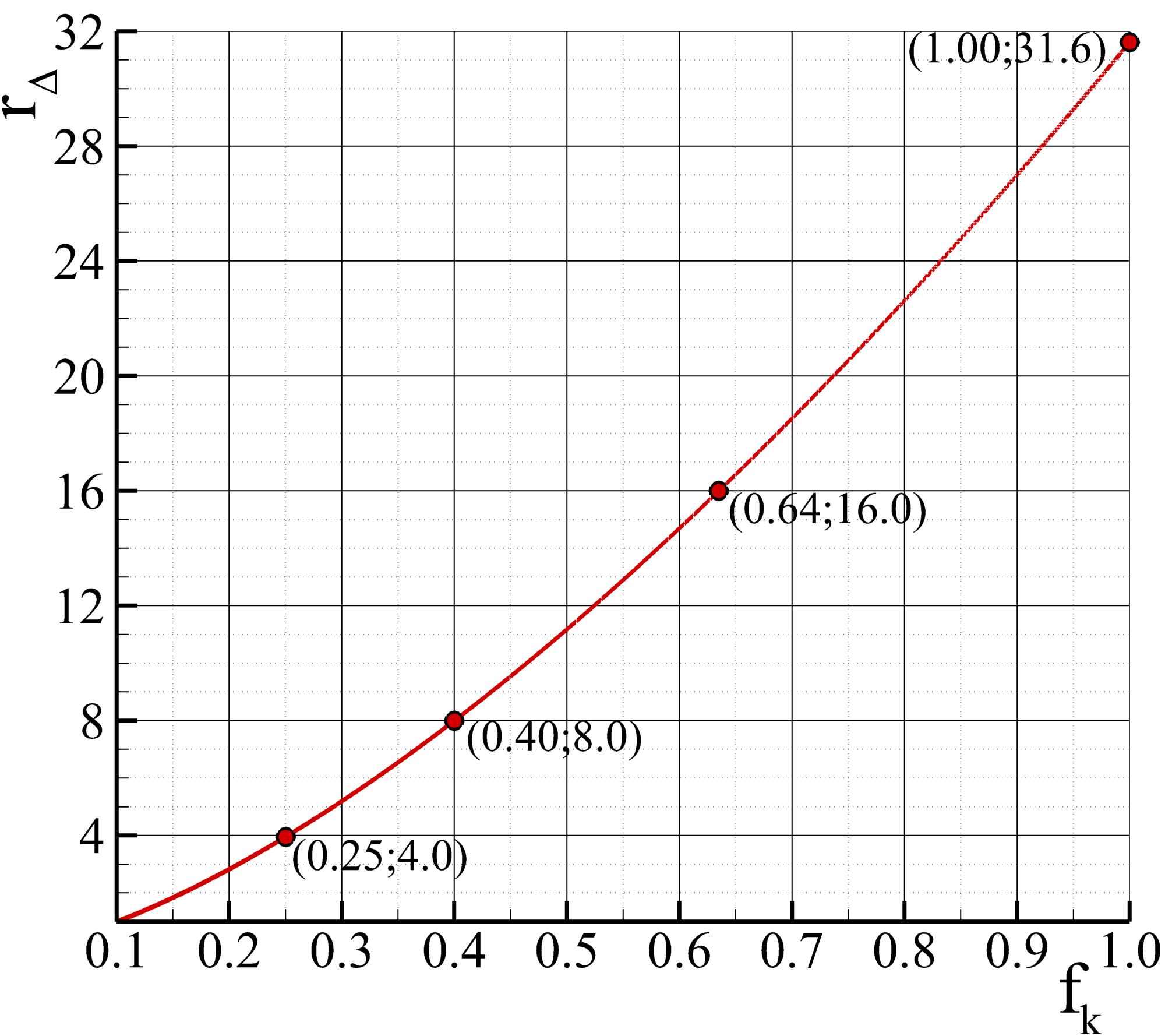}
\caption{Ratio between the minimum spatial grid resolution needed for computations at a given $f_k$ and $(f_k)_\text{ref}=0.10$, $r_\Delta$.}
\label{fig:2.3_1} 
\end{figure}

The second point is ideally addressed through verification exercises \cite{ROACHE_BOOK_1998,TRUCANO_SANDIA_2002,OBERKAMPF_BOOK_2010}. Nevertheless, it is possible to obtain a loose \textit{a-priori} estimate of the maximum physical resolution that a spatio-temporal grid resolution can support, as well as the effect of increasing the physical resolution on the numerical requirements of the simulations. At a minimum, the largest grid cell size, $\Delta$, needs to have the size of the smallest resolved flow scales, $\eta_r$,
\begin{equation}
\label{eq:2.3_1}
\Delta \leq \eta_r \; .
\end{equation}
Adapting the Kolmogorov arguments to resolved fluid motions,
\begin{equation}
\label{eq:2.3_2}
\Delta \sim \left( \frac{\nu^3_u}{\varepsilon_t} \right)^{0.25} \; ,
\end{equation}
assuming high-Re flow ($f_\varepsilon=1.0$), and using equation \ref{eq:2.2_1} for the unresolved field, the following expression for $\Delta$ is obtained,
\begin{equation}
\label{eq:2.3_3}
\Delta \approx C_\mu^{0.75} \frac{k_t^{1.5}}{\varepsilon_t} f_k^{1.5}\; .
\end{equation}
This relation can be rewritten to provide an estimate for the smallest $f_k$ that a given spatial grid resolution can support,
\begin{equation}
\label{eq:2.3_4}
f_k \geq \left(\frac{1}{C_\mu}\right)^{0.5}\left(\frac{ \Delta}{k_t^{1.5}/\varepsilon_t} \right)^{2/3} \; ,
\end{equation}
or to estimate the ratio between the smallest grid size for two values of $f_k$,
\begin{equation}
\label{eq:2.3_5}
r_\Delta = \frac{\Delta_{f_k}}{\Delta_{(f_k)_\text{ref}}}=\left[ \frac{f_k}{(f_k)_\text{ref}}\right]^{1.5}\; ,
\end{equation}
where the subscript ``ref'' denotes a reference $f_k$. This last expression enables the assessment of the relative evolution of $\Delta$ with $f_k$, and it is depicted in figure \ref{fig:2.3_1} for $(f_k)_\text{ref}=0.10$. As expected, $r_\Delta(f_k)$ indicates that the spatial grid resolution increases with the physical resolution ($f_k\rightarrow 0$). For instance, figure \ref{fig:2.3_1} shows that the necessary grid resolution for simulations conducted at $f_k=0.25$ and $0.40$ is approximately 4 and 8 times coarser than at $f_k=0.10$, $r_\Delta=4$ and $8$. Considering that SRS computations are inherently three dimensional and time-dependent, this simple analysis illustrates the potential of bridging formulations to reduce the cost of SRS calculations. Also, it emphasizes the importance of selecting adequate physical resolution for a given problem and quantities of interest. 

The third aspect stems from the fact that the physical resolution often affects the turbulence dependent quantities differently, i.e., a given range of resolved scales is not expected to lead to equal ratios of modeled-to-total of all turbulence dependent quantities. Considering the case of {\color{blue}BHR-LEVM} closure, it is well-known that the spectral properties of $k_t$ and $\varepsilon_t$ are distinct \cite{POPE_BOOK_2000} so that $f_k$ and $f_\varepsilon$ should only be equal in the limit of all scales being modeled or resolved. Although most SRS formulations neglect this crucial feature, PANS can account for its effects by considering the ratio modeled-to-total ratio, $f_\phi$, of each turbulent dependent quantity. However, the magnitude of these parameters needs to be determined.

As previously mentioned, the current version of PANS {\color{blue}BHR-LEVM} relies on $f_k$ and $f_\varepsilon$ to set the model's physical resolution. The parameter $f_k$ can be either defined constant \cite{GIRIMAJI_JAM_2005,GIRIMAJI_AIAA43_2005,LAKSHMIPATHY_JFE_2010,PEREIRA_IJHFF_2018} or dynamically \cite{GIRIMAJI_AIAA43_2005,ELMILIGUI_AIAA22_2004,BASARA_IJHFF_2018,DAVIDSON_JOT_2019} in space and time. Whereas the first approach prevents commutation errors and the entanglement of numerical and modeling errors, the second can enhance the computations' efficiency. In this work, we choose to conduct all PANS simulations at constant $f_k$ to avoid commutation errors and perform robust verification and validation exercises. On the other hand, $f_\varepsilon$ is usually defined as constant and equal to one. This option has been motivated by the fact that most turbulence dissipation at high-Re flows is expected to occur at the smallest and dissipative scales \cite{POPE_BOOK_2000,DAVIDSON_BOOK_2006}. Thus, $f_\varepsilon=1.00$ is generally accepted as a reasonable assumption for bridging models operating between RANS and LES. Yet, this assumption has not been verified.

To confirm its validity for practical turbulent flows, we perform \textit{a-priori} testing to calculate $f_k$ and $f_\varepsilon$ at successively smaller physical resolutions - ranging from DNS ($f_k=0.00$) to RANS ($f_k=1.00$). The selected problem is the forced homogeneous turbulent flow at Taylor Reynolds numbers Re$_\lambda = 140$ and $300$ predicted by Silva et al. \cite{SILVA_JFM_2018} utilizing direct numerical simulations. $k_u$ and $\varepsilon_u$ are computed as follows.  The velocity field of the data sets of Silva et al. \cite{SILVA_JFM_2018} is first filtered by applying the following spatial filtering operator \cite{SILVA_PHD_2001,SILVA_JT_2008},
\begin{equation}
\label{eq:2.3_6}
\langle \Phi  \rangle (\mathbf{x})= \int_{-\Delta/2}^{+\Delta/2}\int_{-\Delta/2}^{+\Delta/2}\int_{-\Delta/2}^{+\Delta/2} \Phi(\mathbf{x}) \ G_\Delta(\mathbf{x}-\mathbf{x'})d\mathbf{x'} \; ,
\end{equation}
where bold symbols denote vectors, and $\Delta$ and $G_\Delta$ are the filter's width and kernel, respectively. In this exercise, we use a box filter so that
\begin{equation}
\label{eq:2.3_7}
G_\Delta(\mathbf{x}-\mathbf{x'})= 
\left\{
\begin{array}{lll}
\Delta^{-1}	&\; , & |\mathbf{x-x'}|< 0.5 \Delta \\
0 			&\; , & \text{otherwise}
\end{array}\right.
\; .
\end{equation}
As discussed in da Silva and Pereira \cite{SILVA_POF_2007}, Borue and Orzag \cite{BORUE_JFM_1998}, and Liu et al. \cite{LIU_JFM_1994}, the utilization of box and Gaussian filters is not expected to influence the results of this class of studies. In contrast, cut-off filters are not suitable for this exercise since practical SRS computations do not rely on such operators \cite{VREMAN_JFM_1994,SILVA_POF_2007}. It is also generally accepted that the box filtering operator is the closest approach to the implicit filtering of finite-difference or -volume discretization schemes utilized in engineering computations \cite{SILVA_JT_2008,SCHUMANN_JCP_1975,ROGALLO_ARFM_1984}. $k_u$ and $\varepsilon_u$ are computed from relation \ref{eq:2_9} \cite{GERMANO_JFM_1992},
\begin{equation}
\label{eq:2.3_8}
{\color{blue}
k_u= 0.5 \left( \langle V_i V_i \rangle - \langle V_i \rangle\langle V_i \rangle\right) \; ,
}
\end{equation}
\begin{equation}
\label{eq:2.3_9}
{\color{blue}
\varepsilon_u= \nu  \left( \left\langle \frac{\partial V_i}{\partial x_j} \frac{\partial V_i}{\partial x_j}\right\rangle -  \left\langle \frac{\partial V_i}{\partial x_j}\right\rangle \left\langle \frac{\partial V_i}{\partial x_j}\right\rangle \right) \; .
}
\end{equation}
These quantities have been calculated with $n=\Delta/\Delta x$ up to $349$, with $\Delta x$ the grid size of the DNS simulations.
\begin{figure}
\centering
\subfloat[$f_k(n )$.]{\label{fig:2.3_2a}
\includegraphics[scale=0.11,trim=0 0 0 0,clip]{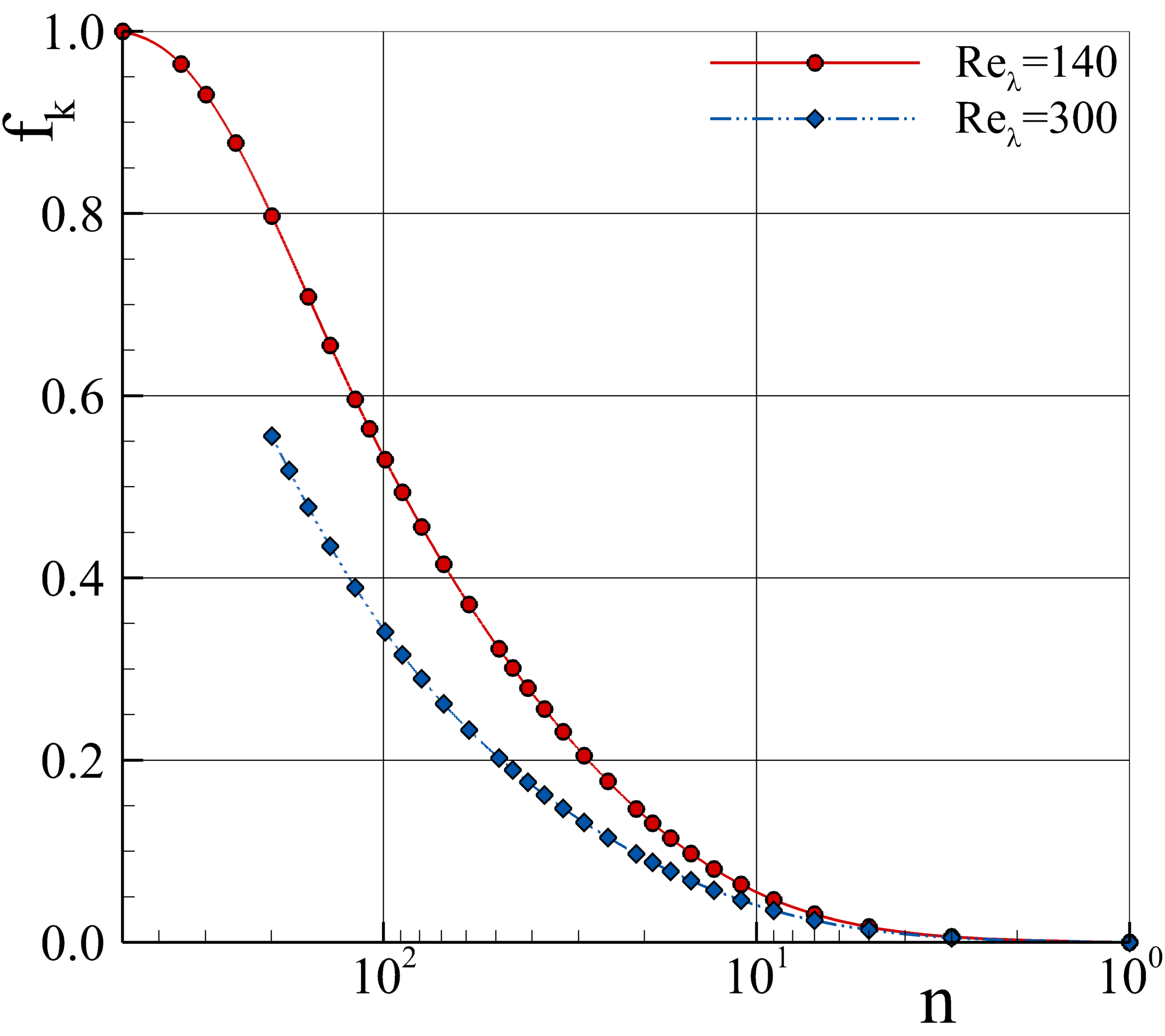}}
\\
\subfloat[$f_\varepsilon(n)$.]{\label{fig:2.3_2b}
\includegraphics[scale=0.11,trim=0 0 0 0,clip]{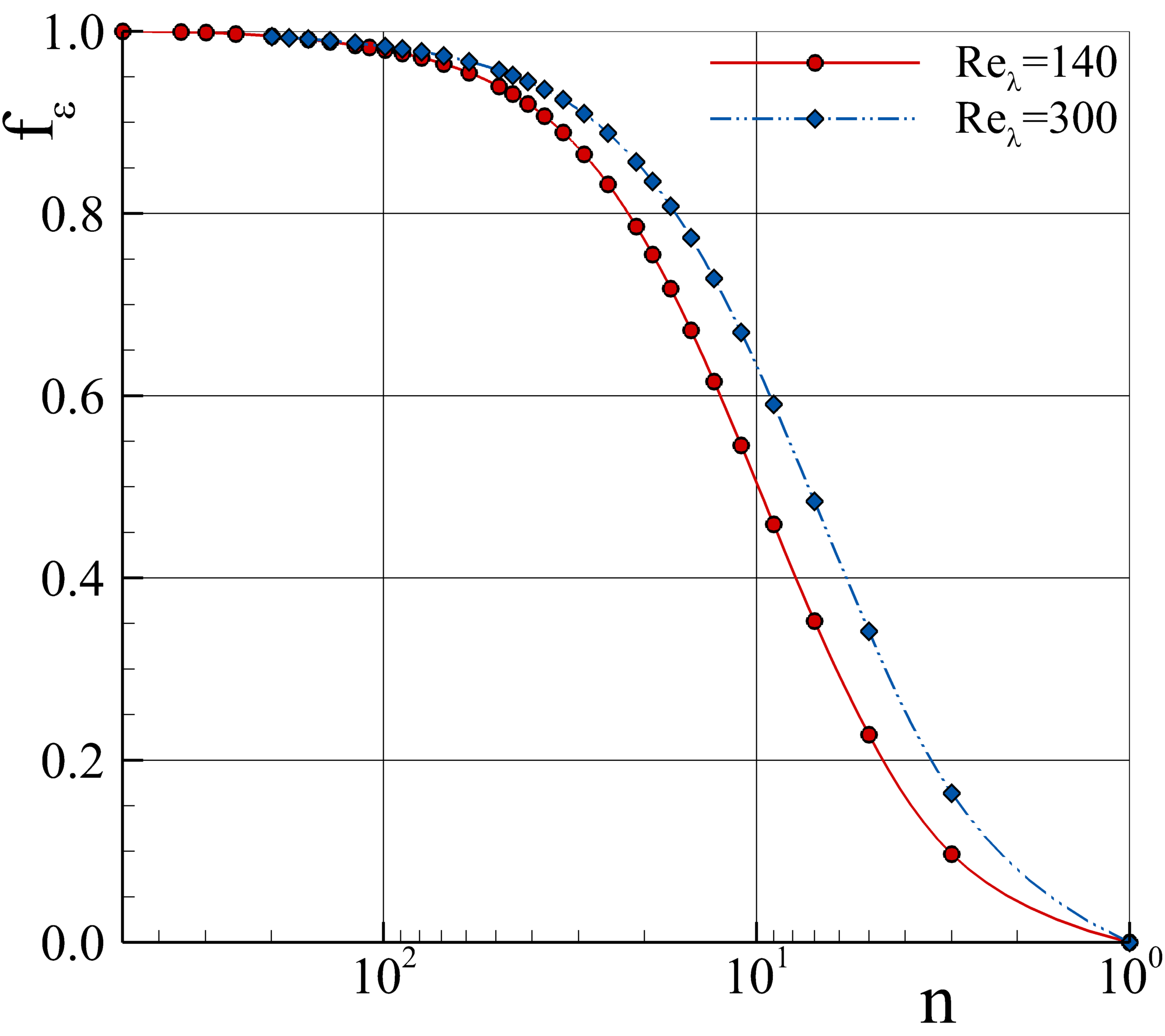}}
\caption{Evolution of $f_k$ and $f_\varepsilon$ with the relative filter size, $n$, at $\mathrm{Re}_\lambda=140$ and $300$.}
\label{fig:2.3_2} 
\end{figure}

The variation of $f_k$ and $f_\varepsilon$ with $n$ is depicted in figure \ref{fig:2.3_2}. Referring to $f_k(n)$, the results indicate that most turbulence kinetic energy is at the largest scales. Also, this tendency becomes more pronounced with $\mathrm{Re}_\lambda$. For instance, the data shows that $\mathrm{Re}_\lambda=140$ and $300$ require $n=29$ and $49$ to filter $20\%$ of $k_t$ ($f_k=0.20$). This outcome highlights the potential of bridging models to reduce the computing cost of simulations of turbulence. It is also interesting to observe that from $f_k=0.00$ to $0.20$, the grid resolution/filter size grows more than one order of magnitude. In contrast, figure \ref{fig:2.3_2b} shows that most turbulence dissipation occurs at the smallest scales and, as such, $f_\varepsilon$ grows more rapidly with $n$ than $f_k$. The data indicates that $f_\varepsilon=0.22$ ($\mathrm{Re}_\lambda=140$) and $0.34$ ($\mathrm{Re}_\lambda=300$) for $n=5$. For $n=199$, this parameter exceeds $0.99$ for both $\mathrm{Re}_\lambda$.

\begin{figure}
\centering
\includegraphics[scale=0.11,trim=0 0 0 0,clip]{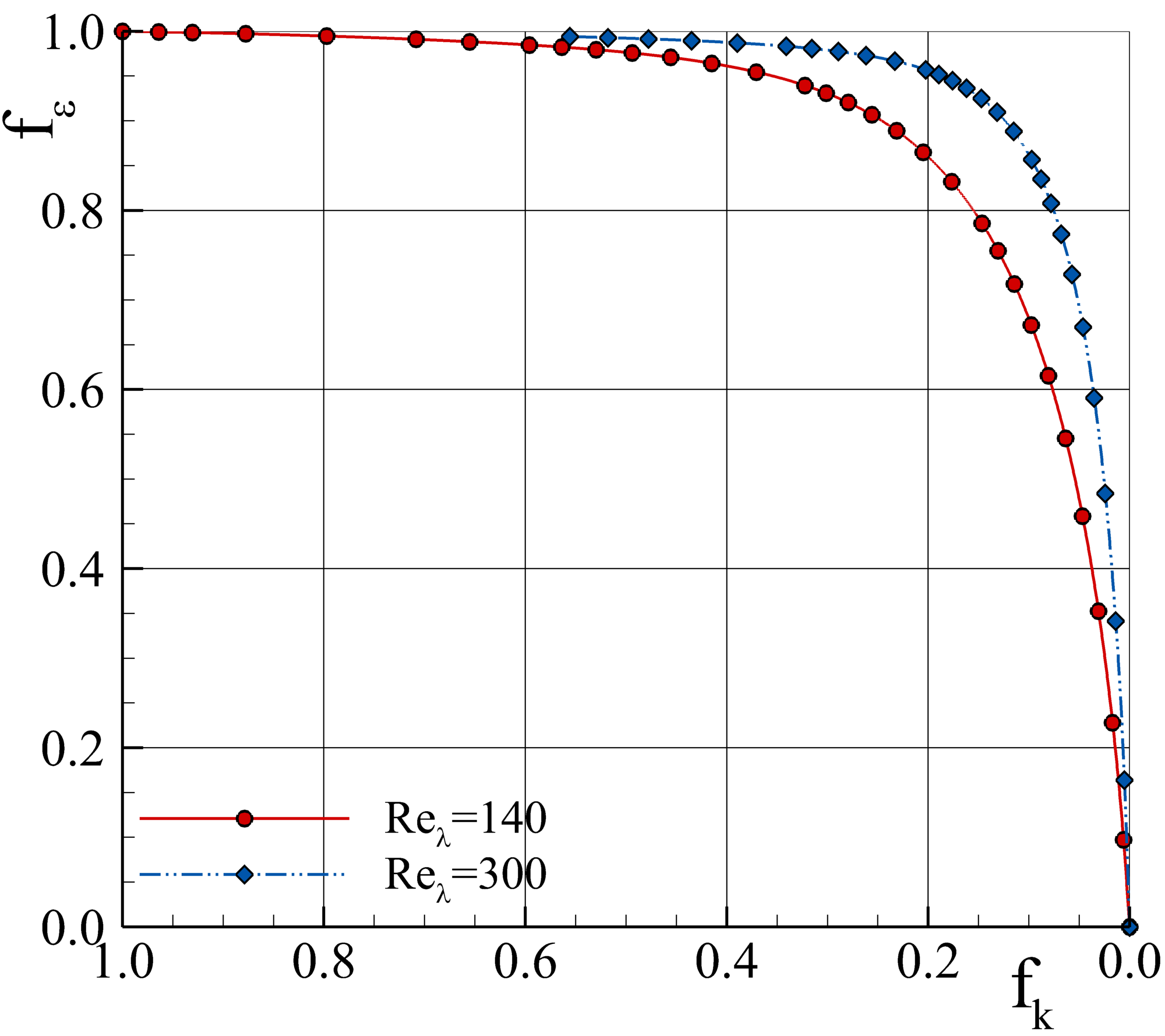}
\caption{Evolution of $f_k$ with $f_\varepsilon$ at $\mathrm{Re}_\lambda=140$ and $300$.}
\label{fig:2.3_3} 
\end{figure}

Next, figure \ref{fig:2.3_3} plots $f_\varepsilon$ as a function of $f_k$ for both Reynolds numbers. It is observed  that once $f_k=0.20$, $f_\varepsilon=0.87$ for $\mathrm{Re}_\lambda=140$, and $f_\varepsilon=0.93$ for $\mathrm{Re}_\lambda=300$. In this manner, and considering that practical SRS are not intended to operate at $f_k<0.20$ due to the associated cost, the data confirm that assuming $f_\varepsilon=1.00$ is a reasonable assumption for practical PANS calculations.  

Notwithstanding these results, it is important to mention that $f_k$ and $f_\varepsilon$ might be affected by the dynamics of transient flow problems, reducing the differences between these parameters at high physical resolutions. In these cases, the exact selection of such parameters would ideally require a dynamic approach. Yet, and in addition to commutation errors, the development of such strategy is complex for transient flows owing to their dependence on the flow properties, time-instant, and zone. For these reasons, we prefer using constant values of $f_k$ and $f_\varepsilon=1.00$ even if at the possible expense of further reducing $f_k$ to compensate calibration deficits at early flow instants.

%
%
%
\section{Problem Setup}
\label{sec:3}
\begin{figure}
\centering
\includegraphics[scale=0.110,trim=0 0 0 0,clip]{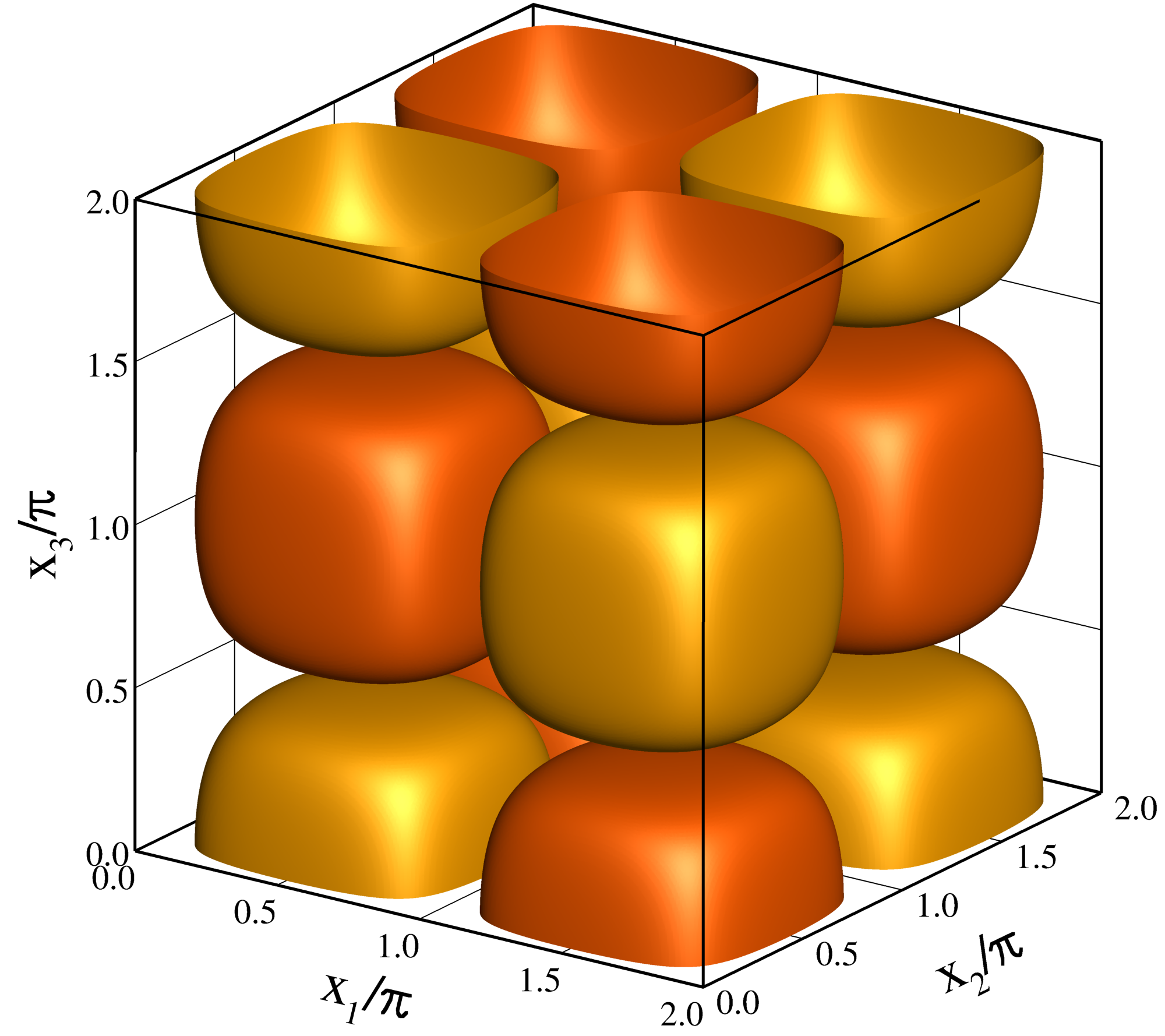}
\caption{Iso-surfaces of the vorticity $x_3$-component at $t=0$.}
\label{fig:3_1} 
\end{figure}
The TGV \cite{TAYLOR_PRSA_1937} is a paradigmatic test-case used to study modeling and simulation of onset, development and decay of turbulence. This problem is initially characterized by several laminar, well-characterized, and single-mode vortices, figure \ref{fig:3_1}. These structures evolve and interact in time, this leading to five major flow stages: \textit{i)} vortex stretching mechanisms generate vortex sheets which gradually get closer; \textit{ii)} vortex sheets roll-up and reconnect \cite{KIDA_ARFM_1994,YANG_JFM_2011}; \textit{iii)} onset of turbulence and subsequent intensification of vorticity; \textit{iv)} coherent structures breakdown and fully developed turbulence; and \textit{v)} turbulence decay.
\begin{table*}
\centering
\setlength\extrarowheight{3pt}
\caption{Initial flow and thermodynamic fluid properties.}
\label{tab:3_1}    
\begin{tabular}{C{1.5cm}C{1.8cm}C{1.8cm}C{1.8cm}C{1.8cm}C{1.8cm}C{1.8cm}C{1.8cm}C{1.8cm}}
\hline 
$\mathrm{Ma}_o$ & $V_o$(cm/s) & $L_o$(cm)  & $\rho_o$ (g/cm$^3$) & $P_o$ (Pa) & $\mu$ (g/(cm.s)) & $\gamma$  & $k_o$ (cm$^2$/s$^2$) & $S_o$ (cm)\\[3pt]
\hline 
$0.28$ & $10^4$ & $1.00$  & $1.178\times 10^{-3}$ & $10^5$ & $3.927\times 10^{-3}$ & {\color{blue}$1.40$} & $10^{-7}$ & $6.136\times 10^{-3}$\\ [3pt]
\hline
\end{tabular}
\end{table*}

The initial velocity field of the TGV problem is given by \cite{TAYLOR_PRSA_1937,BRACHET_JFM_1983},
\begin{equation}
\label{eq:3_1}
V_1(\mathbf{x},t_o)= V_o \sin\left(x_1\right)\cos\left(x_2 \right) \cos \left(x_3\right)\; ,
\end{equation}
\begin{equation}
\label{eq:3_2}
V_2(\mathbf{x},t_o)= -V_o \cos\left(x_1\right)\sin\left(x_2 \right) \cos \left(x_3\right)\; ,
\end{equation}
\begin{equation}
\label{eq:3_3}
V_3(\mathbf{x},t_o)= 0\; ,
\end{equation}
where $V_o$ is the initial velocity magnitude. Figure \ref{fig:3_1} illustrates the initial flow field. The pressure is obtained from solving the Poisson equation,
\begin{equation}
\label{eq:3_4}
P(\mathbf{x},t_o)=P_o + \frac{\rho_o V_o^2}{16} \left[ 2 +\cos \left( 2 x_3 \right) \right] \left[ \cos \left( 2 x_1 \right) +\cos \left( 2 x_2 \right) \right]\; ,
\end{equation}
where $P_o$ and $\rho_o$ stand for the pressure and density at the initial time instant, $t=0$. The calculations are conducted with the compressible xRage solver \cite{GITTINGS_CSD_2008} using an ideal gas equation of state and setting the initial Mach number, $\mathrm{Ma}_o$, equal to $0.28$. This leads to maximum instantaneous and averaged ($L_1$ norm) variations of $\rho$ smaller than $11.0\%$ and $1.4\%$ of $\rho_o$ for $f_k=0.00$, respectively. The Reynolds number, Re=$\rho L_o V_o/\mu$, is set equal to 3000 to replicate the simulations of Brachet et al. \cite{BRACHET_JFM_1983}. The initial flow and thermodynamic fluid properties are summarized in table \ref{tab:3_1}.

xRage utilizes a finite volume approach to solve the compressible and multi-material conservation equations for mass, momentum, energy, and species concentrations. The resulting system of governing equations is resolved through the HLLC \cite{TORO_SW_1994} Riemann solver using a directionally unsplit strategy, direct remap, parabolic reconstruction \cite{COLLELA_JCP_1987}, and the low Mach number correction proposed by Thornber et al. \cite{THORNBER_JCP_2008}. The equations are discretized with second-order accurate methods: the temporal discretization relies on the explicit Runge-Kutta scheme known as Heun's method, whereas the spatial discretization is based on a Godunov scheme. The time-step is defined by imposing a maximum CFL number,
\begin{equation}
{\color{blue}\Delta t = \frac{\Delta x \times \text{CFL}}{ 3  (|V| + c)} }\; ,
\end{equation}
equal to 0.45, where $c$ is the speed of sound. The modeling of high convection-driven flows is performed with a van-Leer limiter \cite{LEER_JCP_1997}, without artificial viscosity. The thermal flux, $q^c$, of the PANS equations is neglected due to its negligible importance to the selected problem. 

In order to mimic the original problem setup \cite{TAYLOR_PRSA_1937}, the computational domain is a cube of width equal to $2\pi$cm (figure \ref{fig:3_1}). The computations are conducted using four grid resolutions, $N_c=128^3$, $256^3$, $512^3$, and $1024^3$ cells. PANS simulations are performed at six different and constant values of $f_k$: 0.00 (no closure), 0.25, 0.35, 0.50, 0.75 and 1.00 (RANS). This strategy enables us to evaluate the dependence of PANS on the range of resolved scales, enable rigorous verification and validation errors, and prevent commutation errors \cite{HAMBA_PF_2011}. $f_\varepsilon$, on the other hand, is set equal to one (see section \ref{sec:2.3}). The initial conditions of PANS {\color{blue}BHR-LEVM} dependent variables are set as $k=k_o f_k$ and $S=S_o f_k^{1.5}/f_\varepsilon$, where $k_o$ and $S_o$ are given in table \ref{tab:3_1}. The simulations are compared against those of Brachet et al. \cite{BRACHET_JFM_1983} and Drikakis et al. \cite{DRIKAKIS_JOT_2007}. It is important to note that these reference studies do not report estimates of numerical uncertainty, and Brachet et al. \cite{BRACHET_JFM_1983} seem to use a computational domain with a length of $\pi L_o$ instead of $2\pi L_o$ and an incompressible flow solver.
%
%
%
\section{Results and Discussion}
\label{sec:4}

We initiate the discussion of the results in Section \ref{sec:4.1} with a succinct description of the temporal evolution of the TGV flow. This constitutes a crucial step to understand the physics of the numerical results.  Afterward, Section \ref{sec:4.2} discusses  the effect of the physical resolution on the numerical and modeling accuracy of PANS calculations. This is accomplished through simple verification and validation exercises using the DNS data of Brachet et al. \cite{BRACHET_JFM_1983} and Drikakis et al. \cite{DRIKAKIS_JOT_2007}. The discussion of the results also includes their physical interpretation in Section \ref{sec:4.3}, and concludes with the comparison of the simulations' cost in Section \ref{sec:4.4}. {\color{blue}Unless mentioned otherwise, all results are normalized by $\nu$, $V_o$, and $L_o$}.
%
%
%
\subsection{Temporal flow evolution}
\label{sec:4.1}

\begin{figure*}[t!]
\centering
\subfloat[$t=0.0$.]{\label{fig4.1_1a}
\includegraphics[scale=0.075,trim=0 0 0 0,clip]{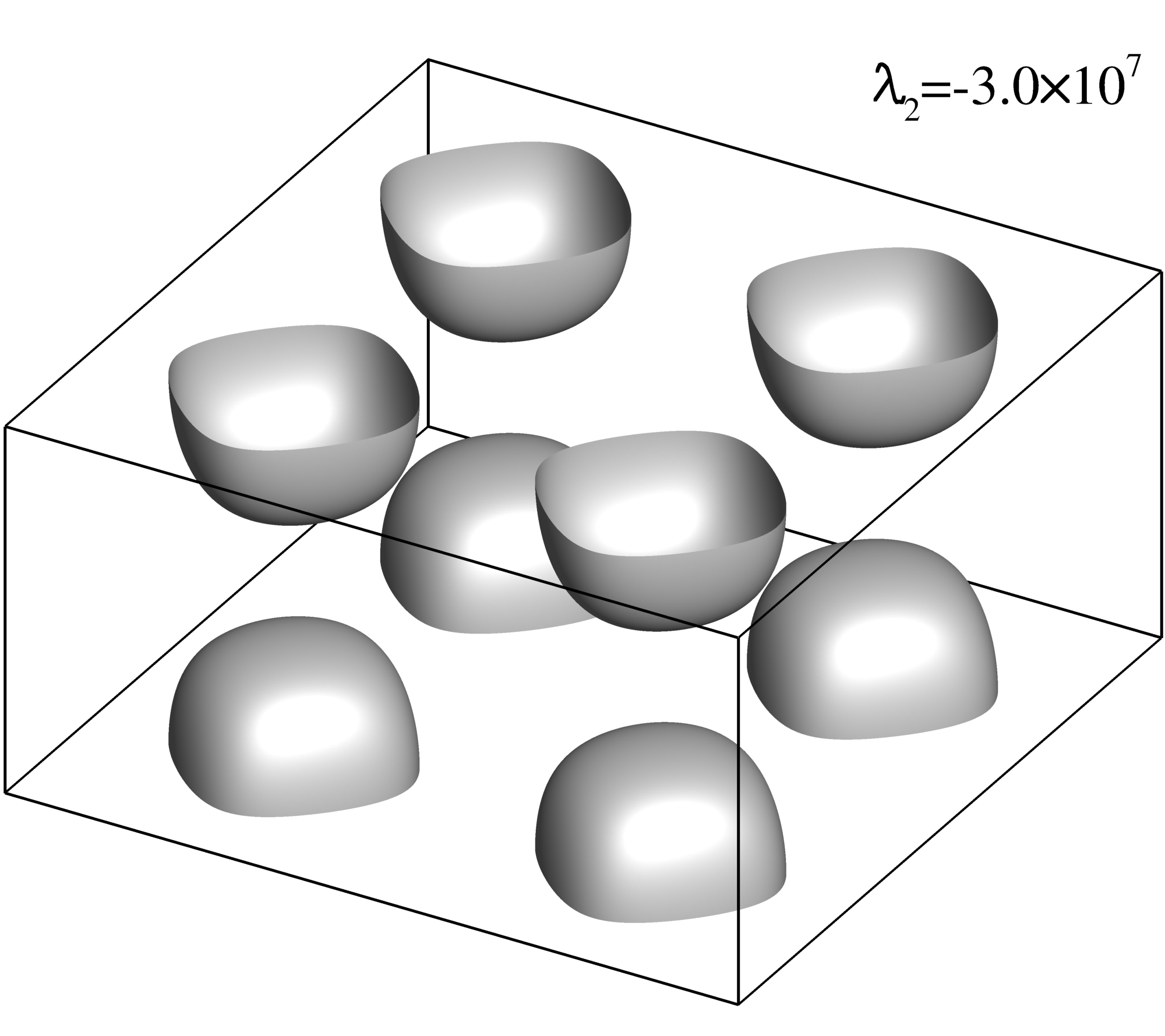}}
~
\subfloat[$t=1.0$.]{\label{fig4.1_1b}
\includegraphics[scale=0.075,trim=0 0 0 0,clip]{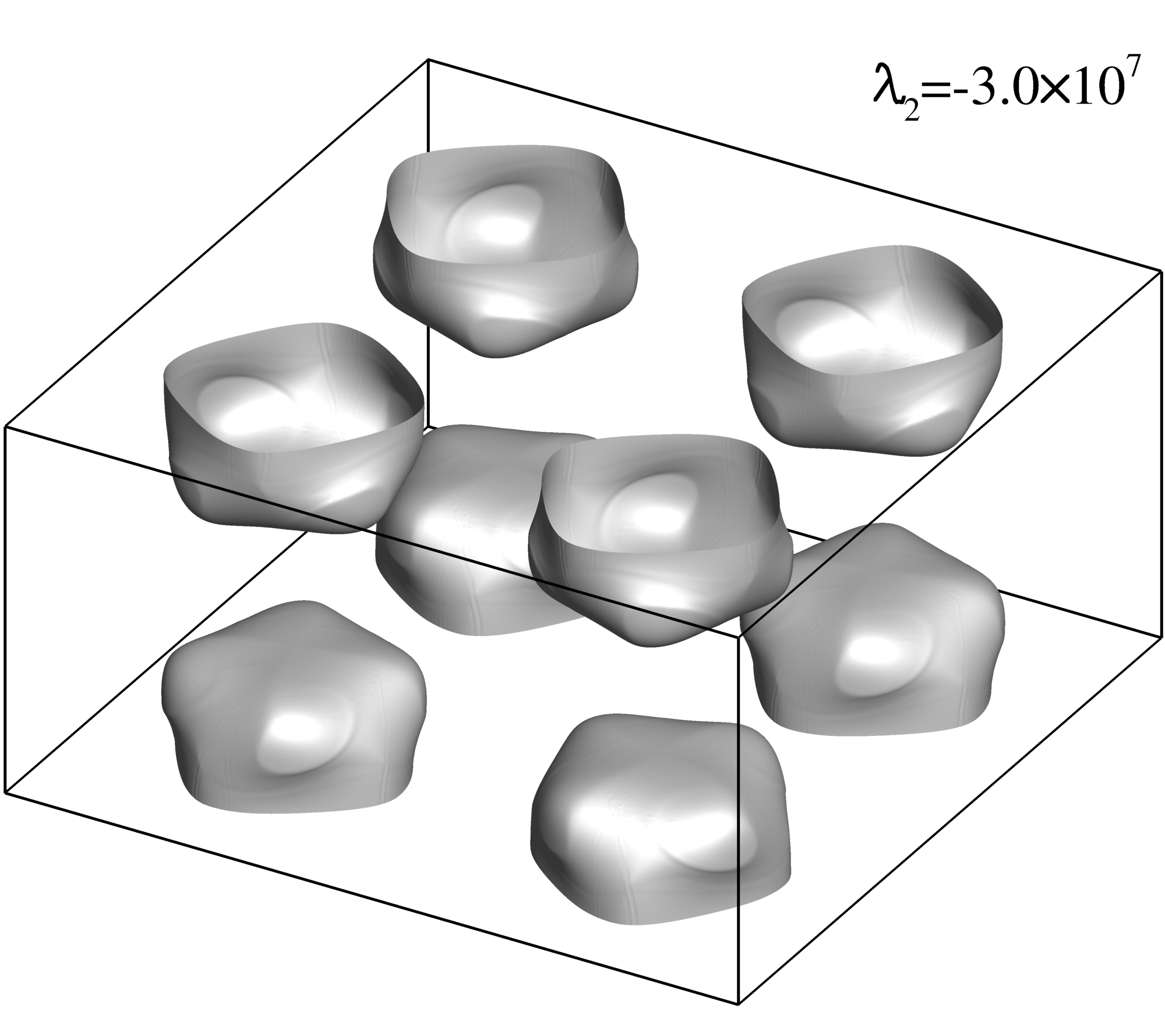}}
~
\subfloat[$t=2.0$.]{\label{fig4.1_1c}
\includegraphics[scale=0.075,trim=0 0 0 0,clip]{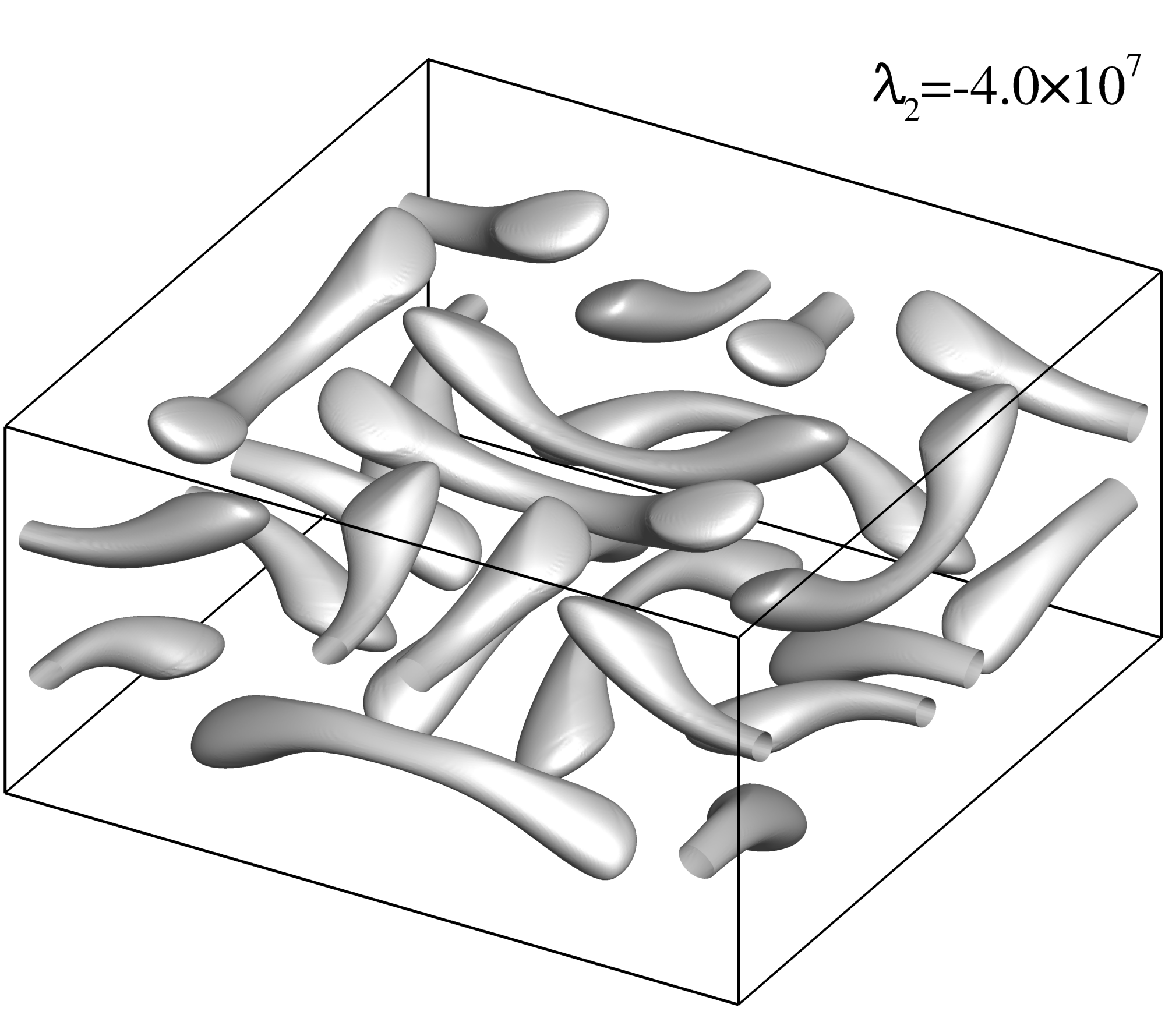}}
\\
\subfloat[$t=3.0$.]{\label{fig4.1_1d}
\includegraphics[scale=0.075,trim=0 0 0 0,clip]{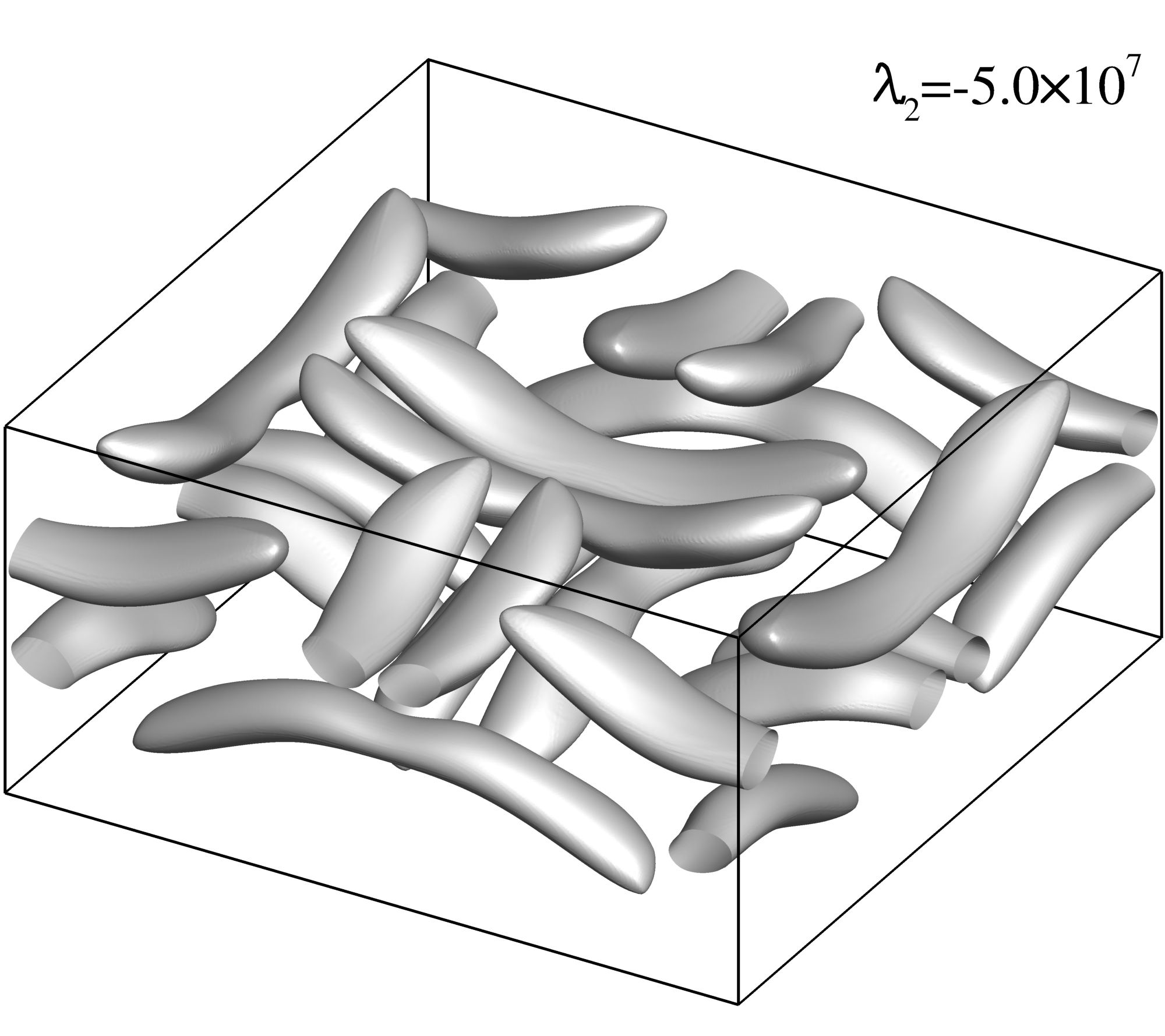}}
~
\subfloat[$t=4.0$.]{\label{fig4.1_1e}
\includegraphics[scale=0.075,trim=0 0 0 0,clip]{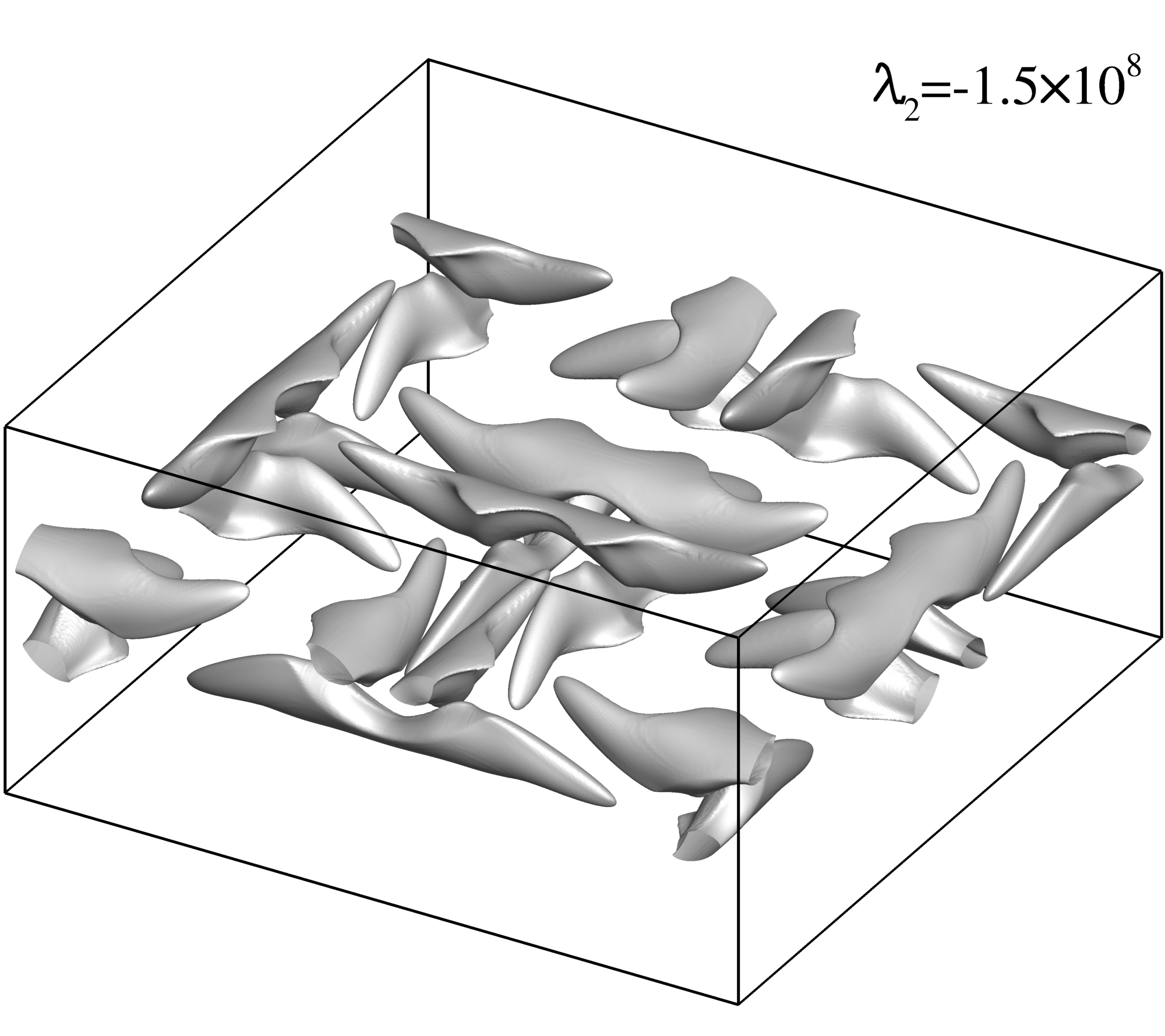}}
~
\subfloat[$t=5.0$.]{\label{fig4.1_1f}
\includegraphics[scale=0.075,trim=0 0 0 0,clip]{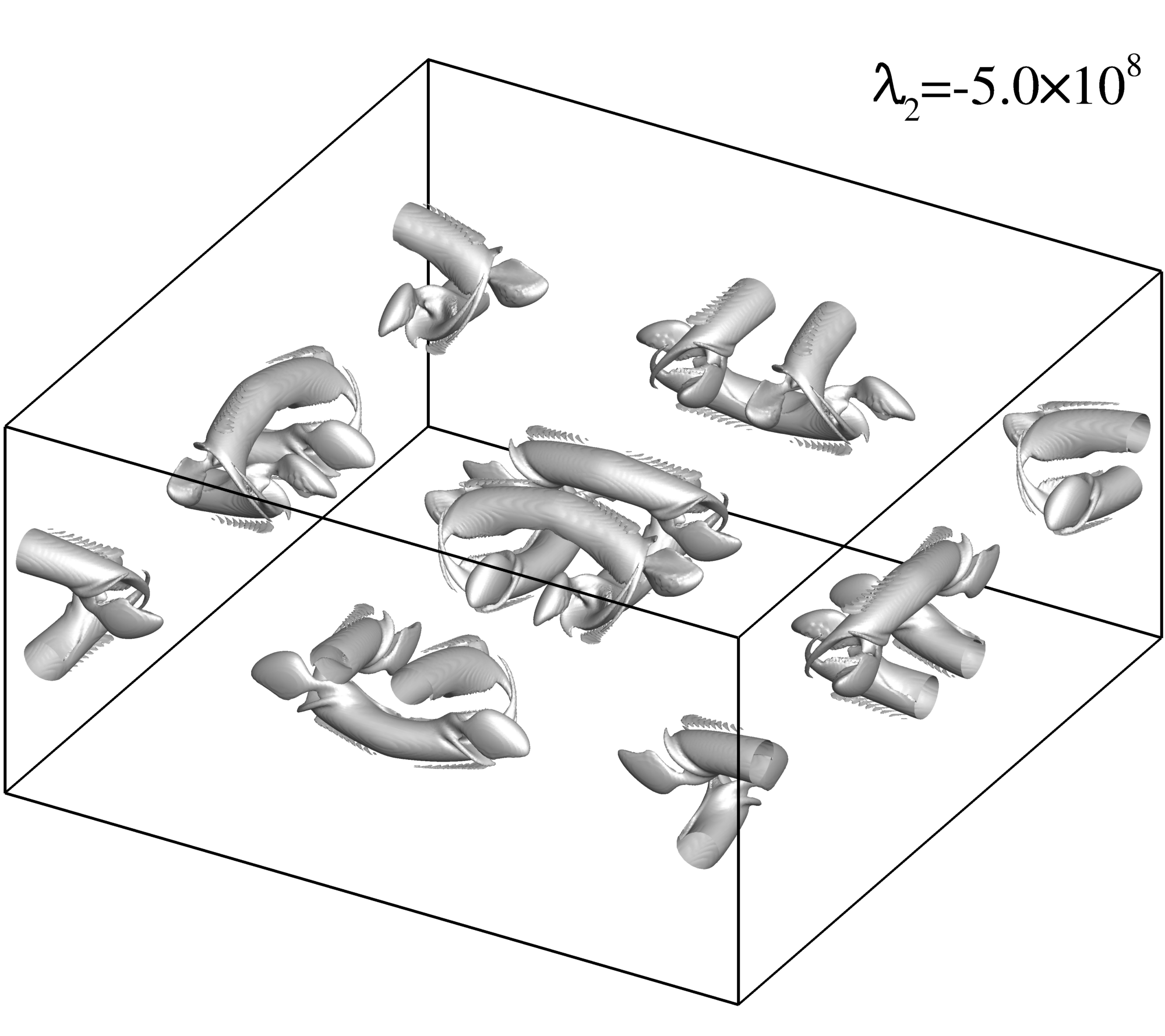}}
\\
\subfloat[$t=6.0$.]{\label{fig4.1_1g}
\includegraphics[scale=0.075,trim=0 0 0 0,clip]{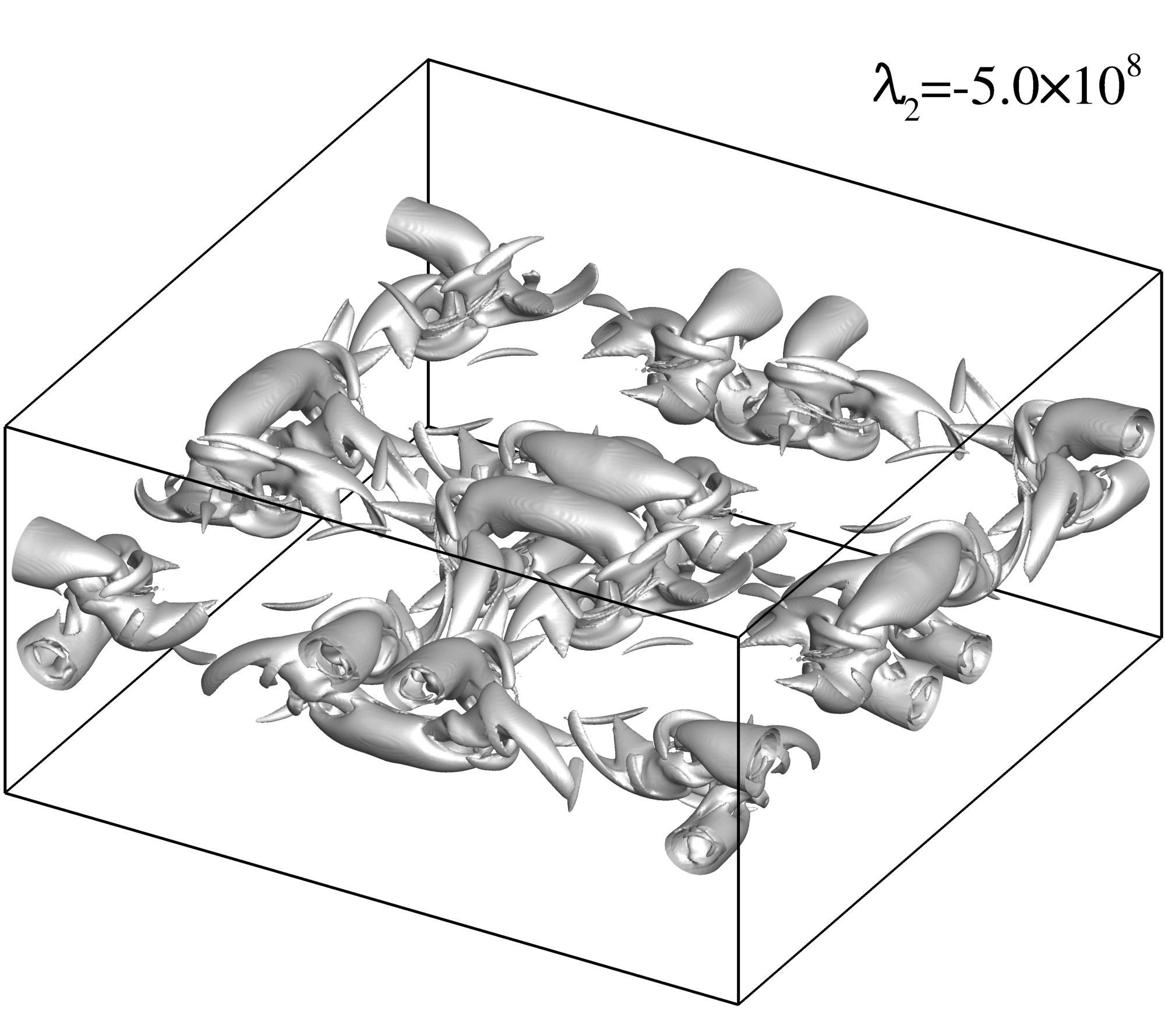}}
~
\subfloat[$t=7.0$.]{\label{fig4.1_1h}
\includegraphics[scale=0.075,trim=0 0 0 0,clip]{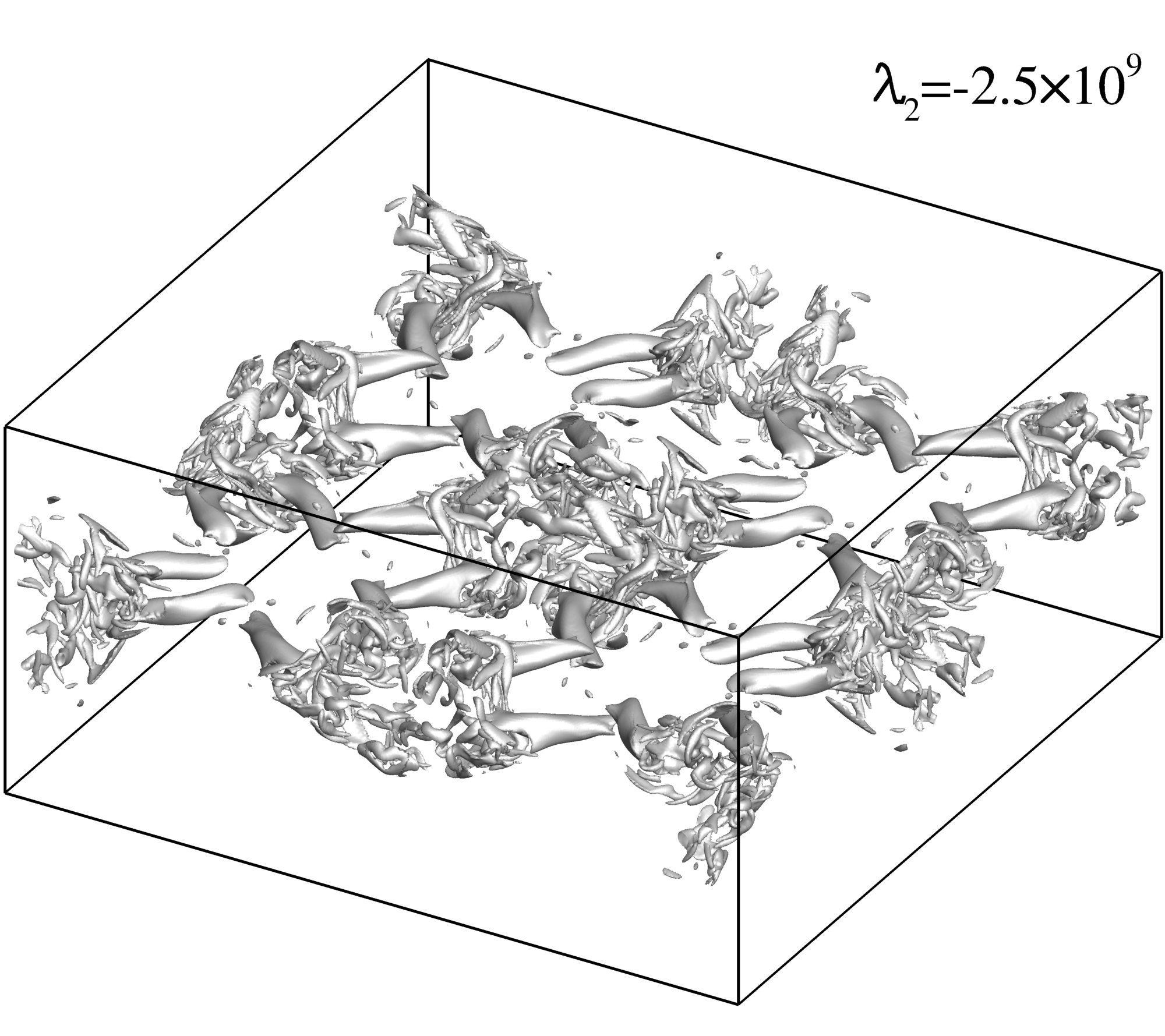}}
~
\subfloat[$t=9.0$.]{\label{fig4.1_1i}
\includegraphics[scale=0.075,trim=0 0 0 0,clip]{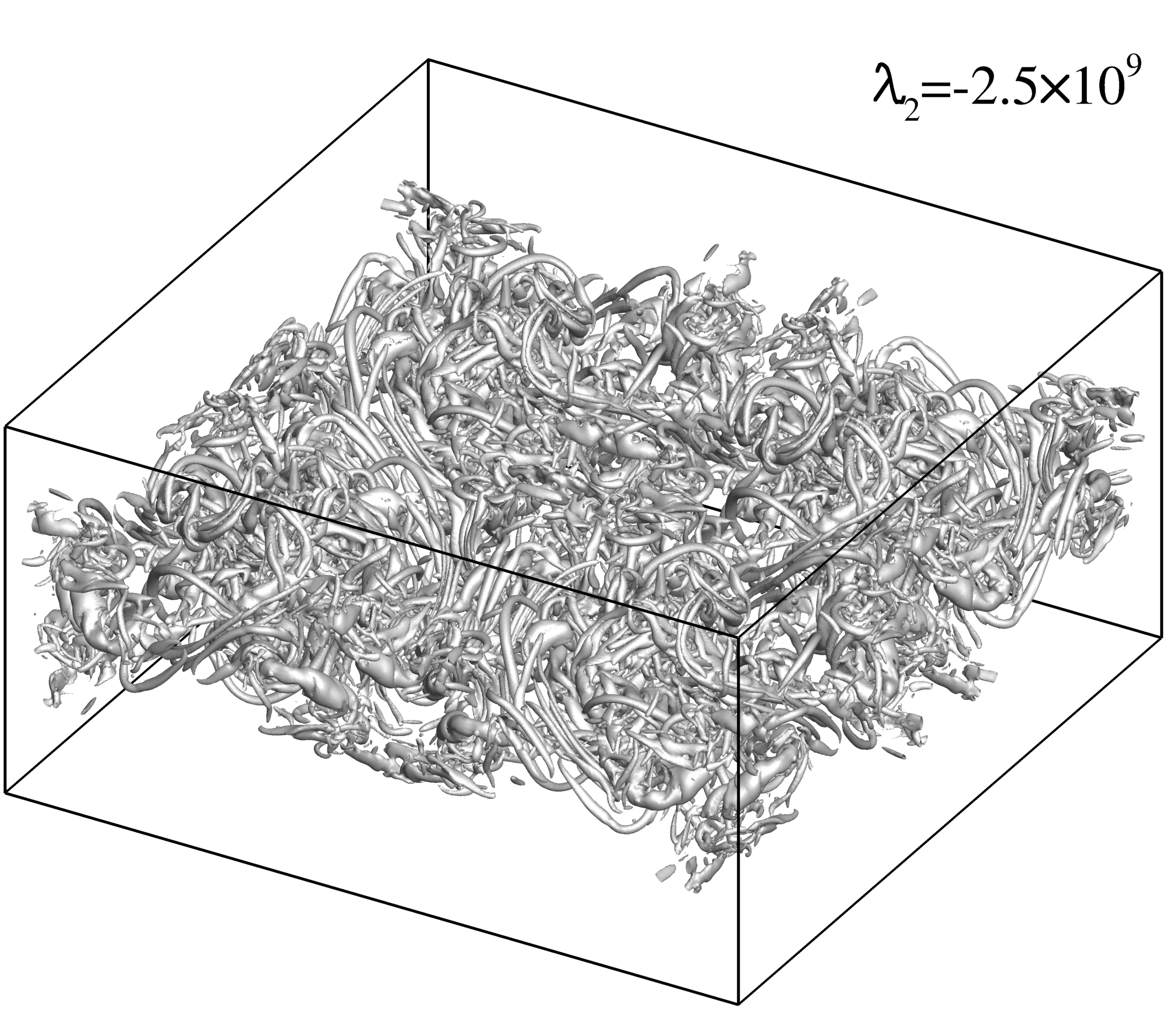}}
\\
\subfloat[$t=12.0$.]{\label{fig4.1_1j}
\includegraphics[scale=0.075,trim=0 0 0 0,clip]{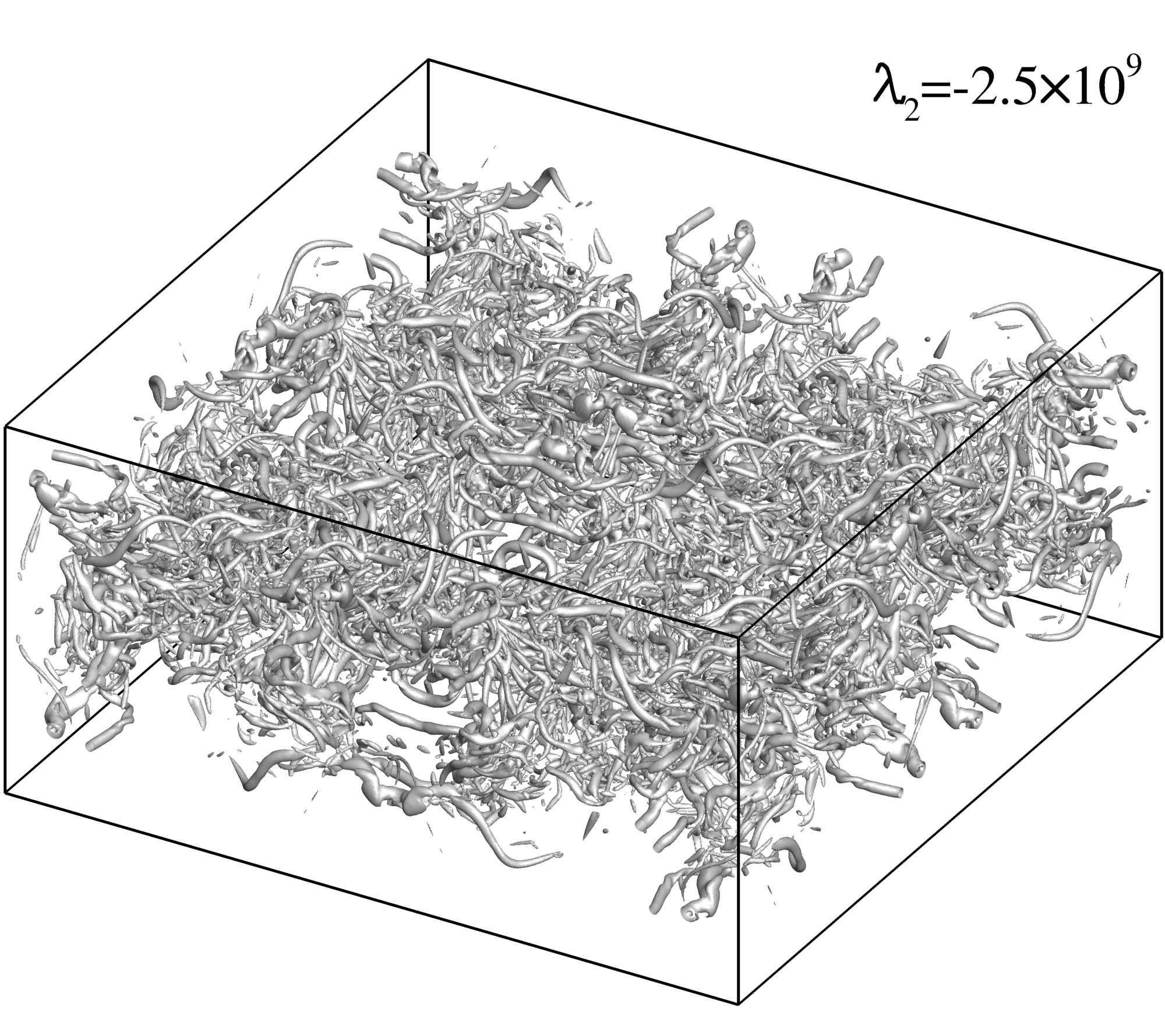}}
~
\subfloat[$t=16.0$.]{\label{fig4.1_1k}
\includegraphics[scale=0.075,trim=0 0 0 0,clip]{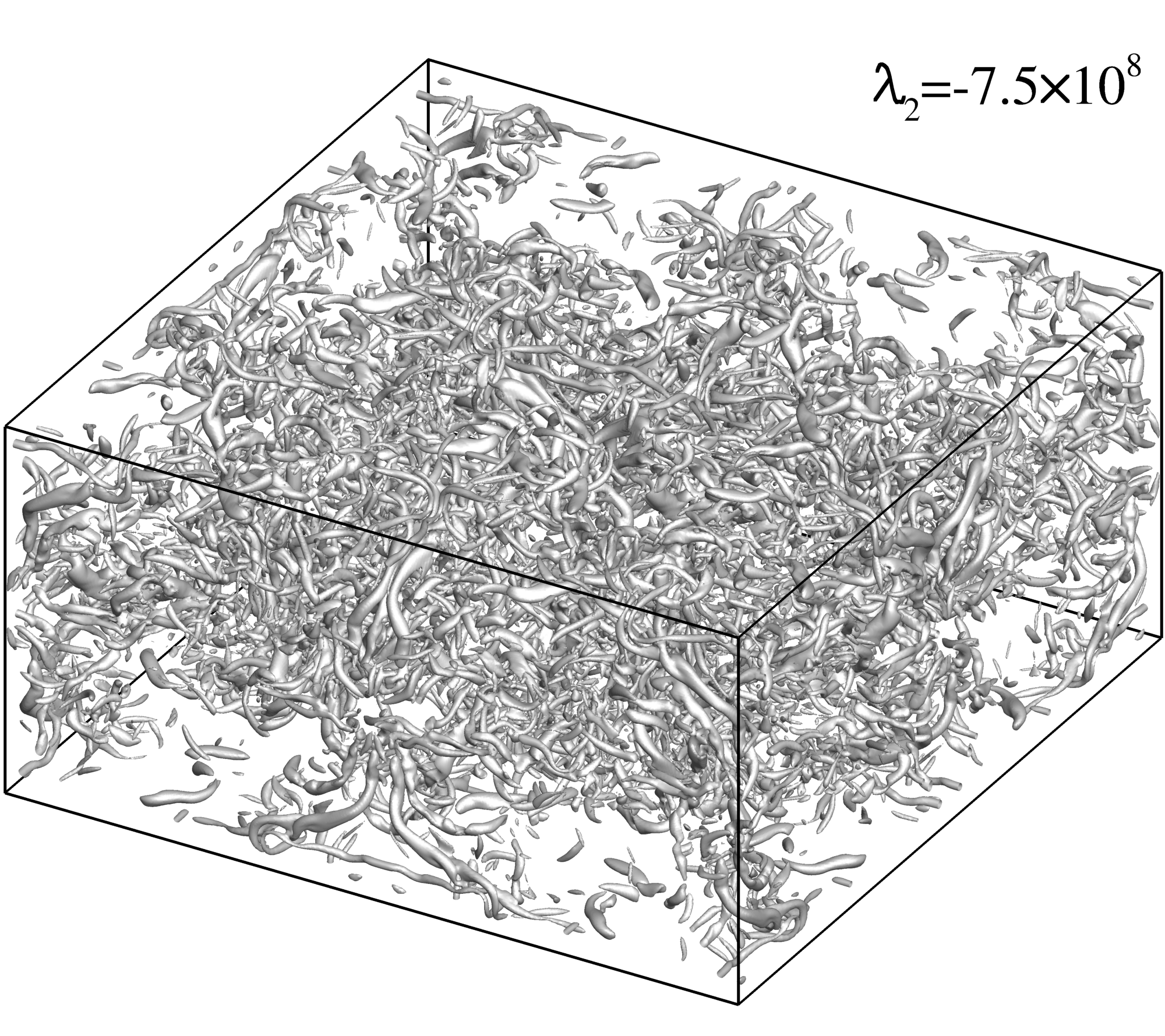}}
~
\subfloat[$t=20.0$.]{\label{fig4.1_1l}
\includegraphics[scale=0.075,trim=0 0 0 0,clip]{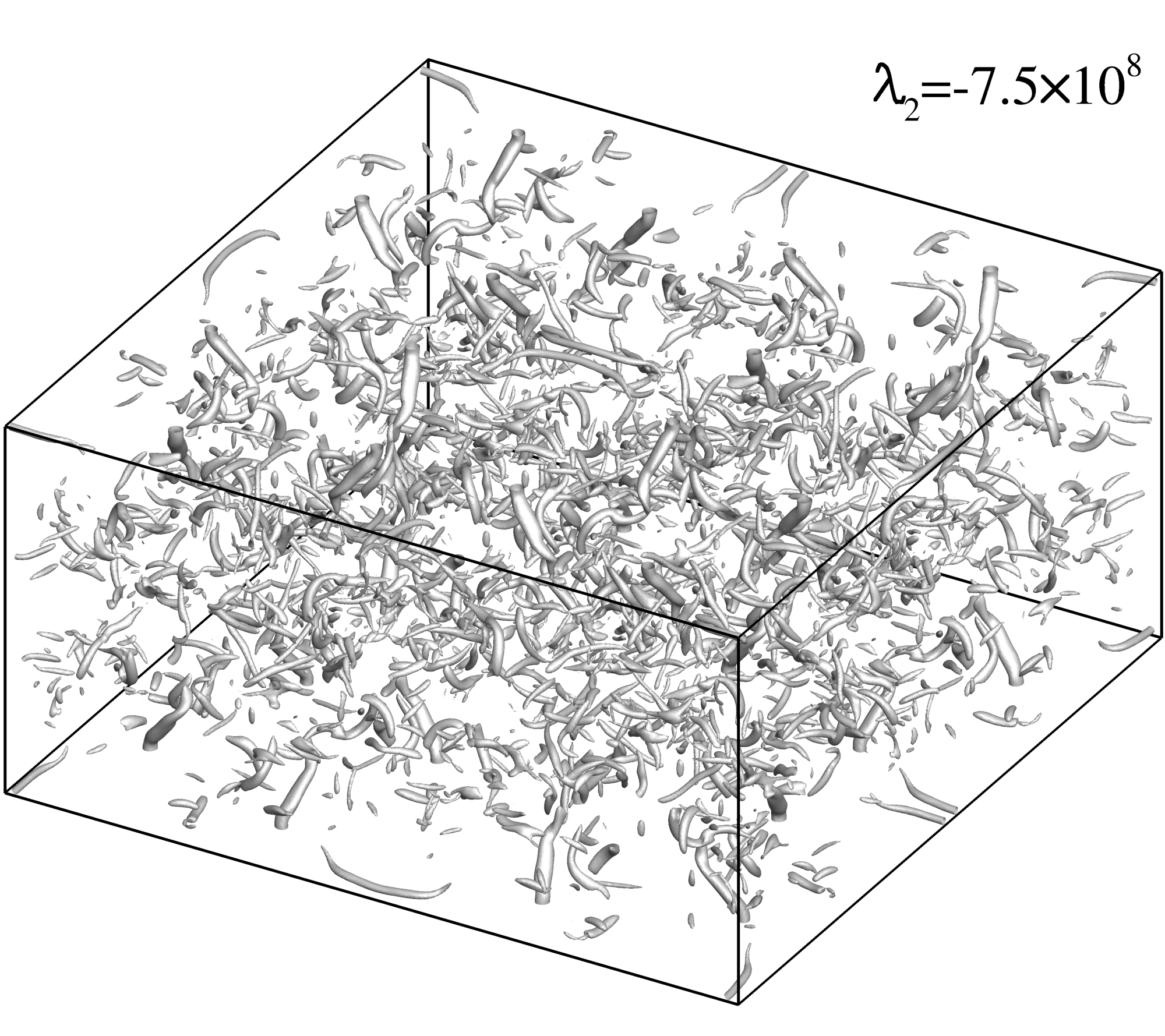}}

\caption{Temporal evolution of the coherent and turbulent structures of the TGV flow predicted with PANS using $f_k=0.00$ and $N_c=512^3$. Vortical structures identified with the $\lambda_2$-criterion {\color{blue}($\mathrm{s^{-2}}$)} \cite{JEONG_JFM_1995}.}
\label{fig:4.1_1}
\end{figure*}

The TGV is a transient flow problem which includes laminar flow, instabilities and coherent structures, onset and development of turbulence, and rapid kinetic energy decay due to {\color{blue}high-intensity turbulence}. To understand the multiple stages of the TGV flow, figure \ref{fig:4.1_1} depicts the temporal evolution of the coherent field using the $\lambda_2$-criterion proposed by Jeong and Hussain \cite{JEONG_JFM_1995}. The present analysis was repeated with the vorticity magnitude and $Q$- \cite{HUNT_REP_1988} criteria to assess its dependence on the method used to identify the vortical structures of the flow \cite{JEONG_JFM_1995}. The results revealed that the three criteria tested do not alter the conclusions of the present analysis. For the sake of clearness and considering the symmetries of the TGV problem, figure \ref{fig:4.1_1} only depicts half of the computational domain, $0\leq x_3 \leq \pi$.

Figure \ref{fig:4.1_1} shows that the TGV flow is initially characterized by the set of laminar, well-defined, and symmetric vortices given by equations (\ref{eq:3_1}-\ref{eq:3_3}). After this initial time instant, $t=0$, the vortical structures interact and develop, this leading the loss of symmetry and deformation observed in figure \ref{fig4.1_1b} ($t=1.0$). In the following instants ($2 \leq t \leq 4$), the vortices continue evolving, and vortex-stretching processes lead to the generation of pairs of long sheet-like structures. These pairs of counter-rotating vortical structures get closer, and the resulting interactions lead to their rapid deformation ($t=4.0$).  From $t=5$ to $6$, it is possible to visualize a complex reconnection \cite{BRACHET_JFM_1983,KIDA_ARFM_1994} mechanism between pairs of counter-rotating vortices. Also, there is a general consensus \cite{CROW_AIAA_1970,BRACHET_JFM_1983,LAPORTE_AIAA_2002,YAO_JFMR_2020,LEWEKE_JFM_1998,LEWEKE_ARFM_2016}
that this process is driven by instabilities such as the Crow, elliptic, and possibly the Kelvin-Helmholtz instabilities. At $t=7.0$, the reconnection of the vortical structures triggers the onset of turbulence. Figure \ref{fig4.1_1h} illustrates the existence of regions featuring bursts of fine-scale turbulent flow. In the next instants, turbulence develops and extends throughout the domain as observed at $t=9$ in the vicinity of $x_3=\pi/2$. This result demonstrates that the turbulent field is not homogeneous due to its dependence on the vertical direction ($x_3$). At later times ($t\ge 12.0$), we can observe fine-scale turbulence in most of the computational domain which suggests fully-developed turbulent flow. This leads to the rapid dissipation of kinetic energy and flow structures observed in figures \ref{fig4.1_1k} to \ref{fig4.1_1l} ($t=12 \rightarrow 20$).

The analysis of figure \ref{fig:4.1_1} also indicates the existence of two distinct periods in the development of the TGV flow: \textit{i)} until $t=9$, the flow is characterized by multiple coherent structures and instabilities which develop in time. {\color{blue}Since most turbulence constitutive relations have not been designed to represent these phenomena, most practical one-point turbulence closures are expected to underperform during this period.} $ii)$ After $t=9$, the flow exhibits fully-developed turbulence features which one-point closures can better model. Considering the properties of these two periods, it is expected that the prediction of the first nine time-units of the TGV flow poses challenges to modeling and simulation. 

We now evaluate the performance of the proposed PANS method predicting the TGV flow, as well as the model's dependence on the physical resolution ($f_k$).
%
%
%
\subsection{Effect of physical resolution}
\label{sec:4.2}

The physical resolution of an SRS model dictates the range of resolved scales and, consequently, the efficiency (balance between accuracy and cost) of the computations. Whereas large physical resolutions may unnecessarily increase the computations' cost, reduced values of this parameter can compromise the accuracy of the predictions. Considering that the physical resolution of PANS {\color{blue}BHR-LEVM} is set through $f_k$ ({\color{blue}and $f_\varepsilon=1.00$ for all cases}), we investigate the dependence of the predictions' accuracy on this parameter. Toward this end, PANS calculations are conducted at six constant values of $f_k$: $0.00$, $0.25$, $0.35$, $0.50$, $0.75$ and $1.00$. Once again, recall that $f_k=0.00$ is equivalent to {\color{blue}DNS (for adequate grid resolution)}, and $f_k=1.00$ to RANS.

\begin{figure*}[t!]
\centering
\subfloat[$f_k=1.00$.]{\label{fig4.2_1a}
\includegraphics[scale=0.085,trim=0 0 300 0,clip]{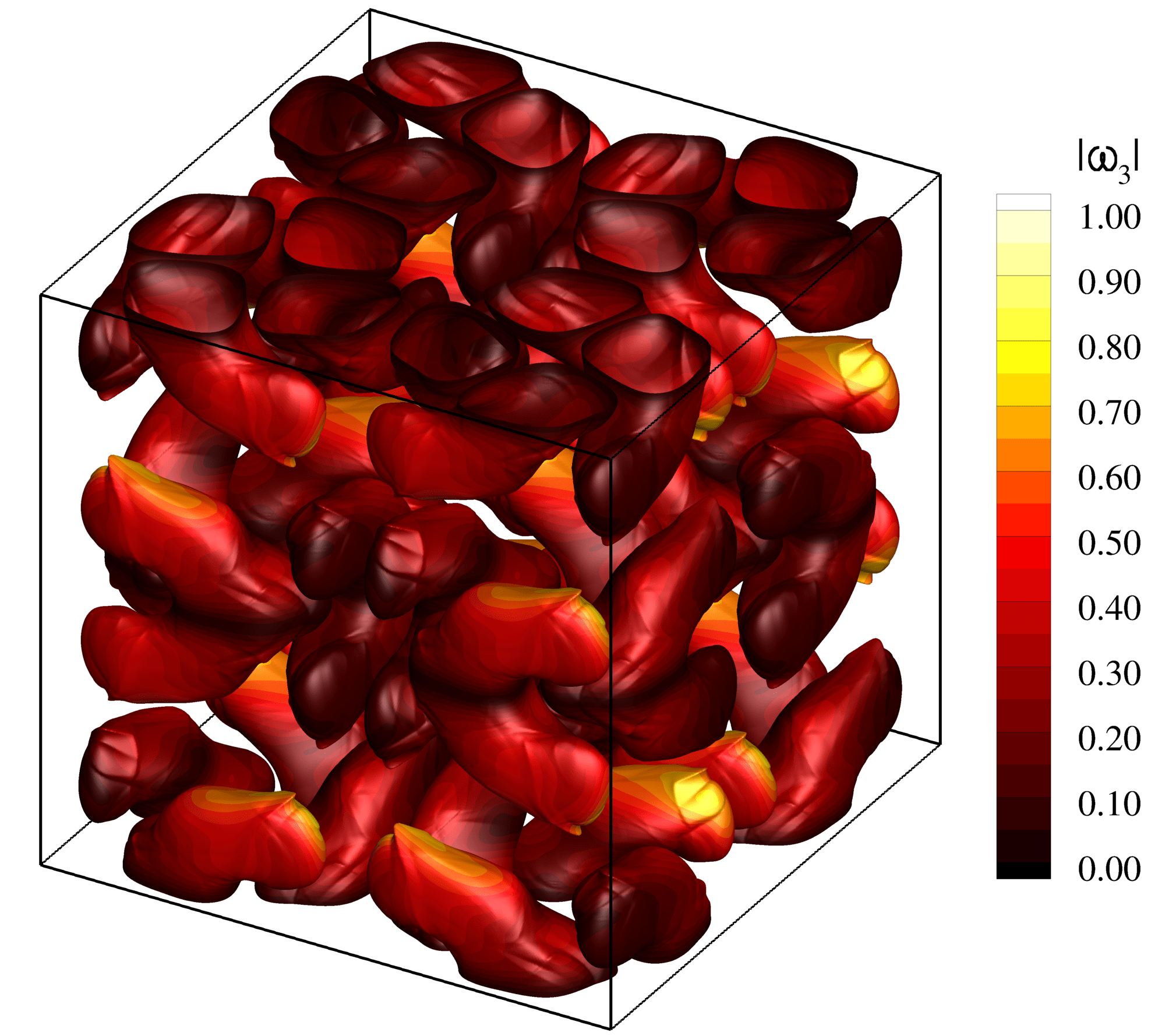}}
~
\subfloat[$f_k=0.75$.]{\label{fig4.2_1b}
\includegraphics[scale=0.085,trim=0 0 300 0,clip]{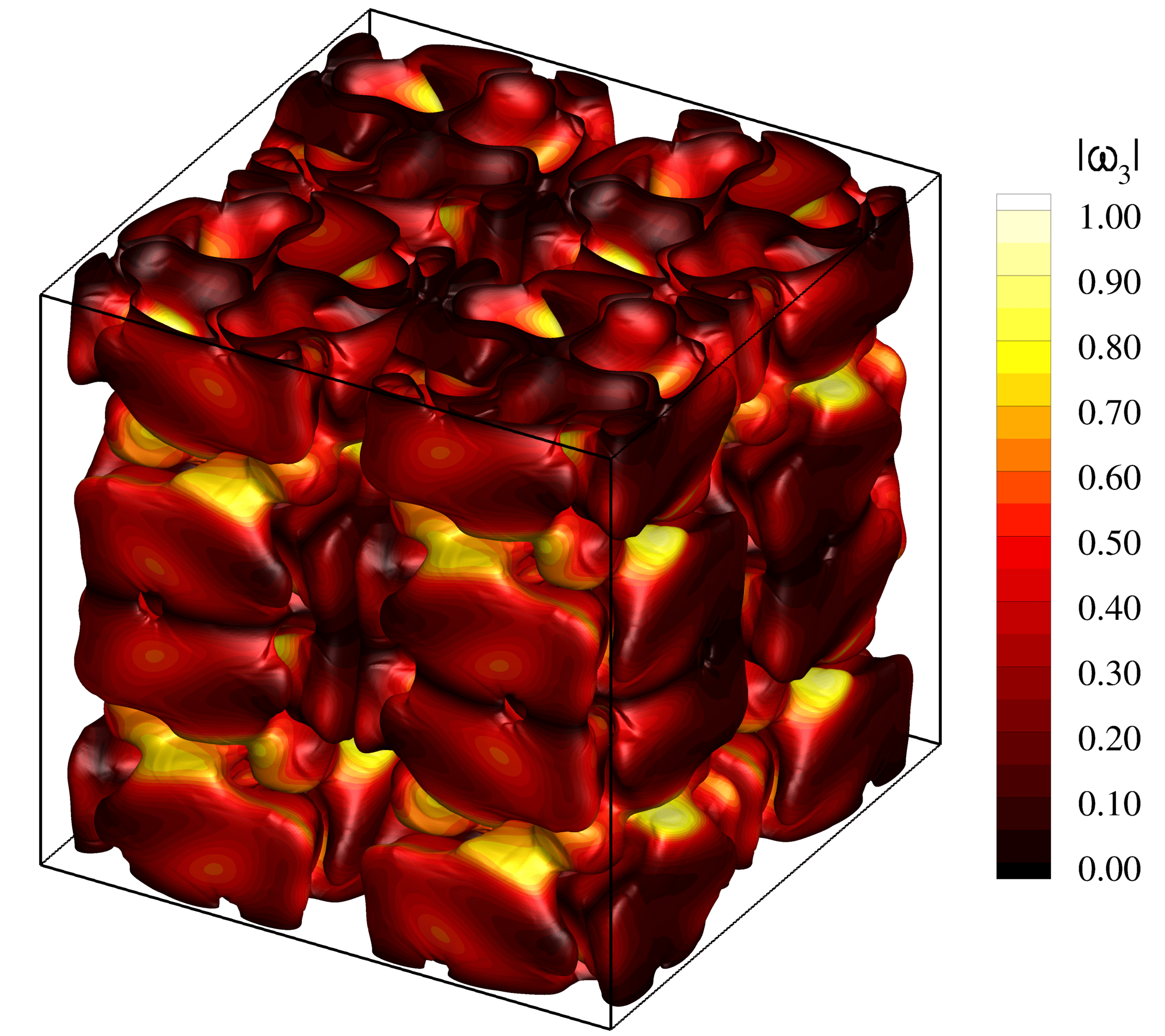}}
~
\subfloat[$f_k=0.50$.]{\label{fig4.2_1c}
\includegraphics[scale=0.085,trim=0 0 300 0,clip]{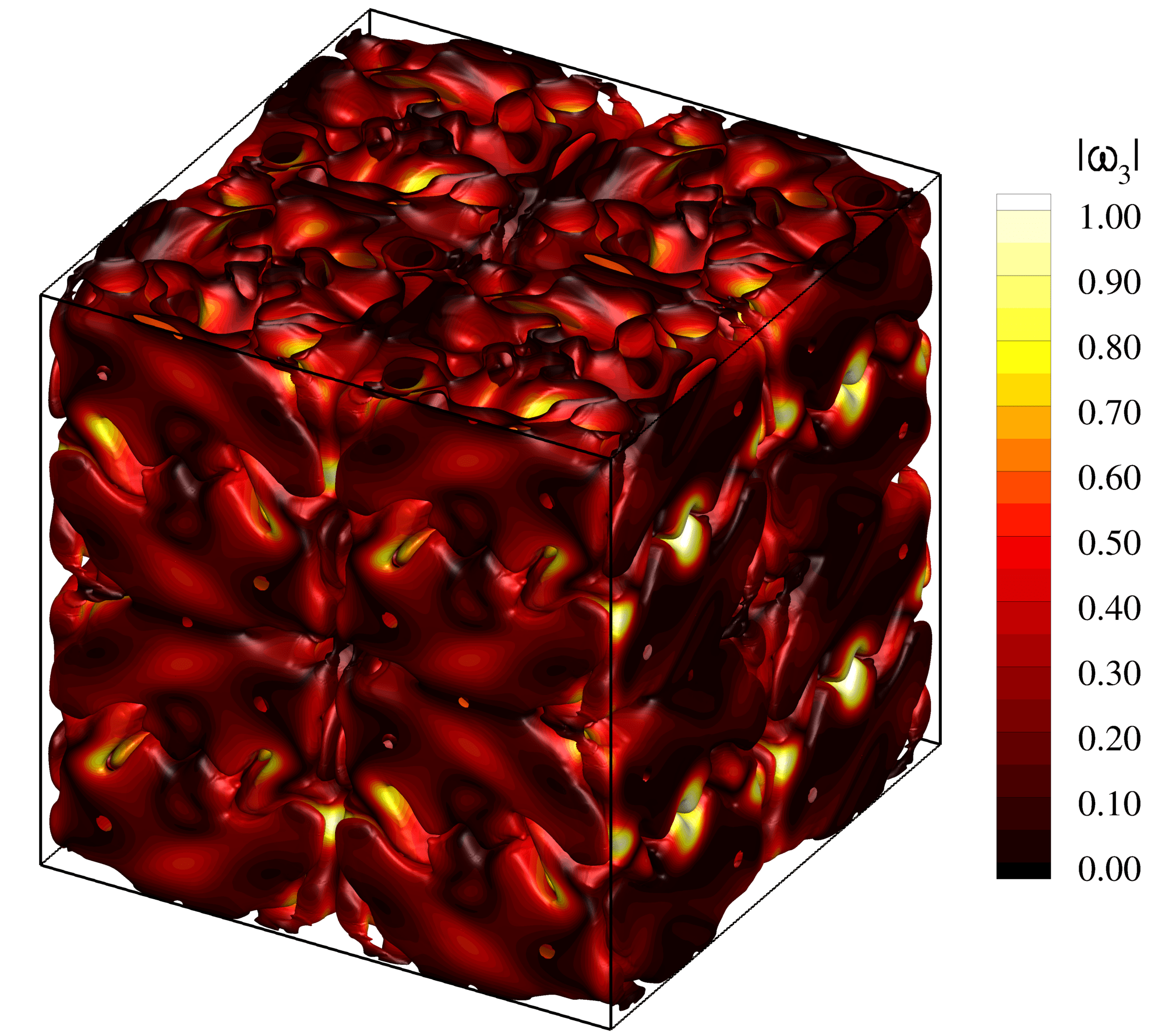}}
\\
\subfloat[$f_k=0.35$.]{\label{fig4.2_1d}
\includegraphics[scale=0.085,trim=0 0 300 0,clip]{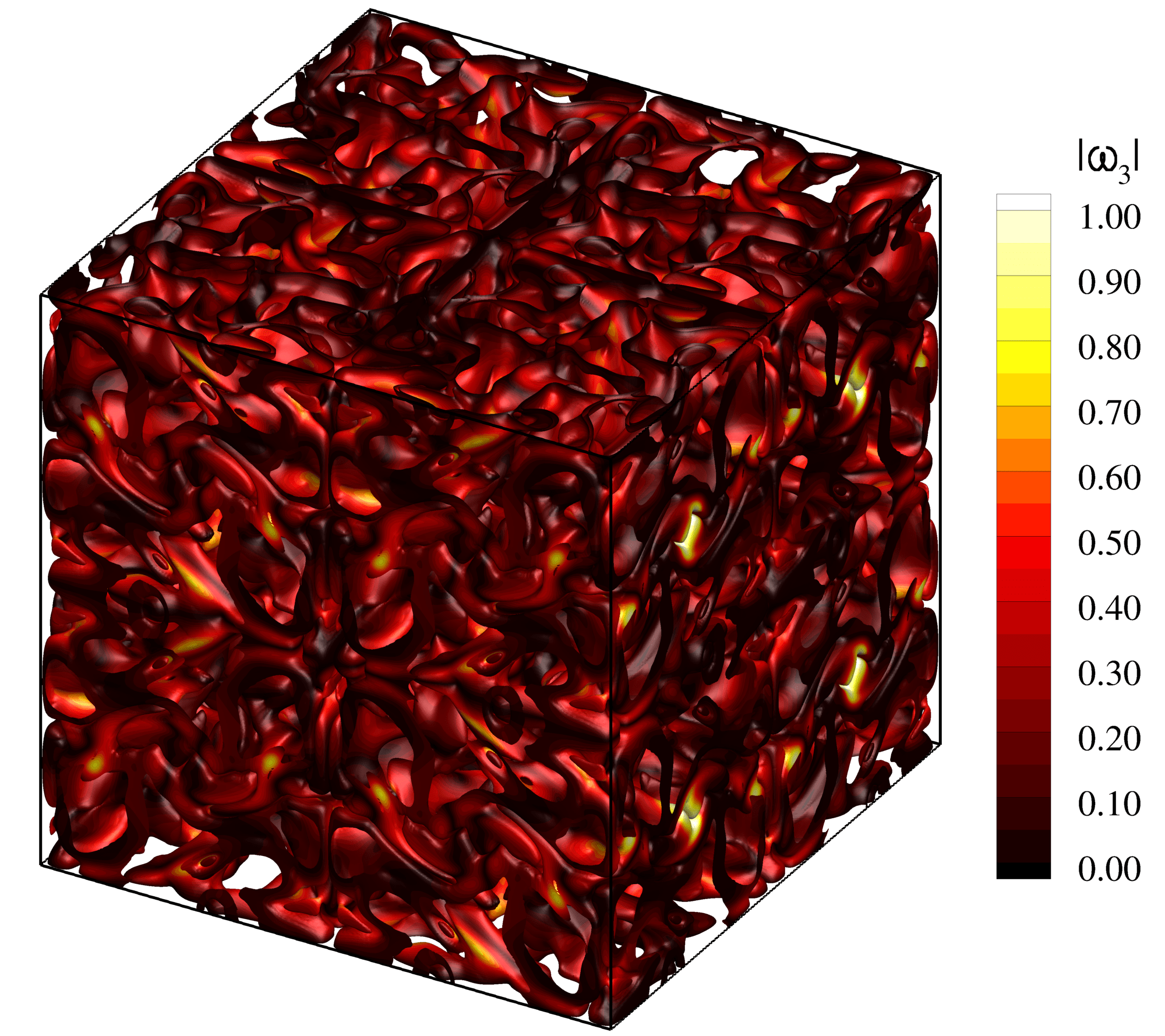}}
~
\subfloat[$f_k=0.25$.]{\label{fig4.2_1e}
\includegraphics[scale=0.085,trim=0 0 300 0,clip]{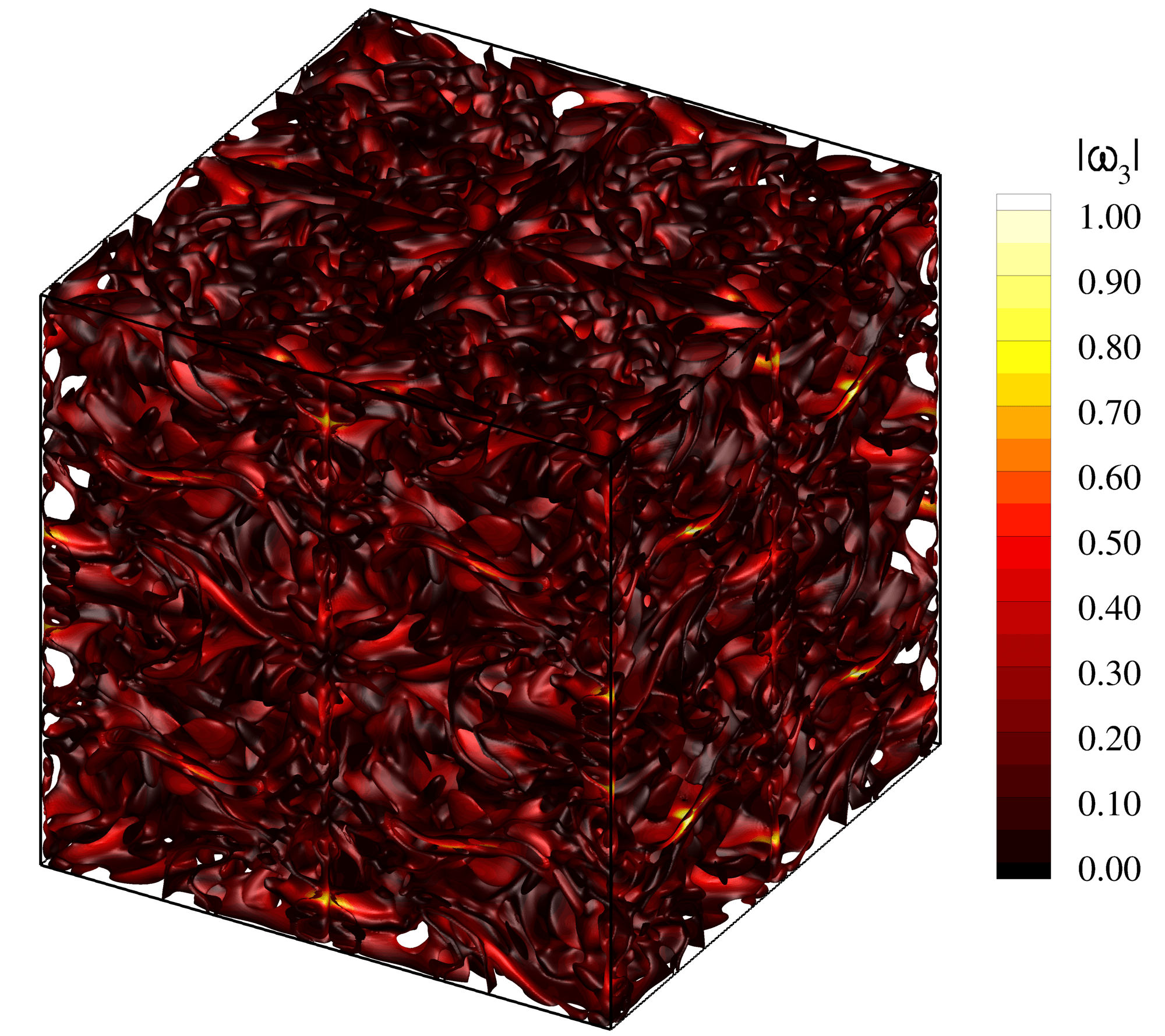}}
~
\subfloat[$f_k=0.00$.]{\label{fig4.2_1f}
\includegraphics[scale=0.085,trim=0 0 300 0,clip]{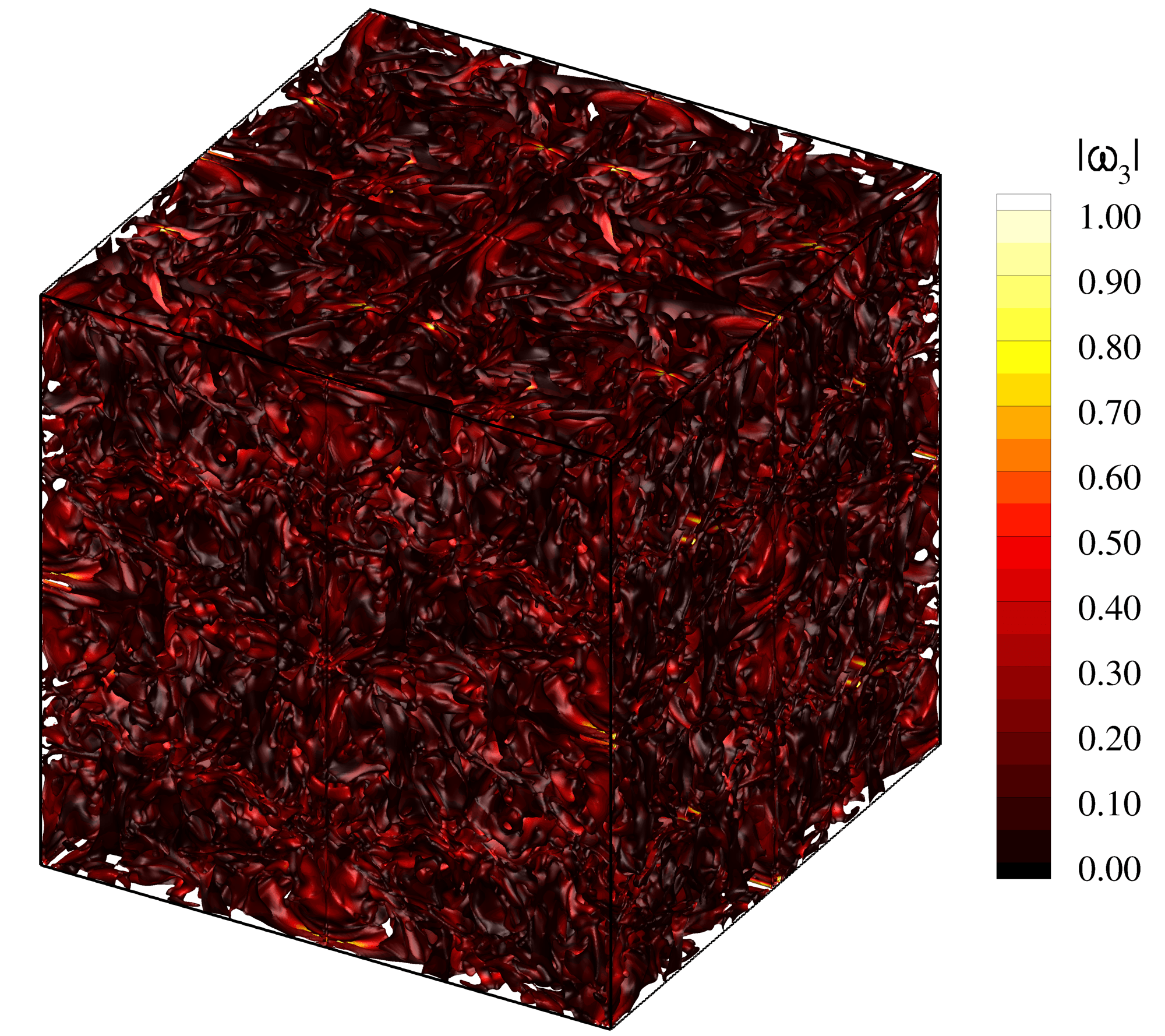}}
\caption{Vertical vorticity, $\omega_3$, iso-surfaces colored by the normalized vorticity magnitude, $|\omega|$, for simulations using different values of $f_k$ and $N_c=512^3$. Results at $t=20$.}
\label{fig:4.2_1}
\end{figure*}

To illustrate the effect of $f_k$ on the flow dynamics, {\color{blue}figure \ref{fig:4.2_1}} depicts the iso-surfaces of the vertical vorticity for different values of $f_k$. As expected, the results indicate that the range of resolved scales grows with $f_k \rightarrow 0$, this increasing the flow unsteadiness. Considering the limiting cases, the data show that the flow predicted at $f_k=1.00$ is dominated by large-scale coherent structures, whereas fine-scale structures predominate at $f_k=0.00$. Most notably, the absence of large-scale structures in the flow field obtained at $f_k=0.00$ is evidence of significant morphological differences between the mean-flow field of these limiting cases.
\begin{figure*}[t!]
\centering
\subfloat[$f_k=1.00$ on $512^3$.]{\label{fig:4.2_2a}
\includegraphics[scale=0.08,trim=0 0 0 0,clip]{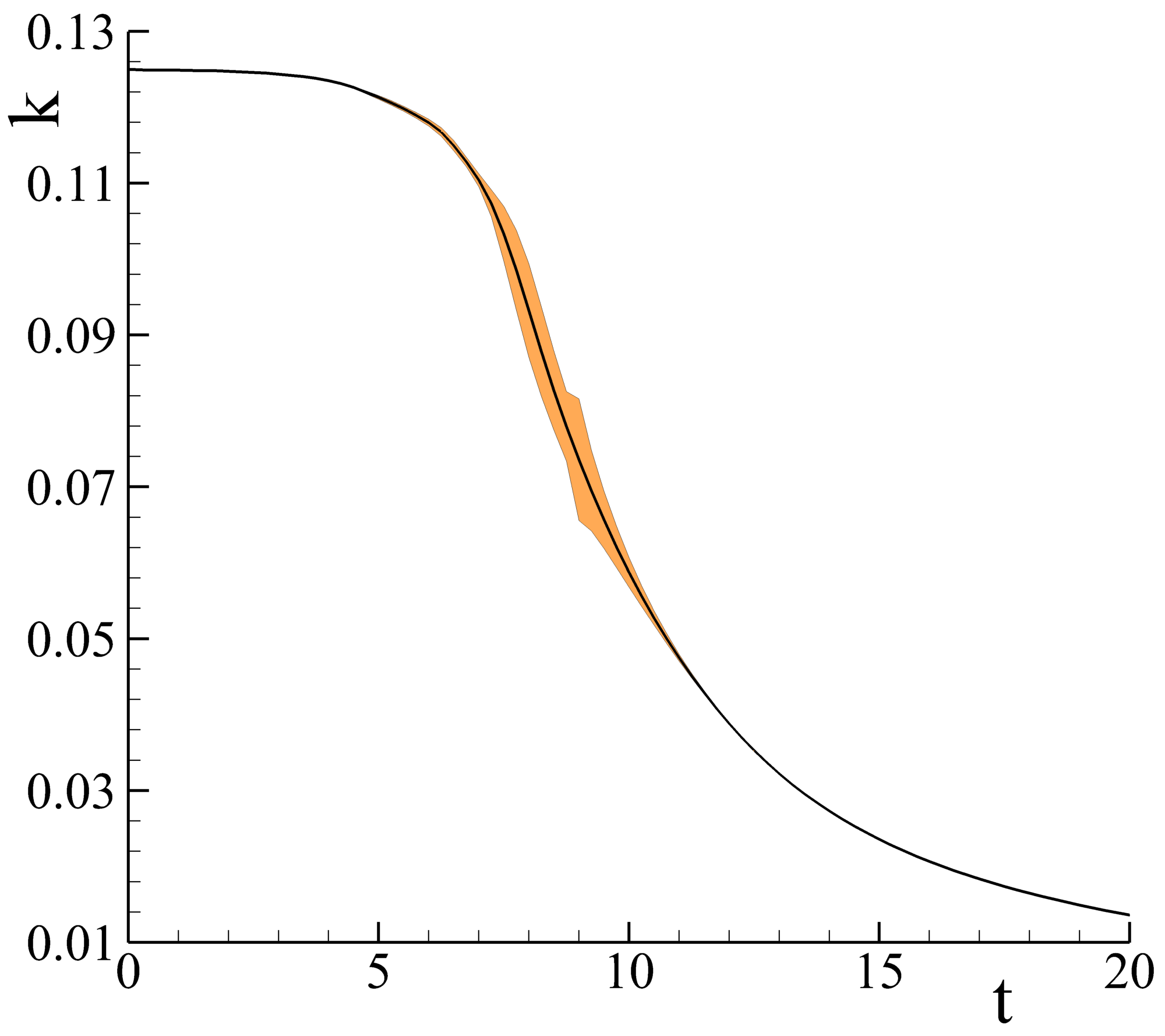}}
~
\subfloat[$f_k=0.75$ on $512^3$.]{\label{fig:4.2_2b}
\includegraphics[scale=0.08,trim=0 0 0 0,clip]{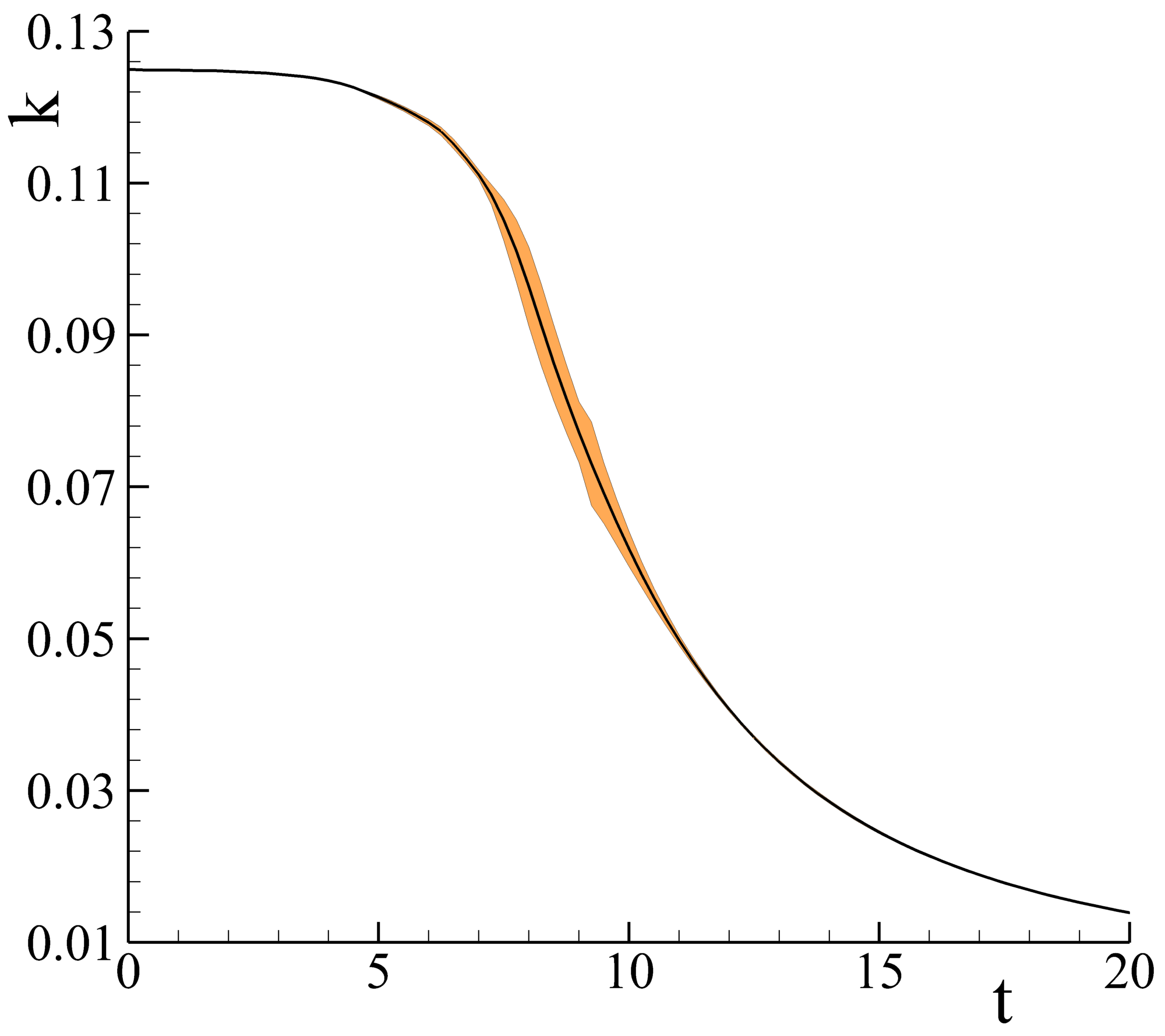}}
~
\subfloat[$f_k=0.50$ on $512^3$.]{\label{fig:4.2_2c}
\includegraphics[scale=0.08,trim=0 0 0 0,clip]{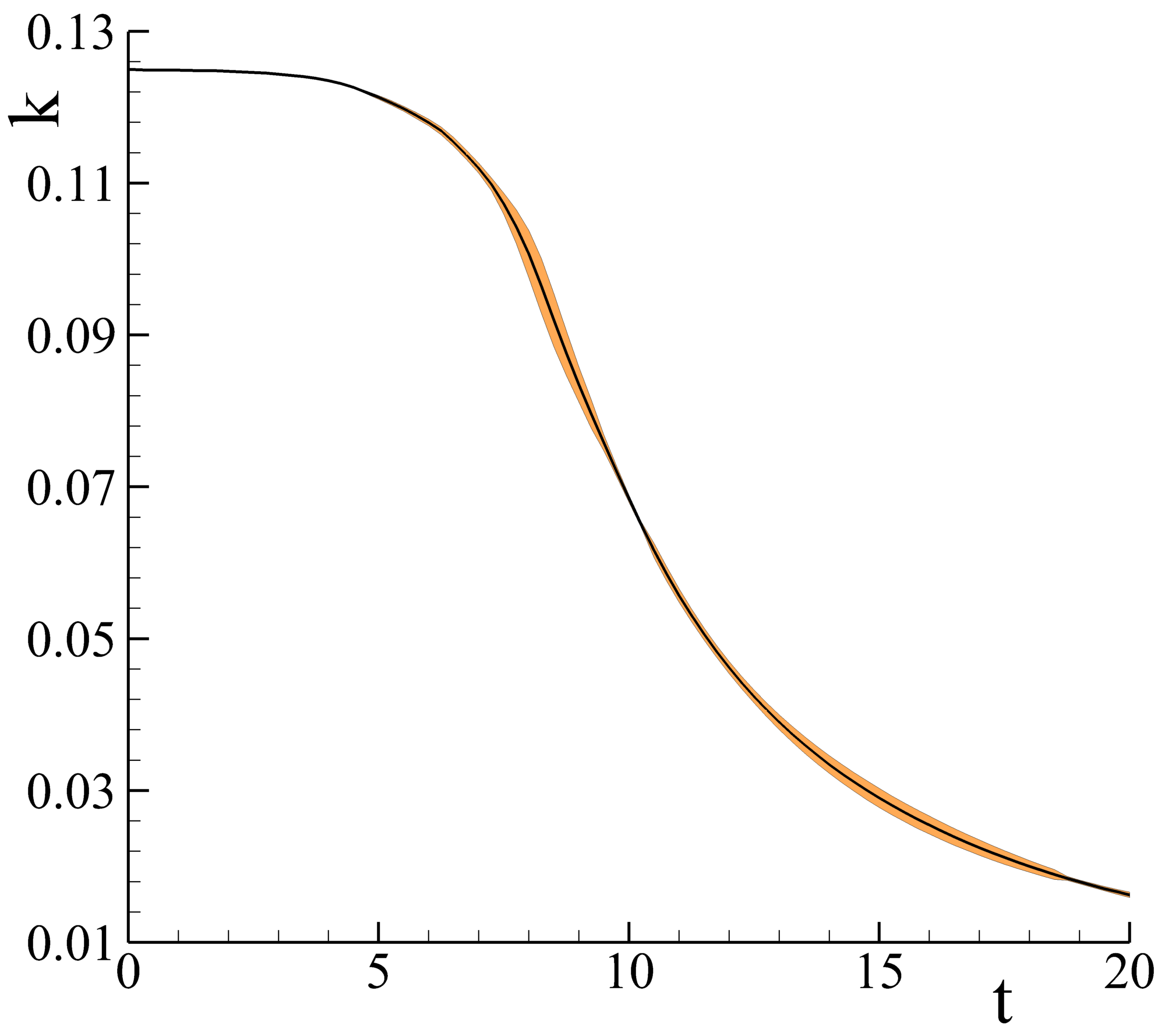}}
\\
\subfloat[$f_k=0.35$ on $512^3$.]{\label{fig:4.2_2d}
\includegraphics[scale=0.08,trim=0 0 0 0,clip]{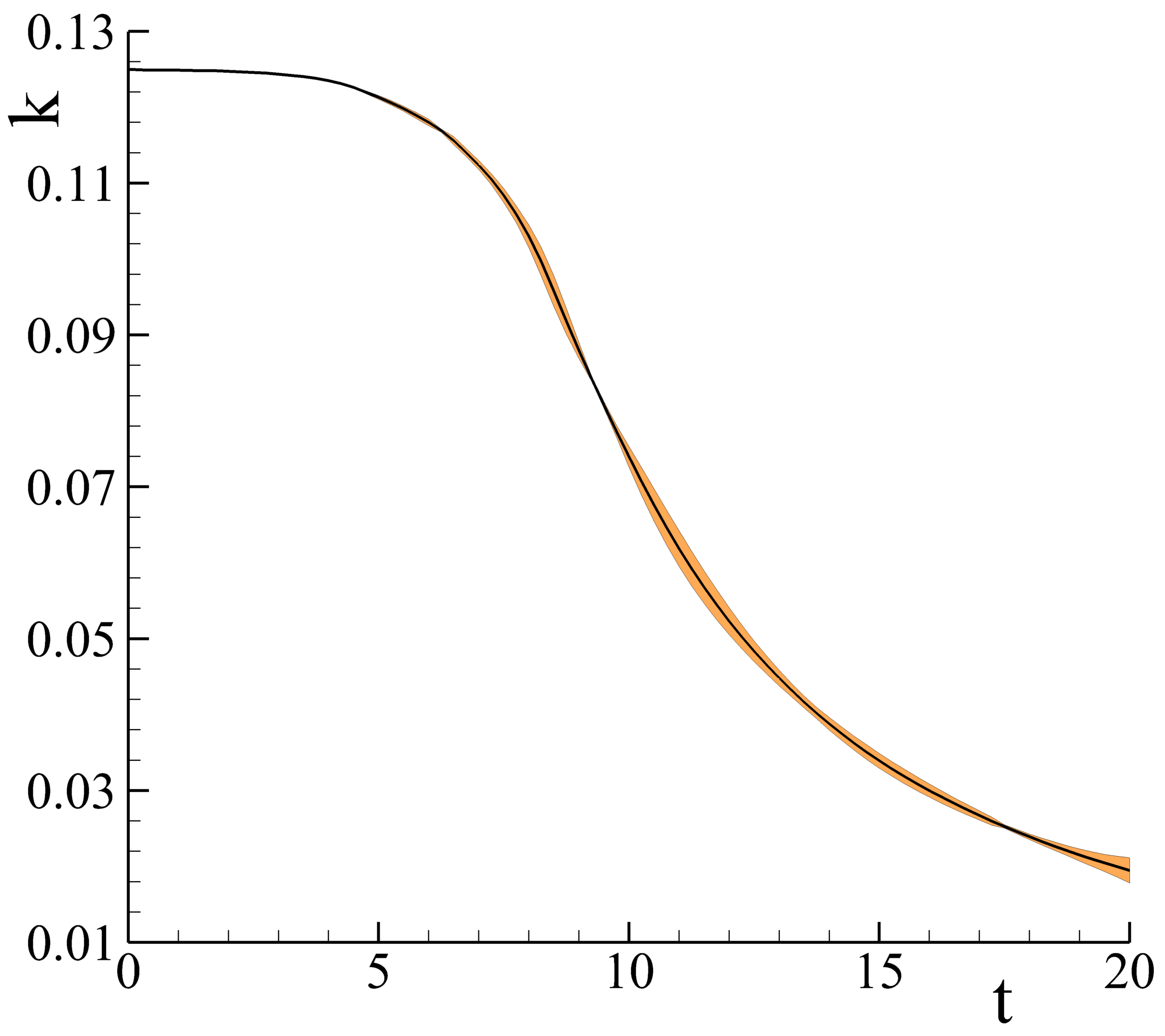}}
~
\subfloat[$f_k=0.25$ on $512^3$.]{\label{fig:4.2_2e}
\includegraphics[scale=0.08,trim=0 0 0 0,clip]{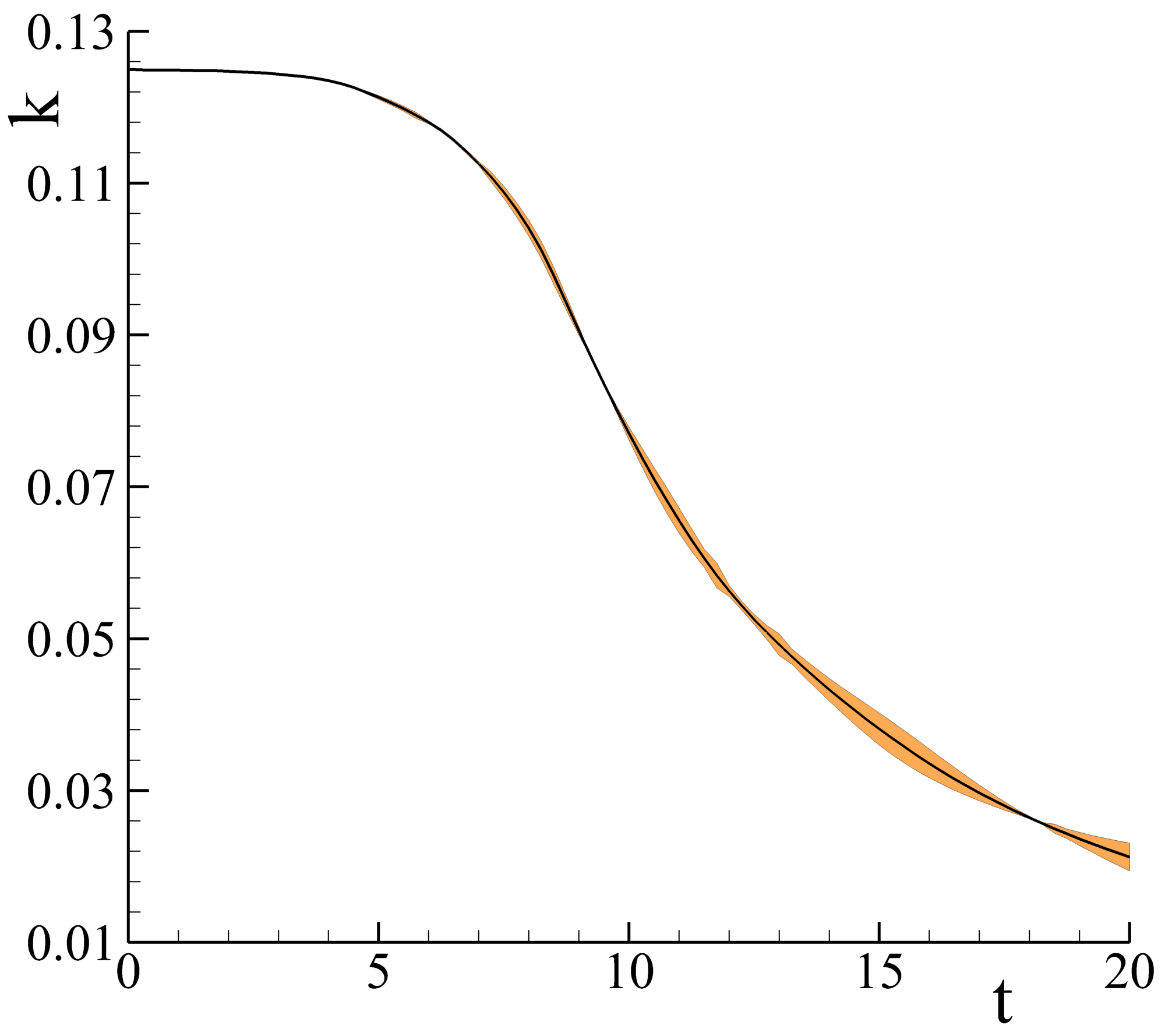}}
\\
\subfloat[$f_k=0.00$ on $512^3$.]{\label{fig:4.2_2f}
\includegraphics[scale=0.08,trim=0 0 0 0,clip]{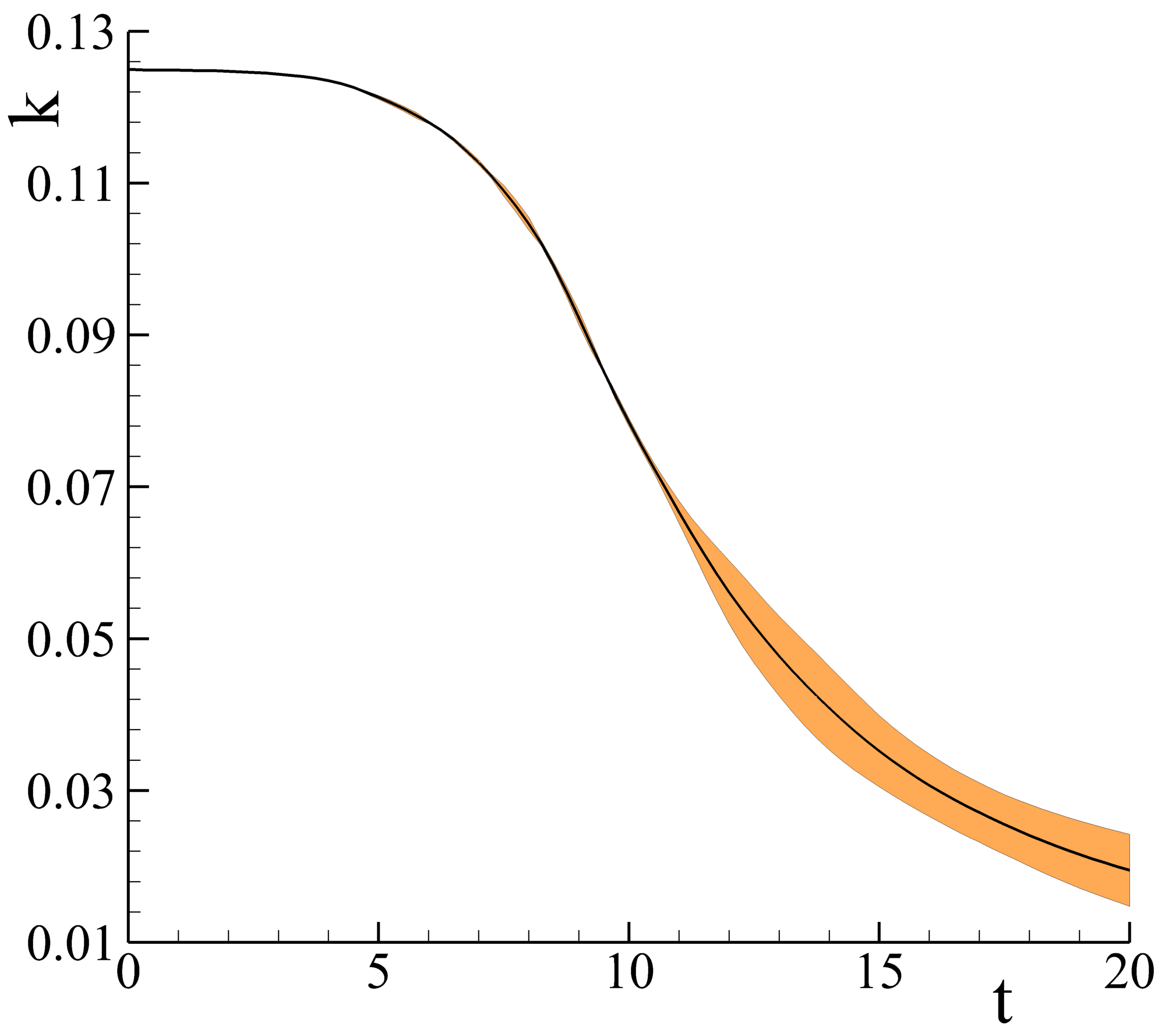}}
~
\subfloat[$f_k=0.00$ on $1024^3$.]{\label{fig:4.2_2g}
\includegraphics[scale=0.08,trim=0 0 0 0,clip]{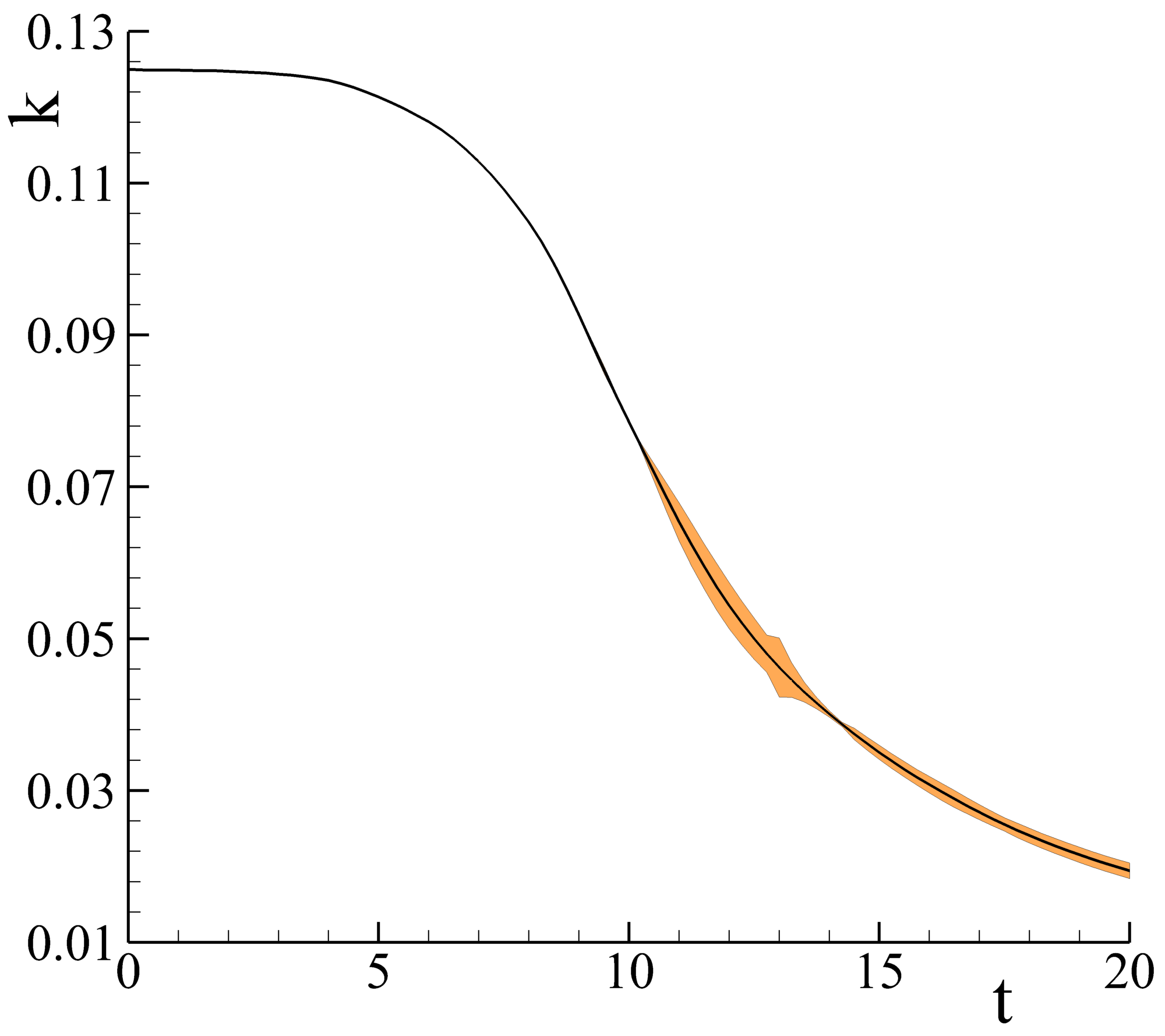}}
\caption{Temporal evolution of the total kinetic energy, $k$, and respective numerical uncertainty, $U_n(k)$, for simulations using different values of $f_k$ and grid resolutions.}
\label{fig:4.2_2}
\end{figure*}

The physical resolution also influences the numerical requirements of the simulations, such as the spatio-temporal grid resolution needed to predict the flow within a given level of numerical uncertainty, $U_n$. We emphasize that the quantification of $U_n$ is crucial to gain confidence in the results and prevent misleading conclusions due to canceling between different components of the computational error \cite{PEREIRA_ACME_2020}. To maintain the level of numerical uncertainty, the cost/grid resolution of the simulations is expected to increase with the physical resolution ($f_k\rightarrow 0$) owing to the resulting steeper gradients.

The numerical accuracy of the simulations at different values of $f_k$ is evaluated in figure \ref{fig:4.2_2}.  This figure depicts the temporal evolution of the mean total flow kinetic energy, $k$,
\begin{equation}
\label{4.2_1}
k = k_r + k_u  \; ,
\end{equation}
and respective numerical uncertainty, $U_n(k)$. Here, $k_r$ and $k_u$ are the resolved and unresolved/modeled components of $k$ obtained from the resolved velocity field ($k_r$) and the closure ($k_u$). The black line in figure \ref{fig:4.2_2} represents the predicted $k(t)$, whereas the orange area $U_n(k)$ estimated with the procedure proposed by E\c{c}a and Hoekstra \cite{ECA_JCP_2014} using the solutions on three grid resolutions: the three coarsest grids for $f_k \ge 0.25$ ($N_c \leq 512^3$), and both the three coarsest and finest ($N_c \ge 256^3$) grids for $f_k=0.00$. The numerical interpretation of $U_n$ is the following. The unavailability of an exact solution for the present problem precludes calculating the simulations' discretization error. Instead, we estimate an uncertainty interval containing the exact solution of the mathematical model within $95\%$ confidence \cite{ECA_JCP_2014}. For the same grid, a reduction of $U_n$ indicates smaller numerical errors. In figure \ref{fig:4.2_2}, the height of the orange area shows the magnitude of $U_n$ at a given time instant. 

Referring to the simulations at $f_k\ge 0.50$, figures \ref{fig:4.2_2a}-\ref{fig:4.2_2c}, these reveal reduced values of $U_n(k)$ for most of the simulated time. Taking the case of $f_k=1.00$ (equivalent to RANS), the results show that the numerical uncertainty does not exceed $3.3\%$ of the predicted value. The exceptions occur between $t=6$ and $10$ where $U_n(k)$ can reach a value of $10.8\%$. This growth originates in the use of the coarsest grid to estimate $U_n$ and the fact that the onset and development of turbulence occur during this period. As discussed below, low-physical resolution ($f_k\ge 0.50$) simulations overpredict turbulence during the onset of turbulence. This stems from well-recognized limitations of one-point closures to predict this phenomenon and leads to a premature and rapid dissipation of $k$, which increases the steepness of the flow gradients. As a result, the numerical uncertainty increases in these time instants.

The refinement of physical resolution to $f_k=0.35$ and $0.25$, figures \ref{fig:4.2_2d}-\ref{fig:4.2_2e}, causes a significant reduction of $U_n(k)$. For these cases, the numerical uncertainty does not exceed $8.7\%$ of $k$ at late times when the flow features high-intensity turbulence. In contrast, the further refinement of physical resolution ($f_k=0.00$) leads to a strong increase of $U_n(k)$ at $t>11.0$, figure \ref{fig:4.2_2f}. To understand this result, recall that during this period, the flow is characterized by high-intensity turbulence. Hence, formulations resolving all flow scales are expected to be computationally more intensive than those at $f_k>0.0$ predicting high-intensity turbulent flow. Results on the grid with $1024^3$ cells confirm this idea by exhibiting similar values of $U_n(k)$ to those of PANS at $f_k=0.25$ on $N_c=512^3$. Nevertheless, $U_n(k)$ can still reach $8.4\%$ at $t=13$, whereas for $f_k=0.25$ does not exceed $2.9\%$. 

Overall, these numerical uncertainties enable the robust evaluation of the PANS model due to their reduced values. The results also demonstrate that to obtain similar numerical uncertainties, simulations at $f_k=0.00$ require a minimum of $N_c=1024^3$, whereas those at $f_k\leq 0.25$ relax the grid resolution requirement to $N_c=512^3$. This is expected to have a significant impact on the computational resources required by the computations. It is important to emphasize that estimates using the coarsest grid, $N_c=128^3$, should overpredict $U_n(\phi)$ due to the observed large discrepancies between solutions on this and the remaining grids (see Pereira et al. \cite{PEREIRA_PRE_2020}).
\begin{figure*}[t!]
\centering
\subfloat[$t=5.0$.]{\label{fig:4.2_3a}
\includegraphics[scale=0.11,trim=0 0 0 0,clip]{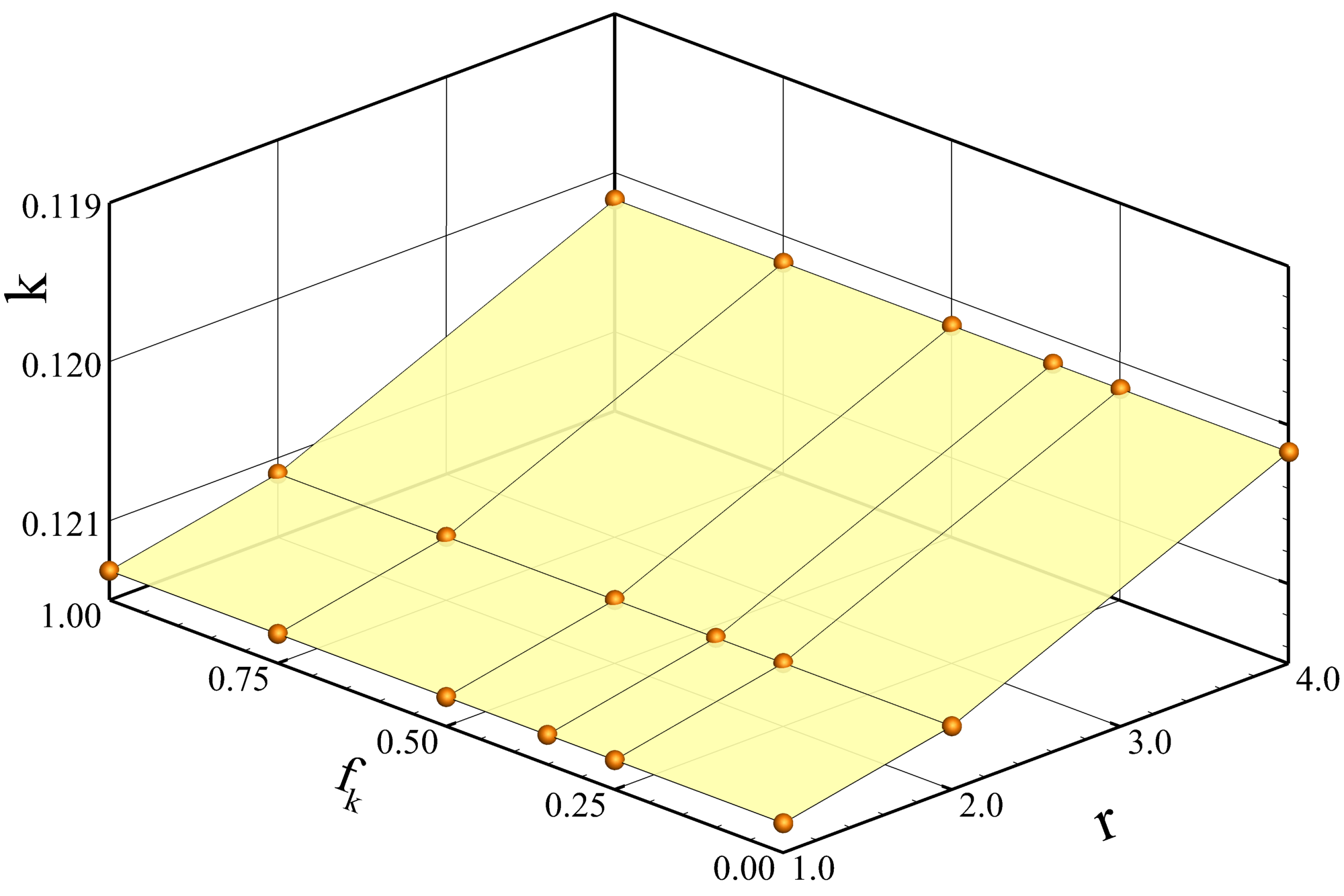}}
~
\subfloat[$t=7.0$.]{\label{fig:4.2_3b}
\includegraphics[scale=0.11,trim=0 0 0 0,clip]{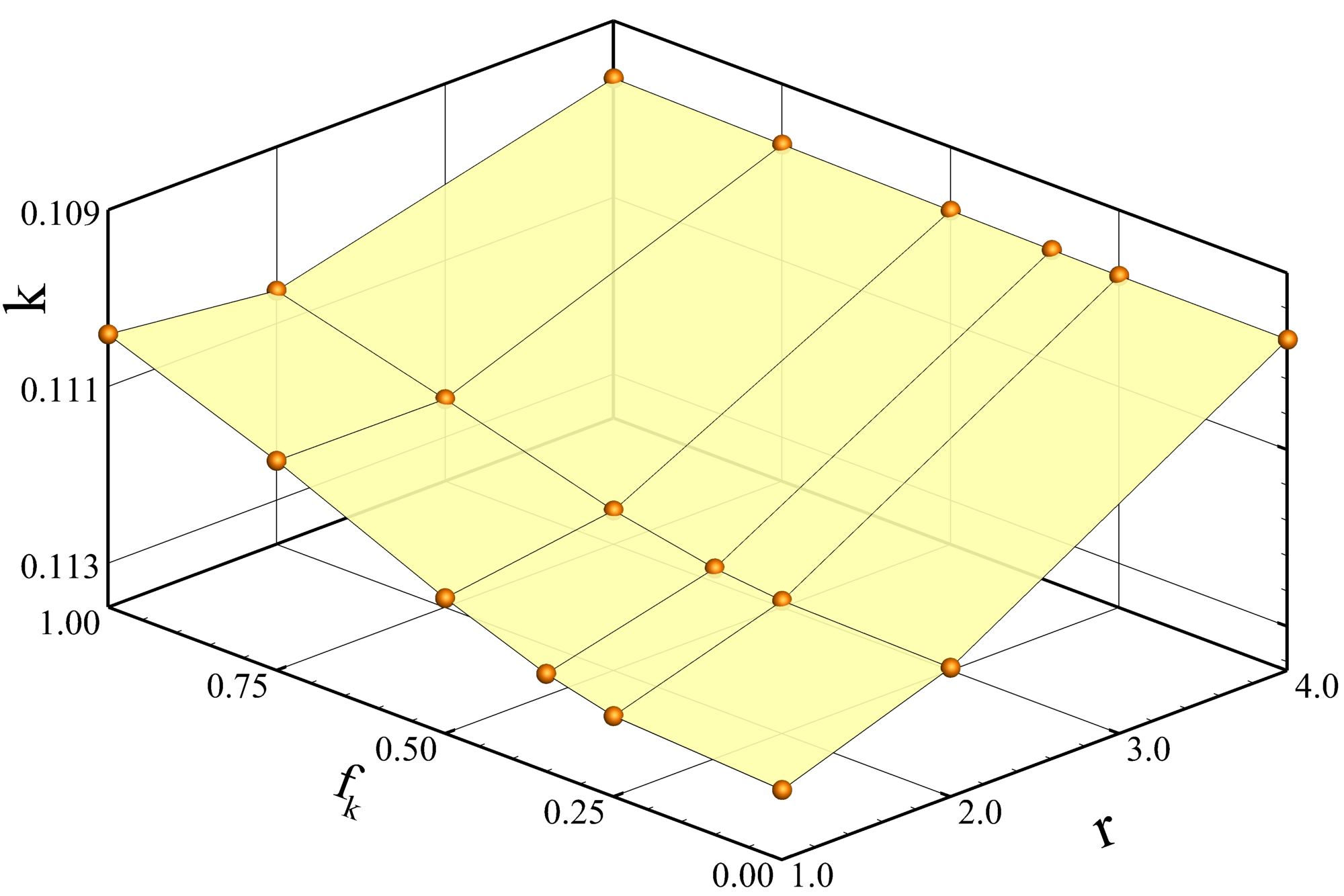}}
\\
\subfloat[$t=9.0$.]{\label{fig:4.2_3c}
\includegraphics[scale=0.11,trim=0 0 0 0,clip]{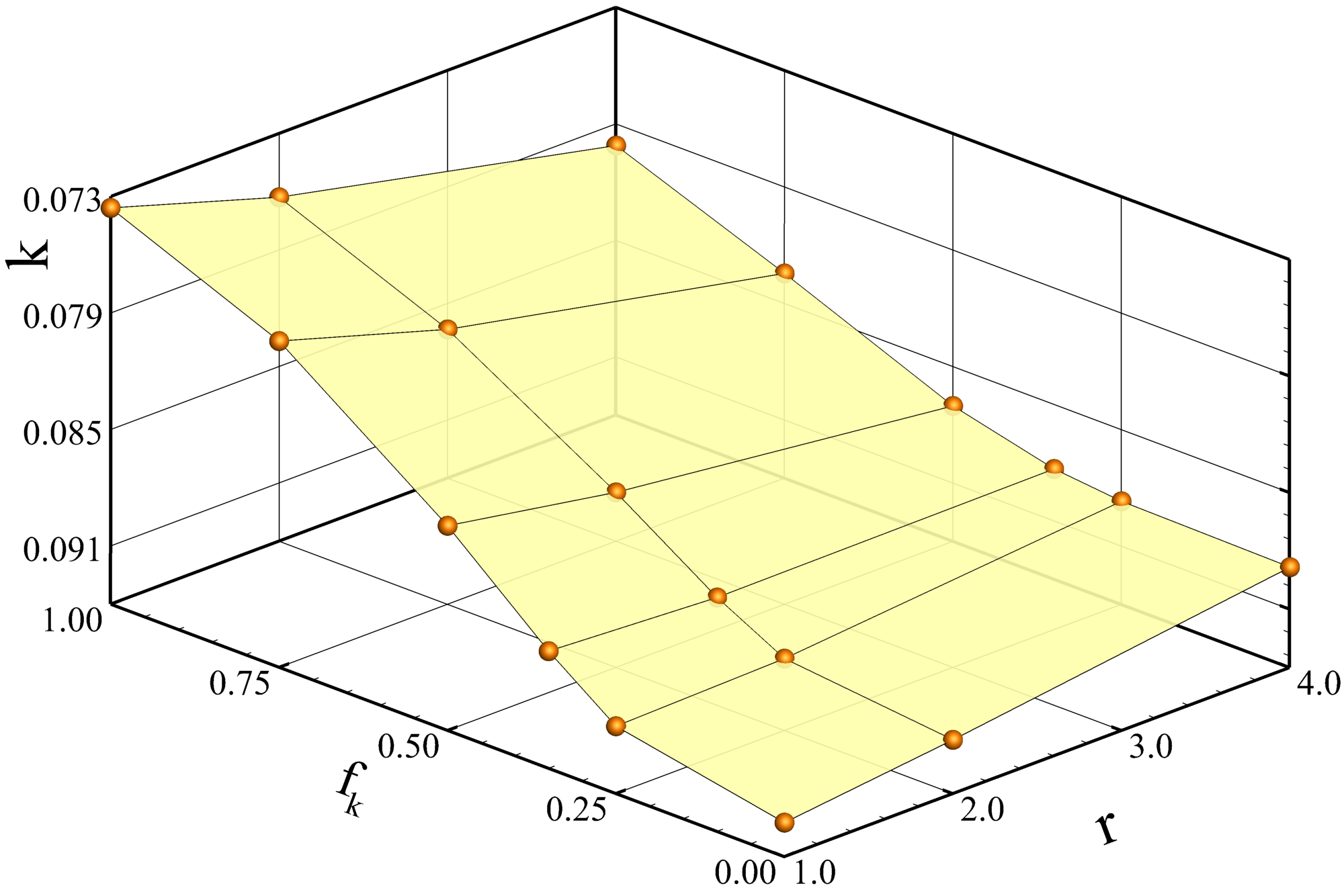}}
~
\subfloat[$t=11.0$.]{\label{fig:4.2_3d}
\includegraphics[scale=0.11,trim=0 0 0 0,clip]{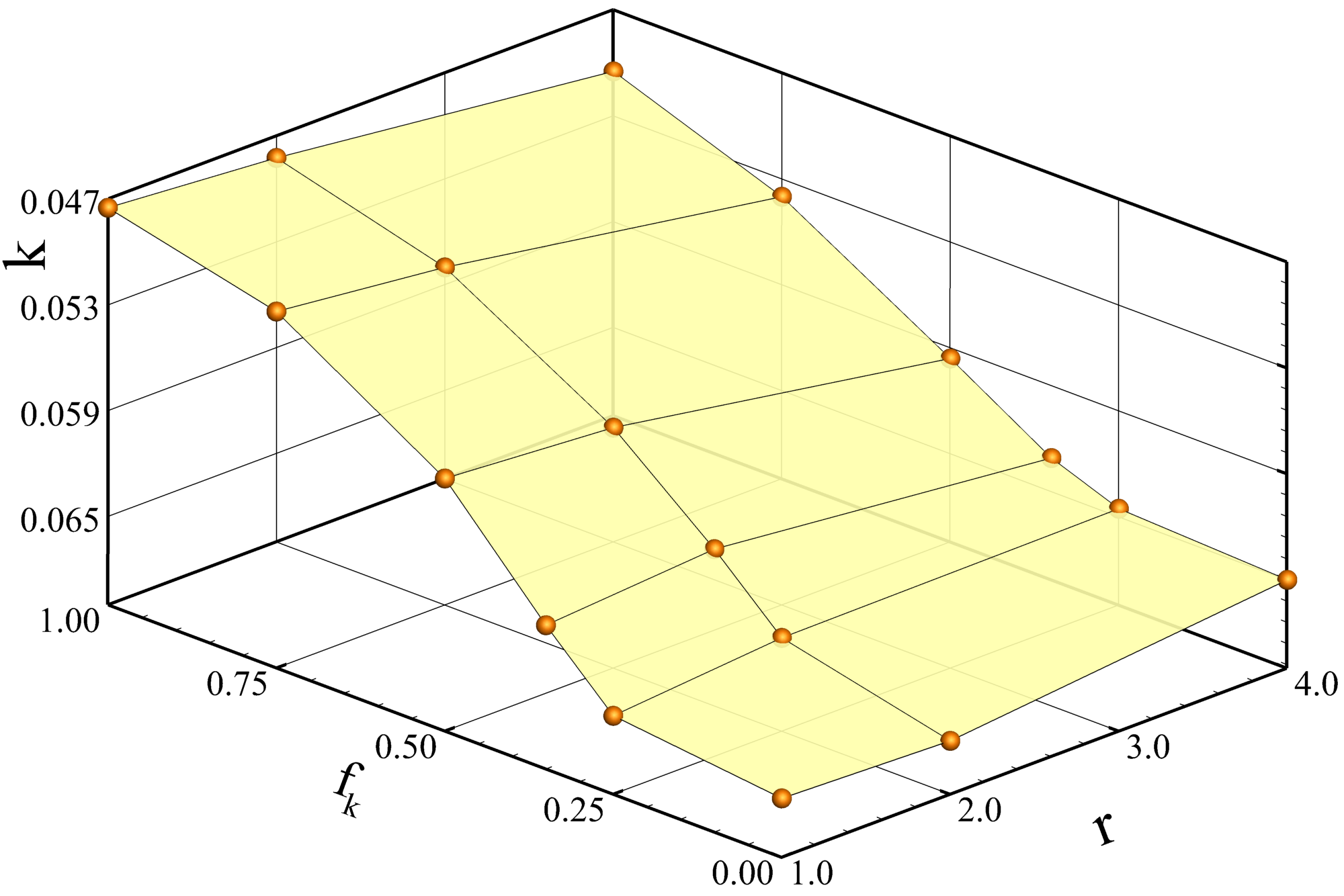}}
\caption{Evolution of the total kinetic energy, $k$, with $r_i$ and $f_k$ for different time instants.}
\label{fig:4.2_3}
\end{figure*}

Next, figure \ref{fig:4.2_3} presents the temporal evolution of $k$ upon grid and physical resolution refinement. The grid resolution is here defined through the grid refinement ratio, $r_i$,
\begin{equation}
\label{4.2_2}
r_i=\frac{\sqrt[3]{(N_c)_1)}}{\sqrt[3]{(N_c)_i)}} \; ,
\end{equation}
where the index $i$ denotes the grid resolution with $i=1$ referring to the finest grid resolution. At $t=5$, figure \ref{fig:4.2_3a}, the results of all PANS simulations converge monotonically upon grid refinement ($r_i \rightarrow 1$) and show that the closure and physical resolution have a negligible impact on the solutions ($\tau^1(V_i,V_j)\approx 0)$. At slightly later time, $t\ge 7$, the solutions start exhibiting an increasingly larger dependence on the physical resolution and convergence with $f_k$ and $r_i$. Figures \ref{fig:4.2_3b}-\ref{fig:4.2_3d} also evidence large discrepancies between low- ($f_k \ge 0.50$) and high- ($f_k<0.50$) physical resolution simulations. Whereas low-physical resolution computations exhibit a strong dependence on $f_k$, high-physical resolution simulations are significantly less dependent on this parameter. This behavior has also been observed for transitional bluff-body flows \cite{PEREIRA_IJHFF_2018,PEREIRA_JCP_2018,PEREIRA_IJHFF_2019,PEREIRA_OE_2019}. 
\begin{figure}[t!]
\centering
\subfloat[$k$ at $1.00 \leq f_k \leq 0.50$.]{\label{fig:4.2_4a}
\includegraphics[scale=0.11,trim=0 0 0 0,clip]{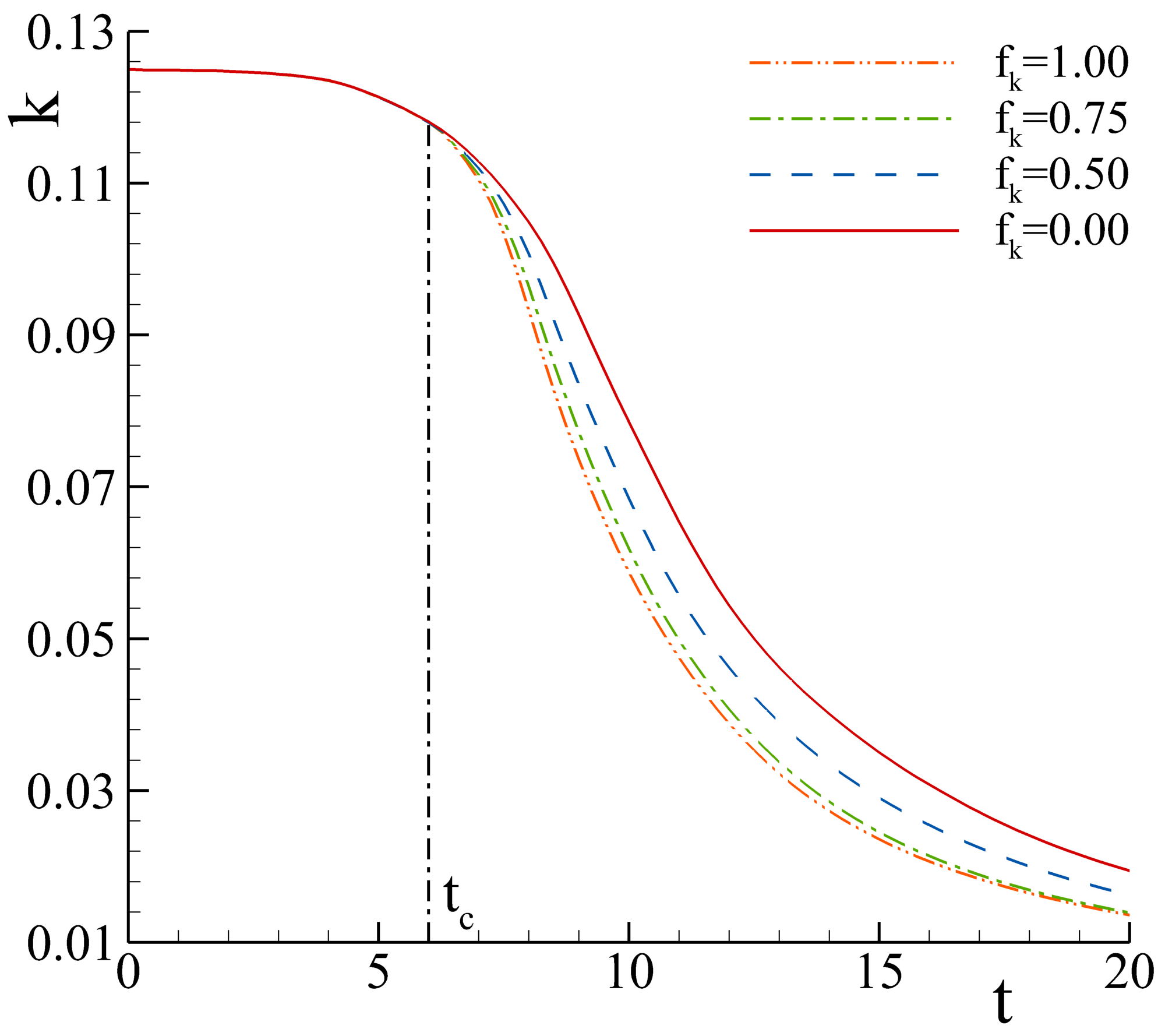}}
\\
\subfloat[$k$ at $0.35 \leq f_k \leq 0.00$.]{\label{fig:4.2_4b}
\includegraphics[scale=0.11,trim=0 0 0 0,clip]{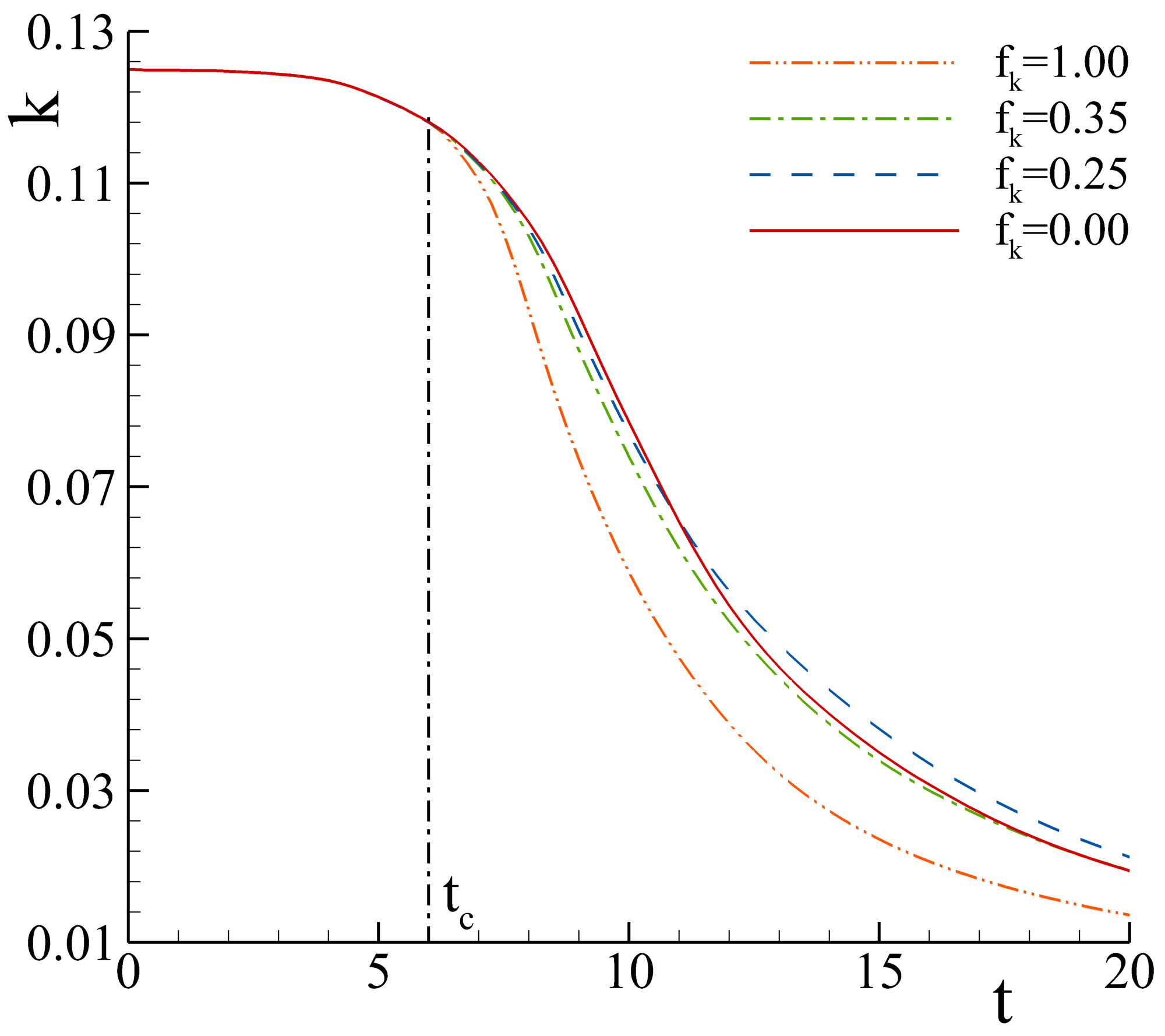}}
\caption{Temporal evolution of the total kinetic energy, $k$, for simulations using different values of $f_k$.}
\label{fig:4.2_4}
\end{figure}

Figure \ref{fig:4.2_4} depicts the temporal evolution of $k$ for simulations conducted at different values of $f_k$ on the finest grid resolution available ($N_c=512^3$ or $1024^3$). The numerical results of low- and high-physical resolution simulations are presented individually in figures \ref{fig:4.2_4a} and \ref{fig:4.2_4b}, respectively. Focusing on the low-physical resolution calculations, the results show that the simulations are independent of $f_k$ until $t_c=t \approx 6$. After this instant, we observe a rapid decay of $k$ (compared to $f_k<0.50$) caused by the growth of the modeled turbulent stresses. Yet, it is essential to recall that $i)$ the flow is expected to be laminar until $t\approx 7$ \cite{BRACHET_JFM_1983}, and $ii)$ the vortex-reconnection mechanism driving the onset of turbulence occurs between $t=5$ and $7$ (figure \ref{fig:4.1_1}). Between low-physical resolution computations, those performed at $f_k=1.00$ and $0.75$ predict similar temporal evolutions of $k$.

On the other hand, the results of high-physical resolution simulations are in close agreement, especially until $t=10$. The discrepancies between different $f_k$'s do not exceed $0.0047$. Compared to low-physical resolution computations, these simulations lead to slower decay rates of $k$ and, consequently, to larger kinetic energy values. The results also show that the solutions do not converge monotonically with $f_k$ after  $t\approx 10$. We attribute this behavior to numerical uncertainty, especially for simulations at $f_k=0.00$. 

\begin{figure}[t!]
\centering
\subfloat[$\varepsilon$ at $1.00 \leq f_k \leq 0.50$.]{\label{fig:4.2_5a}
\includegraphics[scale=0.11,trim=0 0 0 0,clip]{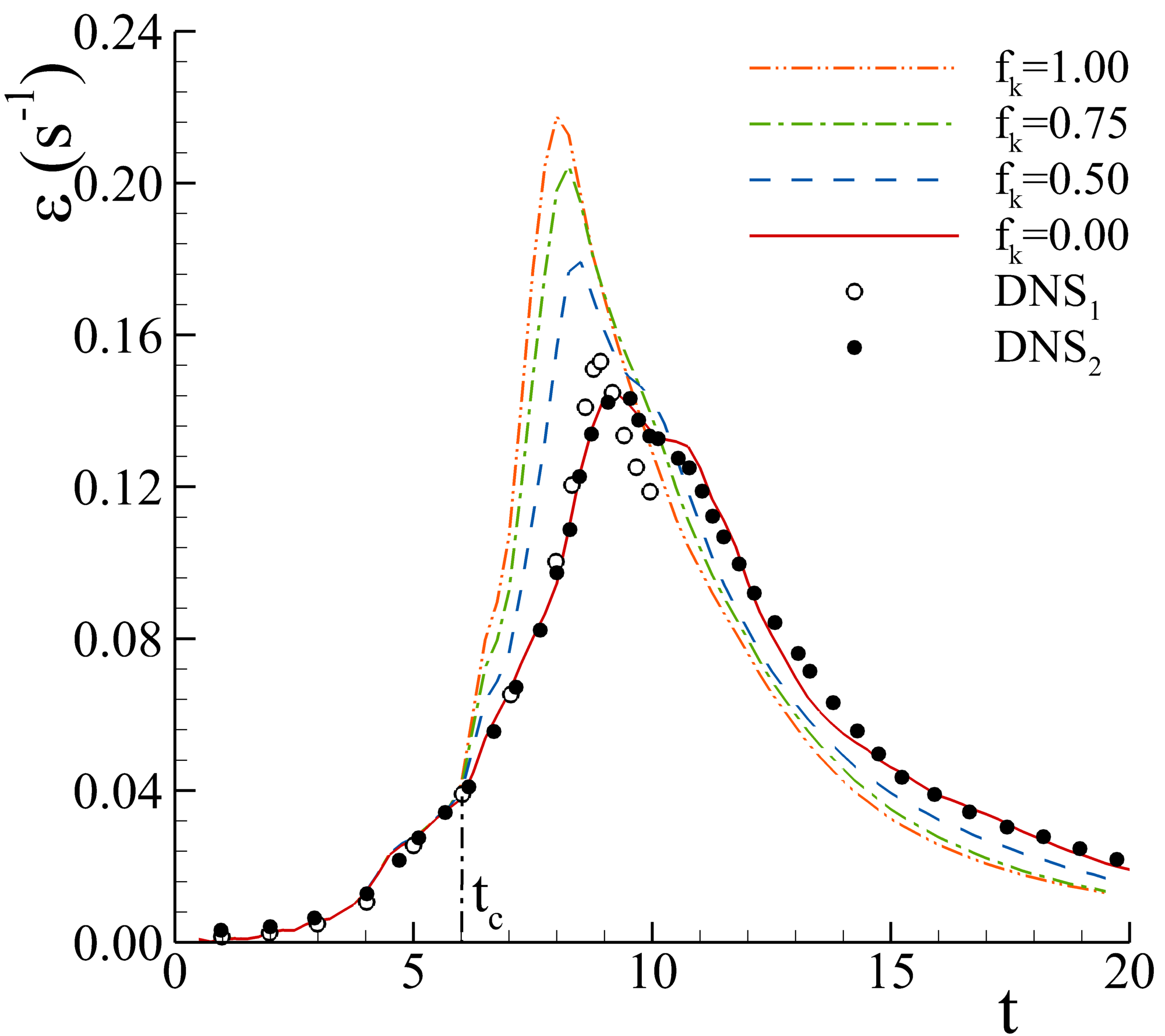}}
\\
\subfloat[$\varepsilon$ at $0.35 \leq f_k \leq 0.00$.]{\label{fig:4.2_5b}
\includegraphics[scale=0.11,trim=0 0 0 0,clip]{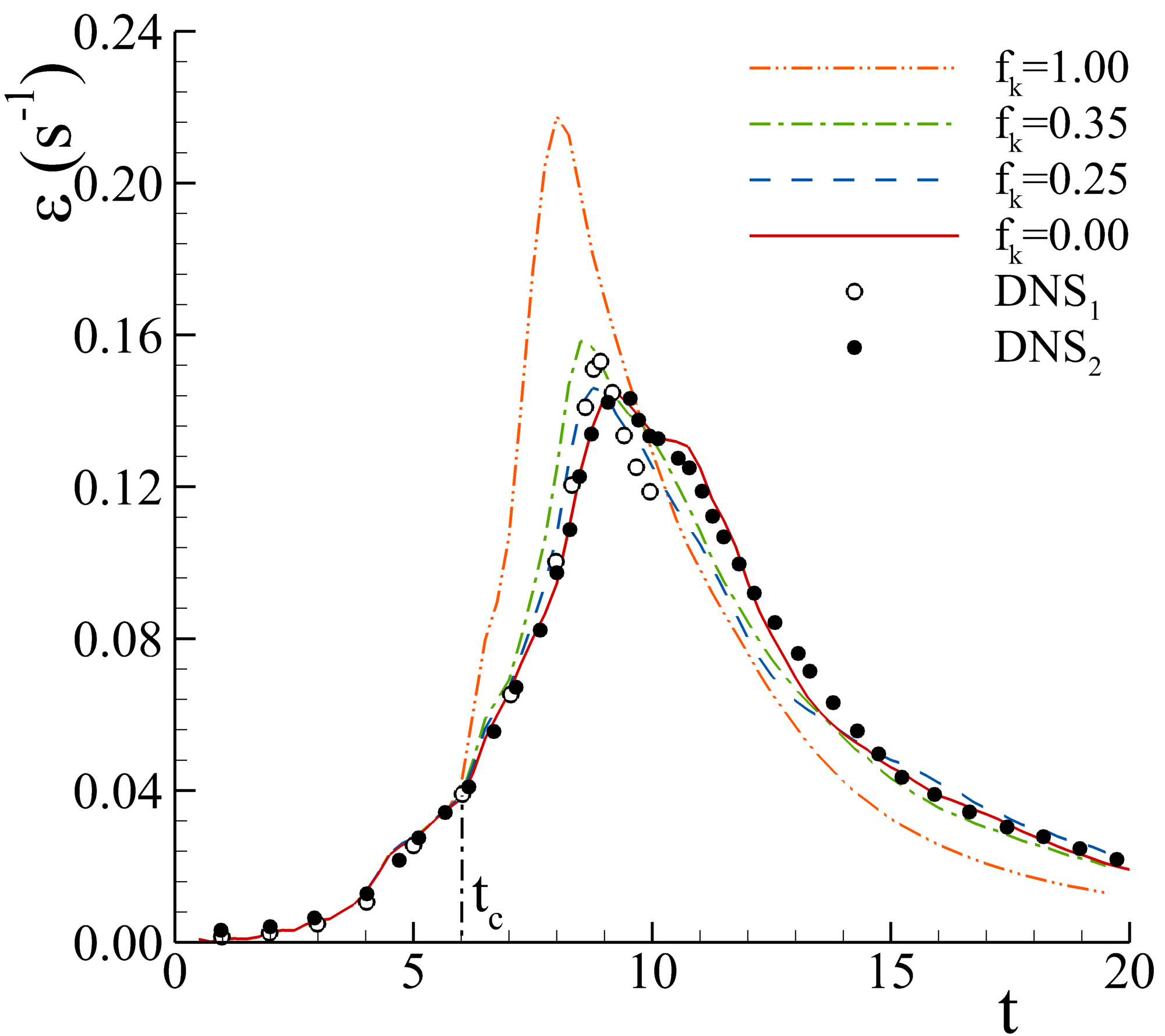}}
\caption{Temporal evolution of the total kinetic energy dissipation, $\varepsilon \times 10$ {\color{blue}$\mathrm{(s^{-1})}$}, for simulations using different values of $f_k$.}
\label{fig:4.2_5}
\end{figure}

Now, we turn our attention to the temporal evolution of the kinetic energy dissipation, $\varepsilon$,
\begin{equation}
\label{4.2_3}
\varepsilon = -\frac{\partial k}{\partial t} \; ,
\end{equation}
and its dependence on $f_k$. Figure \ref{fig:4.2_5} compares the predictions against the DNS of Brachet et al. \cite{BRACHET_JFM_1983} {\color{blue}($\mathrm{DNS_1}$)} and Drikakis et al. \cite{DRIKAKIS_JOT_2007} {\color{blue}($\mathrm{DNS_2}$)}. As mentioned in \cite{BRACHET_JFM_1983,DRIKAKIS_JOT_2007}, these reference DNS simulations are likely underesolved at late times and were conducted at different Mach numbers (Ma). Whereas the simulations of Brachet et al. \cite{BRACHET_JFM_1983} are incompressible (Ma$=0$), those of Drikakis et al. \cite{DRIKAKIS_JOT_2007} are characterized by an initial $\mathrm{Ma}_o=0.28$, which matches that of the present PANS calculations. Considering the differences in Ma and the study of Virk et al. \cite{VIRK_JFM_1995}, we can expect small discrepancies in $\varepsilon$ between the two datasets.

The results of figure \ref{fig:4.2_5} illustrate that the predictions are independent of the physical resolution ($f_k$) until $t=t_c$. Recall that this is the moment when the vortex-reconnection process {\color{blue}occurs} (figure \ref{fig:4.1_1}). After this time instant, low- ($f_k\ge 0.50$) and high- ($f_k<0.50$) physical resolution computations exhibit distinct behaviors. Low-physical resolution simulations predict the peak of dissipation prematurely and overpredict its magnitude. These tendencies become more pronounced as the physical resolution is coarsened ($f_k \rightarrow 1.00$). For instance, it is observed that the peak of $\varepsilon$ obtained at $f_k=1.00$ can reach $0.218$, whereas the reference DNS report values between $0.143$ and $0.153$ (values extracted graphically from \cite{BRACHET_JFM_1983,DRIKAKIS_JOT_2007}). Compared to the value predicted by Drikakis et al. \cite{DRIKAKIS_JOT_2007}, this represents an increase of approximately $52.0\%$. Also, the maximum $\varepsilon$ occurs at $t\approx 8$ for the simulation at $f_k=1.00$, whereas the reference DNS report values between $9.0$ and $9.3$. These results suggest that low-physical resolution formulations lead to premature onset of turbulence and consequent turbulence overprediction, causing the rapid decay of $k$ observed in figure \ref{fig:4.2_4a}.

In contrast, high-physical resolution computations are in good agreement with the reference numerical experiments. For example, the results of the simulations at $f_k\leq 0.25$ show that the maximum value of $\varepsilon$ occurs at $t=8.8-9.3$ and ranges from $0.145$ to $0.146$, whereas the DNS studies report $(\varepsilon)_{\max}$ between $0.143$ and $0.153$ at $t=8.8-9.3$. The largest differences between PANS and DNS simulations occur at late times, $t>10$. These small discrepancies are likely caused by numerical uncertainty and Ma effects. 
%
%
%
\subsection{Physical interpretation of the results}
\label{sec:4.3}
%
The accuracy of the numerical results shows strong dependence on the physical resolution ($f_k$) of the model. Whereas high-physical resolution simulations ($f_k < 0.50$) reveal a good agreement with the DNS studies  \cite{BRACHET_JFM_1983,DRIKAKIS_JOT_2007}, those performed at low-physical resolution ($f_k\ge 0.50$) lead to large discrepancies with the reference data. The modeling and physical reasons for this outcome are now examined.
%
%
\subsubsection{Effective ratio of modeled-to-total turbulent kinetic energy}
\label{sec:4.3.1}
%
\begin{figure}[t!]
\centering
\includegraphics[scale=0.11,trim=0 0 0 0,clip]{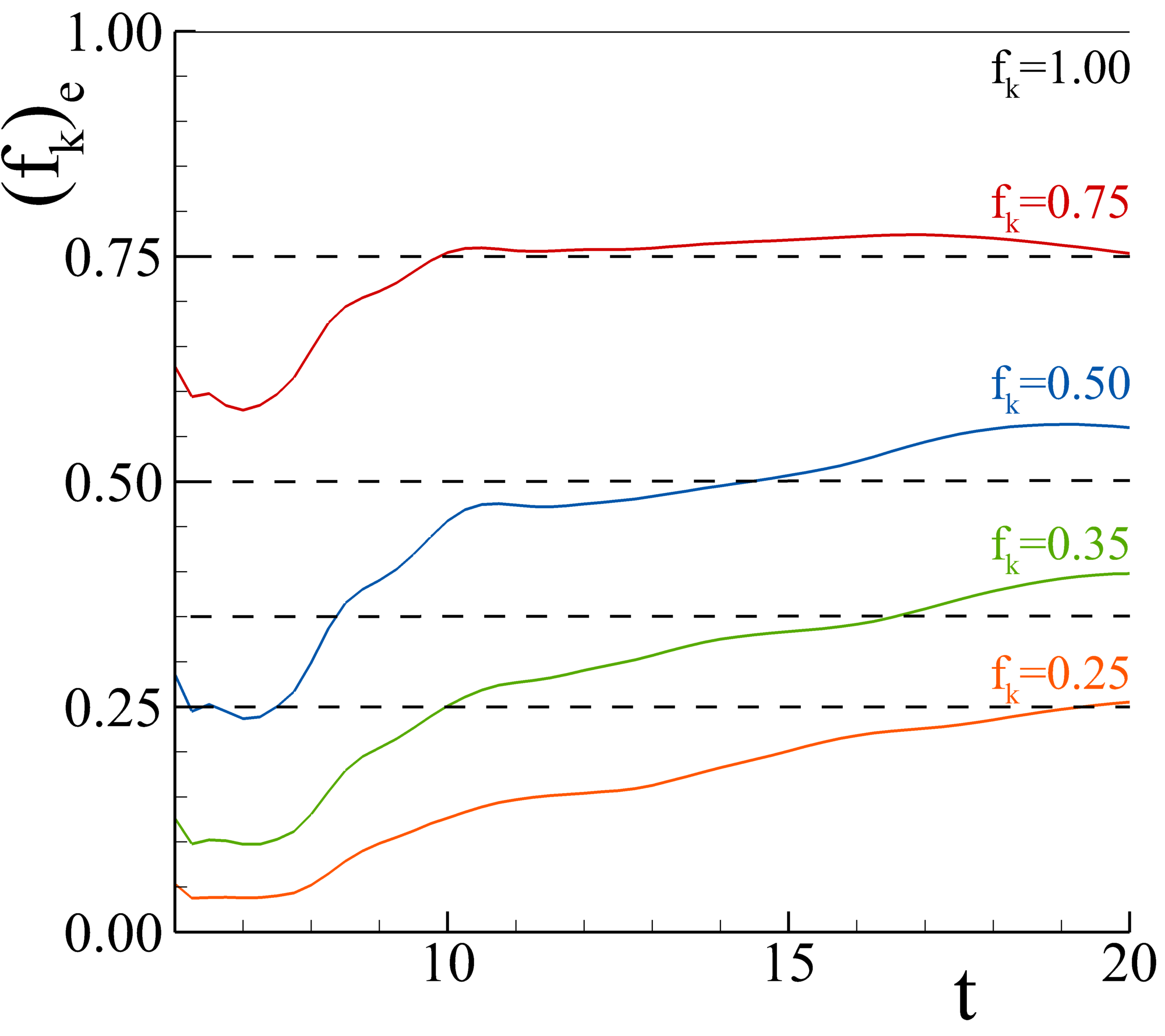}
\caption{Temporal evolution of the effective $f_k$, $(f_k)_e$, for simulations using different values of $f_k$.}
\label{fig:4.3.1_1}
\end{figure}

In the PANS {\color{blue}BHR-LEVM} model, the physical resolution is set through the modeled-to-total ratio of turbulence kinetic energy $f_k$ ($f_\varepsilon =1.00$). Yet, this parameter may not match the effective $f_k$,
\begin{equation}
\label{4.3.1_1}
(f_k)_e = \frac{k_u}{{k_u}_{(f_k=1.00)}} \; ,
\end{equation}
of the simulations in cases where the closure misrepresents the turbulent kinetic energy field. {\color{blue}This is often the case in problems characterized by massive flow separation or turbulence onset since RANS closures ($f_k=1.00$) tend to overpredict the $k_u$ of such flows}.

To analyze this aspect, figure \ref{fig:4.3.1_1} depicts the temporal evolution of $(f_k)_e$ for computations at different values of $f_k$. This figure only considers the period in which the solutions are dependent on $f_k$, $t\ge t_c$. The results indicate that $(f_k)_e$ is initially significantly smaller than the prescribed $f_k$ (for cases at $f_k<1.00$). This stems from the fact that the onset of turbulence delays with the refinement of $f_k$ (see Section \ref{sec:4.3.3}). After this period ($t>8-9$), $(f_k)_e$ grows for all cases and gets closer to the prescribed value. Among all $f_k$, only the case at $f_k=0.75$ converges toward the prescribed value during the simulated time. This is not observed for the remaining cases, highlighting that $f_k\ne (f_k)_e$ and reaffirming the dependence of the turbulent kinetic energy on the physical resolution. {\color{blue} Also, the pronounced differences between $f_k$ and $(f_k)_e$ observed during the development of turbulence ($t\leq 9$) indicate that low-physical resolution simulations overpredict turbulence during this period, explaining the results of Section \ref{sec:4.2}. We emphasize that transient flows are highly sensitive to history effects, and high-resolution PANS simulations showing $f_k \approx (f_k)_e$ would indicate that the RANS closure can accurately represent the mean-flow field of the selected problem. This is often observed in statistically steady flows.}
\begin{figure}[t!]
\centering
\includegraphics[scale=0.11,trim=0 0 0 0,clip]{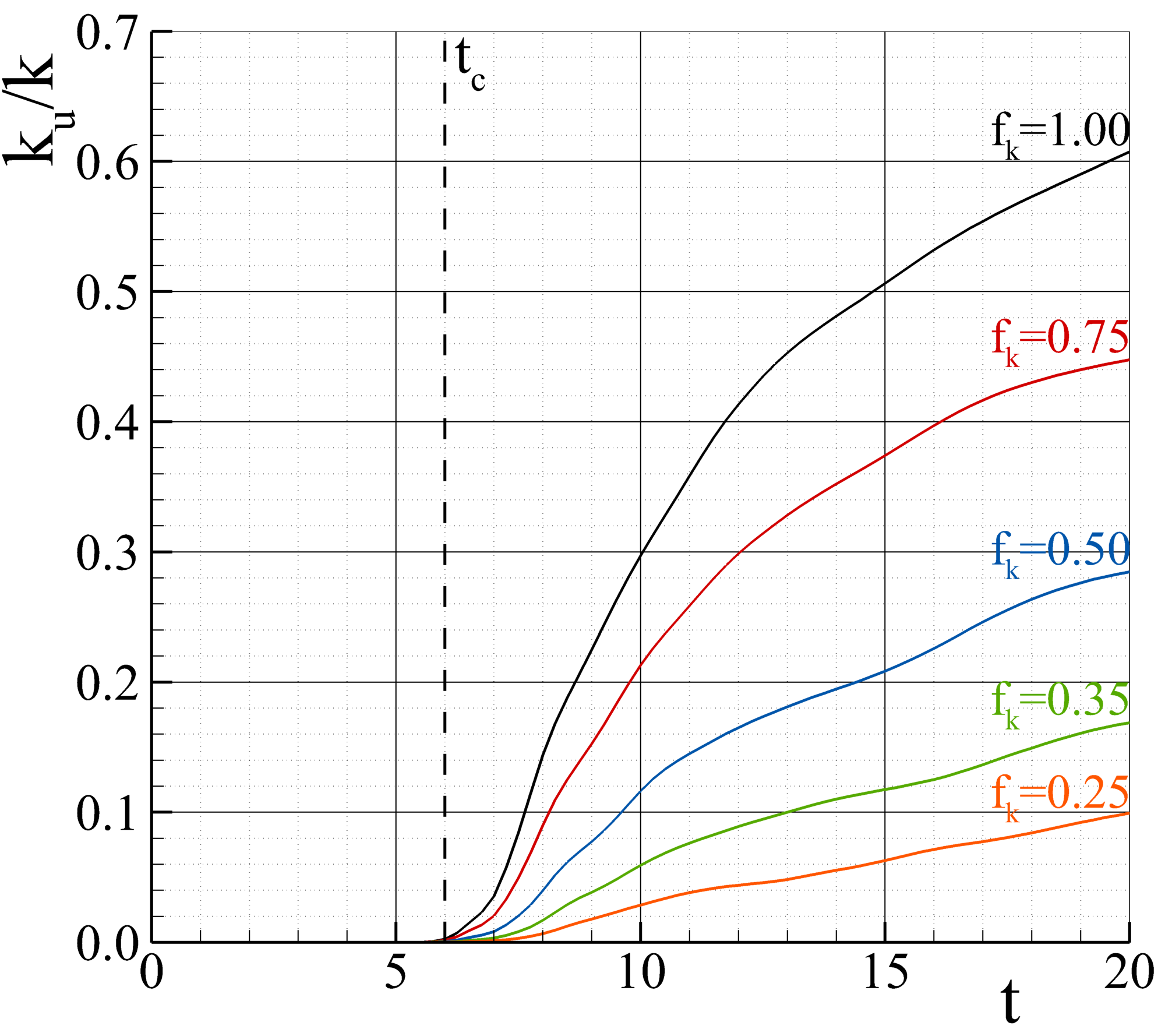}
\caption{Temporal evolution of the ratio modeled-to-total kinetic energy, $k_u/k$, for simulations using different values of $f_k$.}
\label{fig:4.3.1_2}
\end{figure}

Next, figure \ref{fig:4.3.1_2} depicts the ratio of unresolved-to-total kinetic energy, $k_u/k$, to assess the modeled fraction of $k$. {\color{blue} Note that $k_u$ refers to the unresolved turbulence kinetic energy, whereas $k$ comprises a laminar and turbulent component.} The data show that $k_u/k$ is nearly zero until $t=t_c$. This explains the observed independence of $k$ and $\varepsilon$ solutions (figures \ref{fig:4.2_4} and \ref{fig:4.2_5}) from $f_k$. After this instant, this quantity presents a rapid growth, which rate is closely dependent on $f_k$. 
Also, and considering the two periods of the development of the TGV flow (see Section \ref{sec:4.1}), it is possible to infer the following: \textit{i)} during the first period, $t\leq 9$, the magnitude of $k_u$ is relatively small compared to $k$.  This indicates that even small fractions of $k'$ have a strong impact on the flow physics and accuracy of the simulations. \textit{ii)} At later times, $t>9$, the contribution of $k_u$ grows significantly. Nevertheless, at $t=20$, $k_u/k$ still does not exceed $16.9\%$ and $60.8\%$ for simulations using $f_k=0.35$ and $1.00$, respectively.

From a modeling perspective, the most significant result of figure \ref{fig:4.3.1_2} is the fact that $k_u/k$ predicted with $f_k=1.00$ is approximately six times larger than at $f_k=0.25$. This outcome is caused by the early onset and overprediction of turbulence by low-physical resolution simulations. It is also interesting to note that the ratio $k_u/k$ for high-physical resolution simulations only starts increasing at $t\ge 7$, i.e., after the vortex-reconnection processes observed in figure \ref{fig:4.1_1}. Recall that this is the time instant when we first observe bursts of {\color{blue}fine-scale turbulence}.
%
%
\subsubsection{Effective Reynolds number}
\label{sec:4.3.2}
%
One of the main consequences of modeling the turbulent field is the inherent reduction of the effective Reynolds number, $\mathrm{Re_e}$,
\begin{equation}
\label{4.3.2_1}
\mathrm{Re_e} \equiv \frac{V_oL_o}{\nu+\nu_u} \; ,
\end{equation}
of the computations. Although lower physical resolutions reduce this quantity and relax the simulations' cost, an excessively large decrease of $\mathrm{Re_e}$ may compromise the accuracy of the simulations. Pereira et al.  \cite{PEREIRA_JCP_2018,PEREIRA_OE_2019,PEREIRA_IJHFF_2019} demonstrated the importance of this quantity to the prediction of flows around circular cylinders in the sub-critical regime \cite{WILLIAMSON_ARFM_1996,ZDRAVKOVICH_BOOK_1997}. These studies show that an excessive reduction of $\mathrm{Re_e}$ can suppress the instabilities and coherent structures driving the flow dynamics, leading to poor results. Since the TGV flow exhibits a strong dependence on the magnitude of Re \cite{BRACHET_JFM_1983}, we now evaluate the effect of the physical resolution on $\mathrm{Re_e}$.
\begin{figure}[t!]
\centering
\includegraphics[scale=0.11,trim=0 0 0 0,clip]{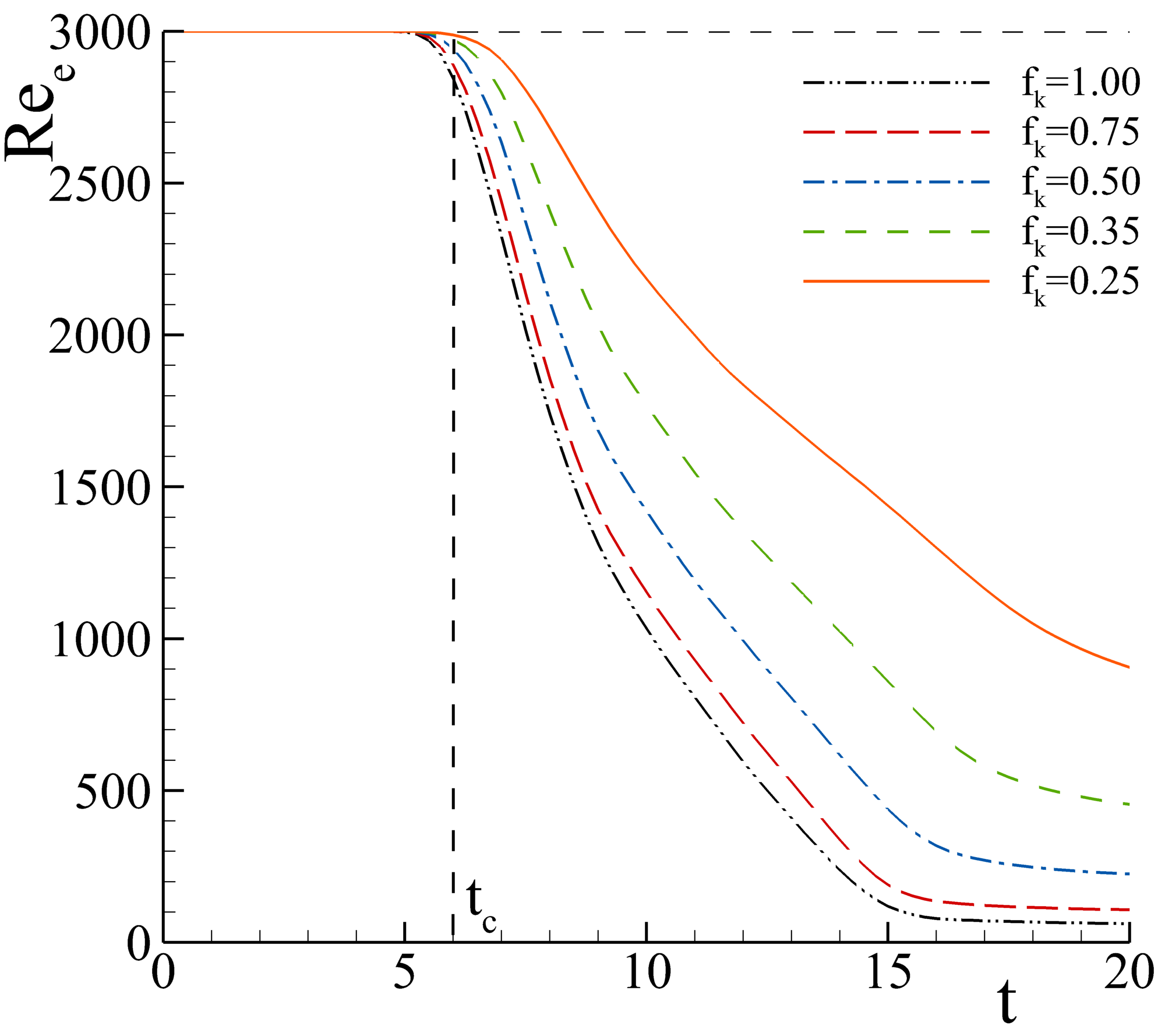}
\caption{Temporal evolution of the effective Reynolds number, $\mathrm{Re_e}$, for simulations using different values of $f_k$.}
\label{fig:4.3.2_1}
\end{figure}

The temporal evolution of $\mathrm{Re_e}$ is depicted in figure \ref{fig:4.3.2_1} for different values of $f_k$. The results show that this quantity is independent of the physical resolution until $t=5$. Also, only high-physical resolution simulations maintain $\mathrm{Re_e}\approx 3000 = \mathrm{Re}$ until $t=t_c$. After this instant, the magnitude of $\mathrm{Re_e}$ experiences a rapid decay, which is particularly pronounced for low-physical resolution simulations. The results also show that the beginning and steepness of the $\mathrm{Re_e}$ reduction depends on $f_k$. For example, it is observed that the averaged value of $\mathrm{Re_e}$ at $t=9$ is equal to $2033$ and $1313$ for simulations at $f_k=0.35$ and $1.00$, respectively. At later times, $t>15$, $\mathrm{Re_e}$ starts becoming independent of time. This stems from a nearly constant averaged value of $\nu_u$, which suggests that the turbulent field is becoming fully-developed. These features occur later as the physical resolution refines. 
%
%
\subsubsection{Coherent field}
\label{sec:4.3.3}

Section \ref{sec:4.1} has shown that the development of the TGV flow comprises two distinct periods. In the first, the flow evolves from a laminar to a turbulent state, driven by multiple coherent structures and instabilities. On the other hand, the second period is characterized by fully-developed, high-intensity, and decay of turbulence. Comparing the features of the two periods, it is possible to infer that the first poses more challenges to modeling and simulation {\color{blue}because one-point turbulence closures have not been designed for transitional flow.} The numerical results have confirmed this idea by showing that the discrepancies between simulations at different physical resolutions begin at $t=t_c$. This instant coincides with the vortex-reconnection {\color{blue}process}.
\begin{table}[b!]
\centering
\setlength\extrarowheight{3pt}
\caption{Maximum value of the turbulent-to-molecular viscosity, $(\nu_u/\nu)_{\max}$, registered for different instants and values of $f_k$.}
\label{tab:4.3.3_1}  
\begin{tabular}{C{1.0cm}C{1.0cm}C{1.0cm}C{1.0cm}C{1.0cm}C{1.0cm}C{1.0cm}}
\hline 
$t$ & $0.00$ & $0.25$ & $0.35$ & $0.50$ & $0.75$ & $1.00$ \\
\hline 
$6.0$ 	& $0.0$	& $0.3$	& $0.7$ 	& $1.7$ 	& $4.4$ 	& $8.6$ \\ 
$6.5$ 	& $0.0$ 	& $1.1$	& $2.7$ 	& $6.3$ 	& $16.0$ 	& $30.1$\\ 
$7.0$ 	& $0.0$ 	& $1.3$	& $2.9$ 	& $6.6$ 	& $17.8$ 	& $35.5$\\ 
\hline
\end{tabular}
\end{table}
\begin{figure*}[t!]
\centering
\subfloat[$t=6.0$ at $f_k=0.00$.]{\label{fig:4.3.3_1a}
\includegraphics[scale=0.13,trim=0 0 0 0,clip]{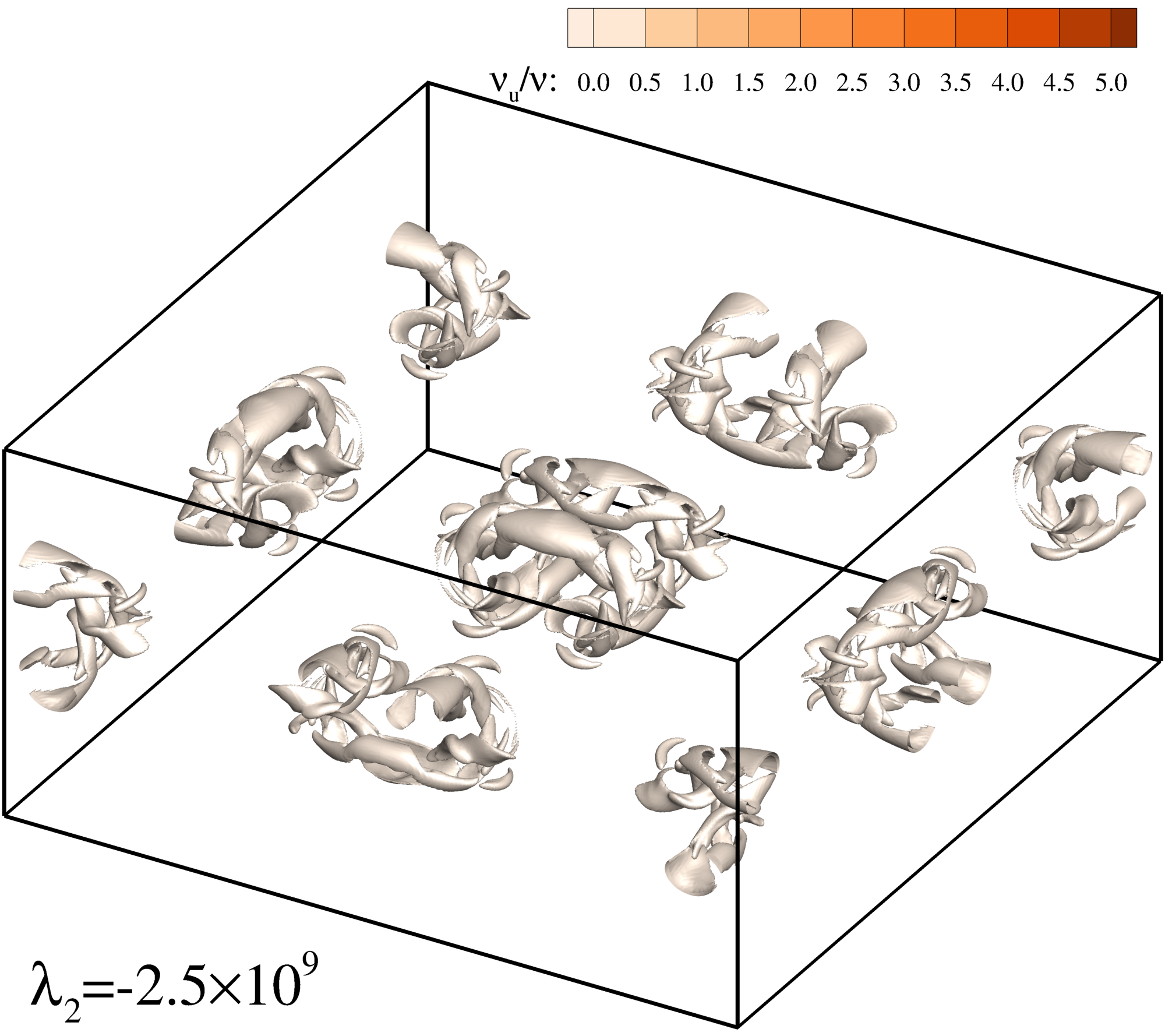}}
\hspace{5mm}
~
\subfloat[$t=6.0$ at $f_k=0.25$.]{\label{fig:4.3.3_1b}
\includegraphics[scale=0.13,trim=0 0 0 0,clip]{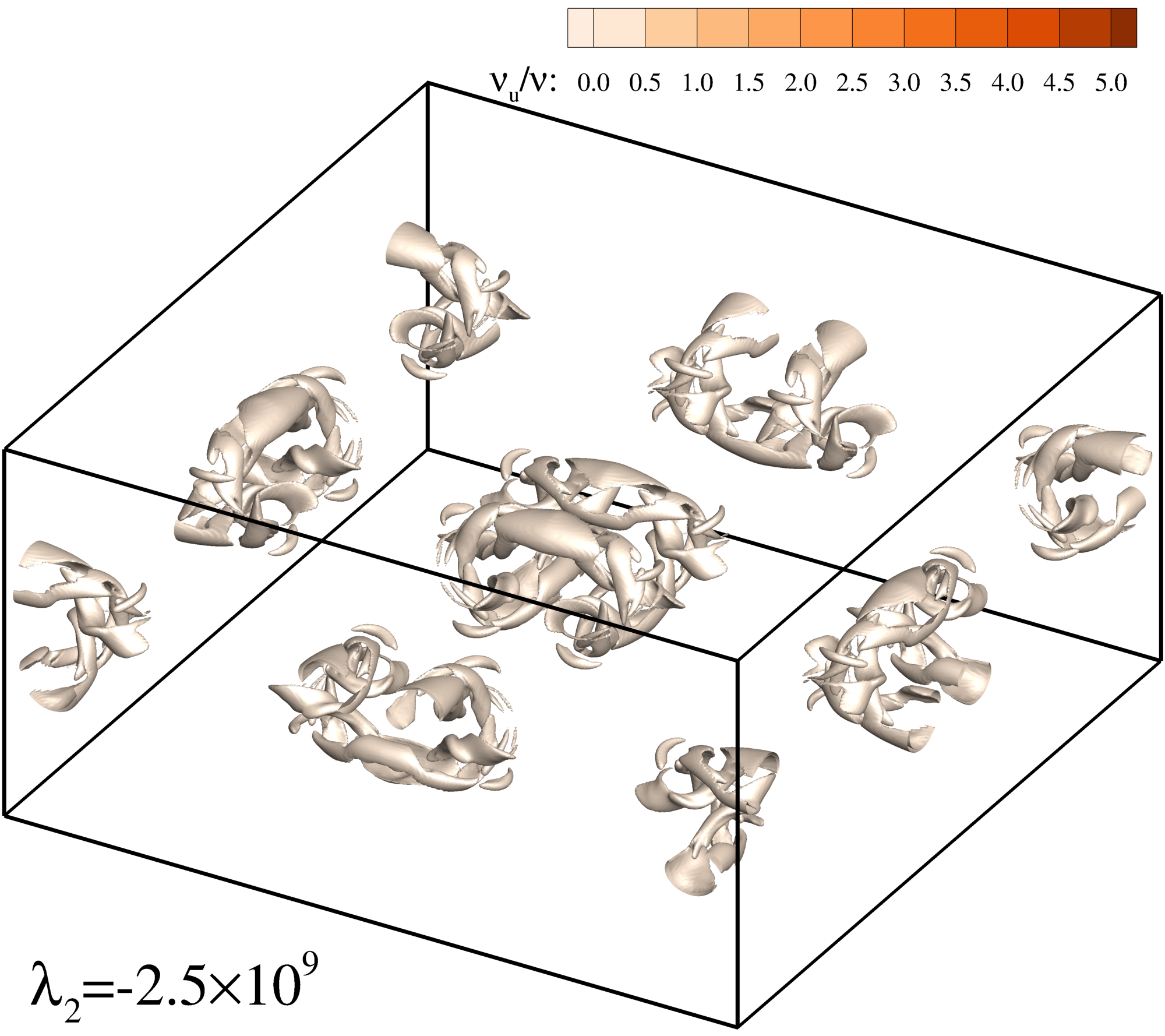}}
\\
\subfloat[$t=6.5$ at $f_k=0.00$.]{\label{fig:4.3.3_1c}
\includegraphics[scale=0.13,trim=0 0 0 0,clip]{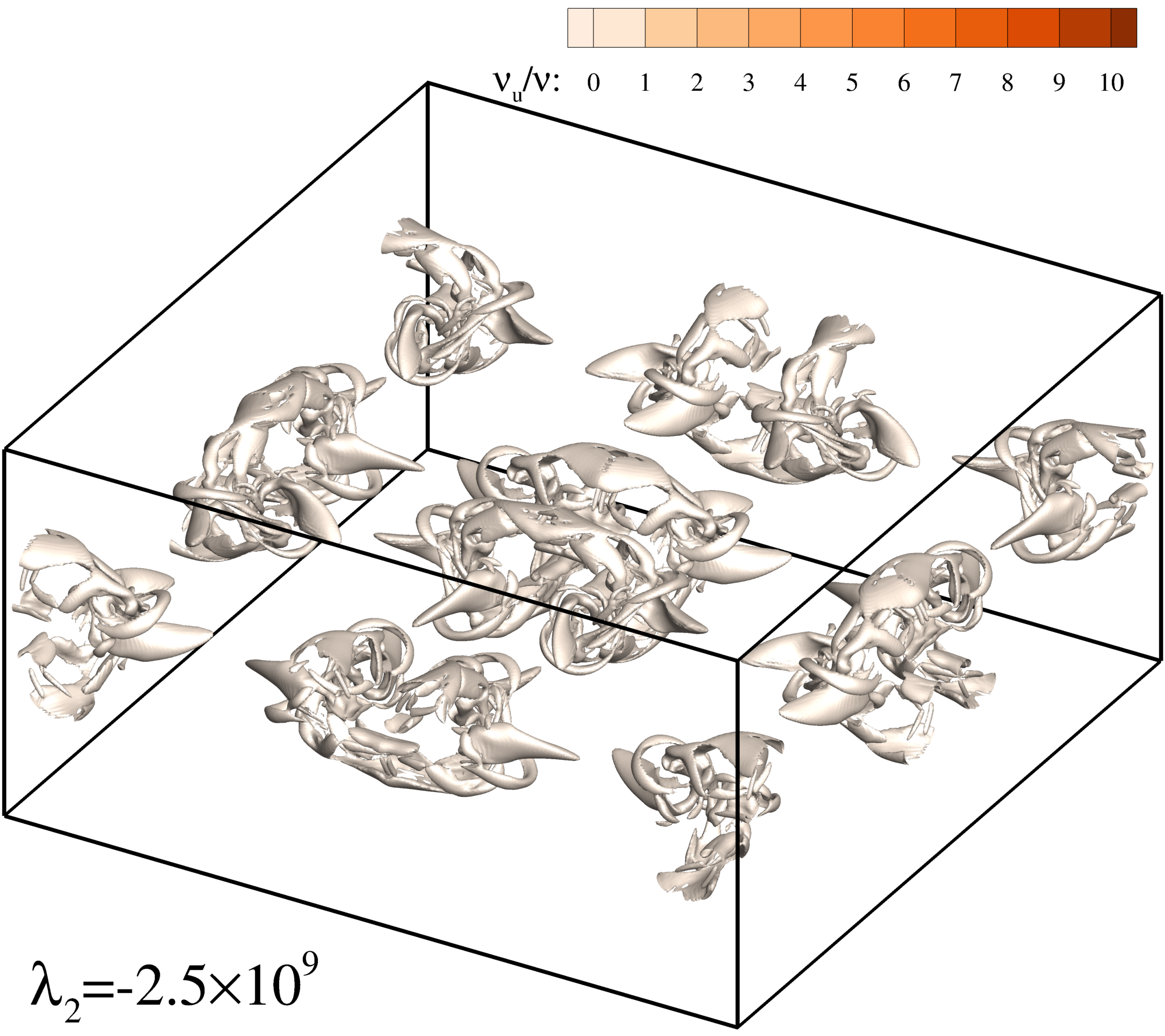}}
\hspace{5mm}
~
\subfloat[$t=6.5$ at $f_k=0.25$.]{\label{fig:4.3.3_1d}
\includegraphics[scale=0.13,trim=0 0 0 0,clip]{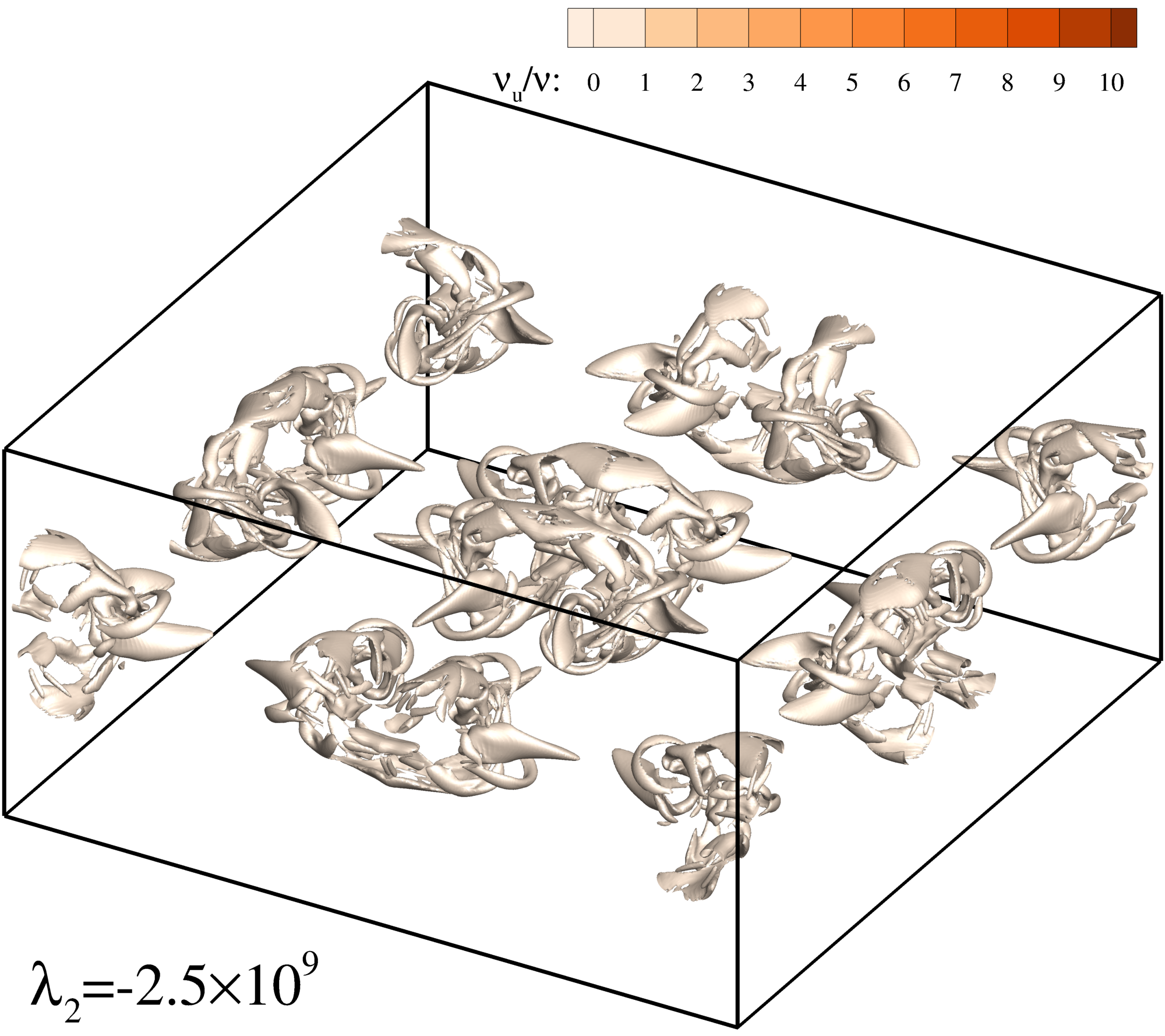}}
\\
\subfloat[$t=7.0$ at $f_k=0.00$.]{\label{fig:4.3.3_1e}
\includegraphics[scale=0.13,trim=0 0 0 0,clip]{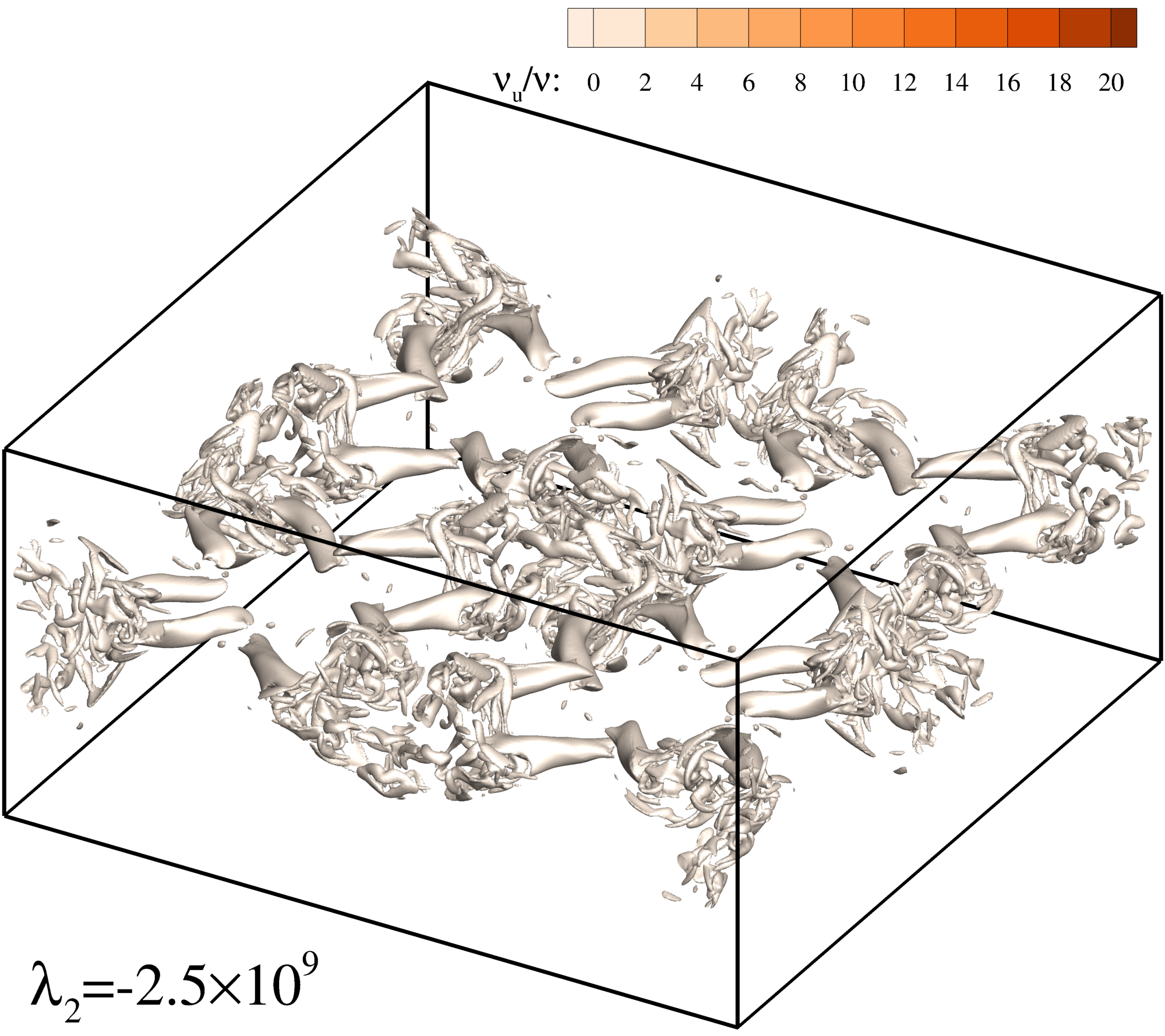}}
\hspace{5mm}
~
\subfloat[$t=7.0$ at $f_k=0.25$.]{\label{fig:4.3.3_1f}
\includegraphics[scale=0.13,trim=0 0 0 0,clip]{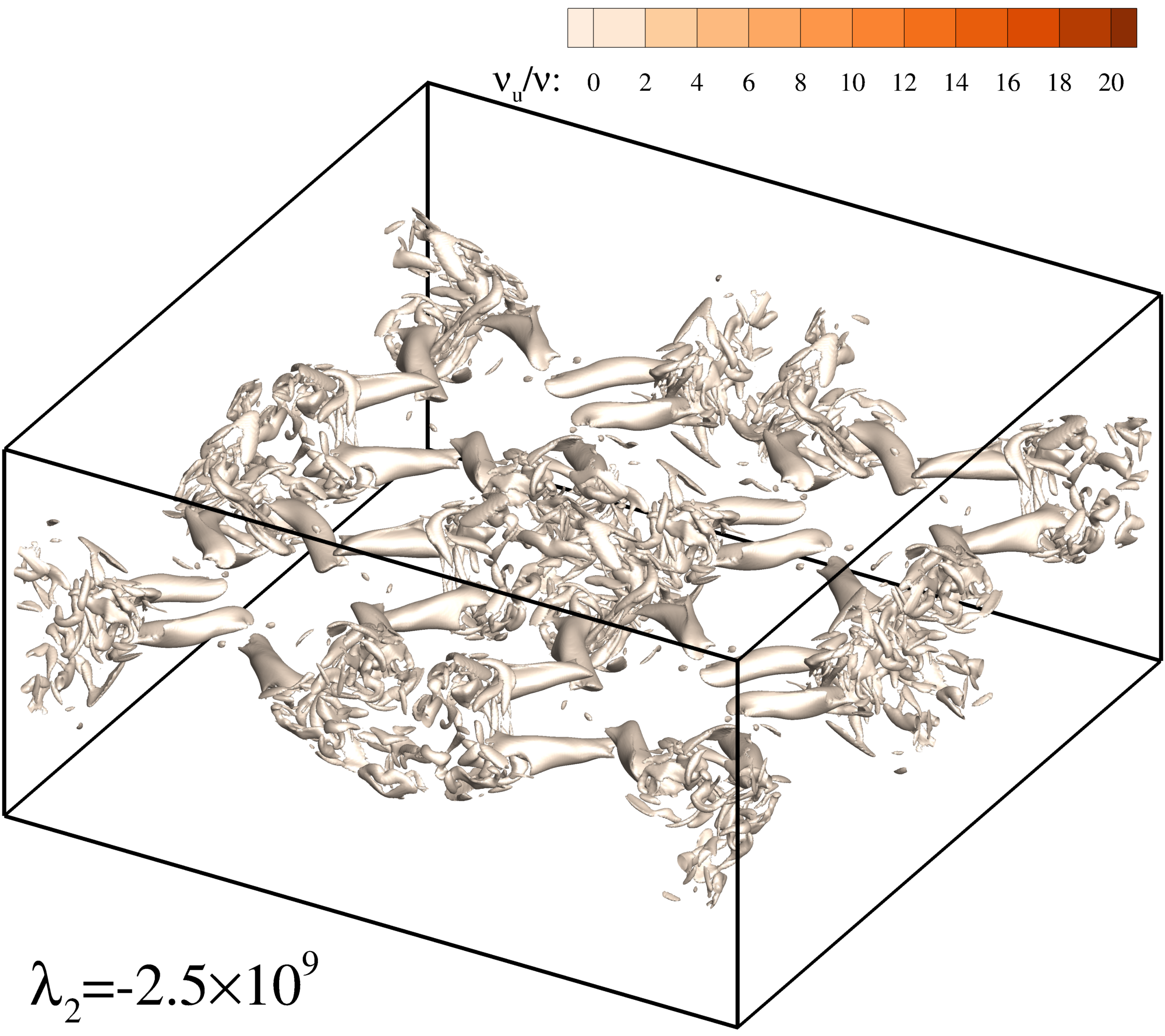}}
\caption{Temporal evolution of the vortical structures of the TGV flow for simulations at $f_k=0.00$ and $0.25$. Structures captured with $\lambda_2$-criterion ($\mathrm{s^{-2}}$) \cite{JEONG_JFM_1995}.}
\label{fig:4.3.3_1}
\end{figure*}

\begin{figure*}[t!]
\centering
\subfloat[$t=6.0$ at $f_k=0.35$.]{\label{fig:4.3.3_2a}
\includegraphics[scale=0.13,trim=0 0 0 0,clip]{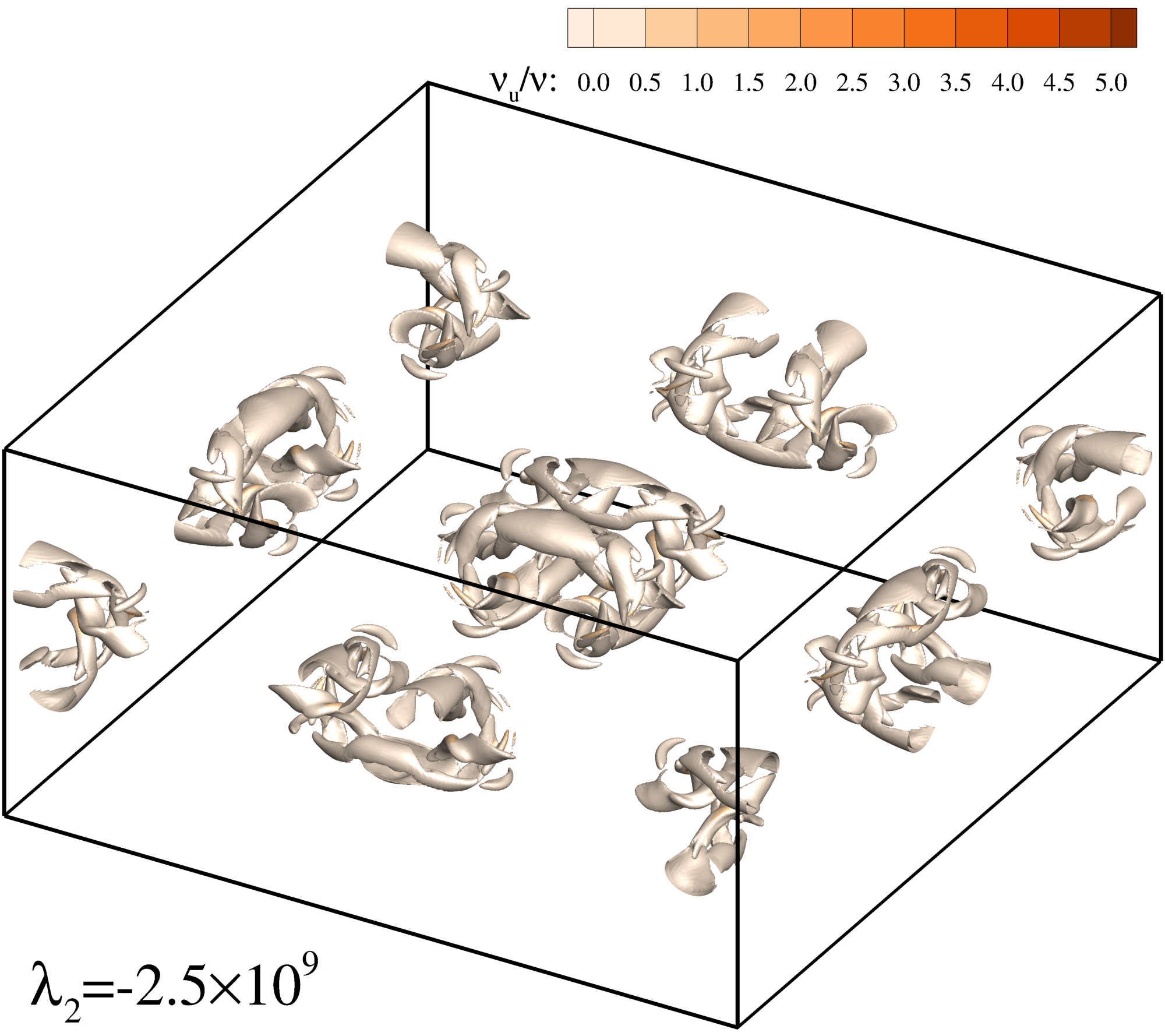}}
\hspace{5mm}
~
\subfloat[$t=6.0$ at $f_k=0.50$.]{\label{fig:4.3.3_2b}
\includegraphics[scale=0.13,trim=0 0 0 0,clip]{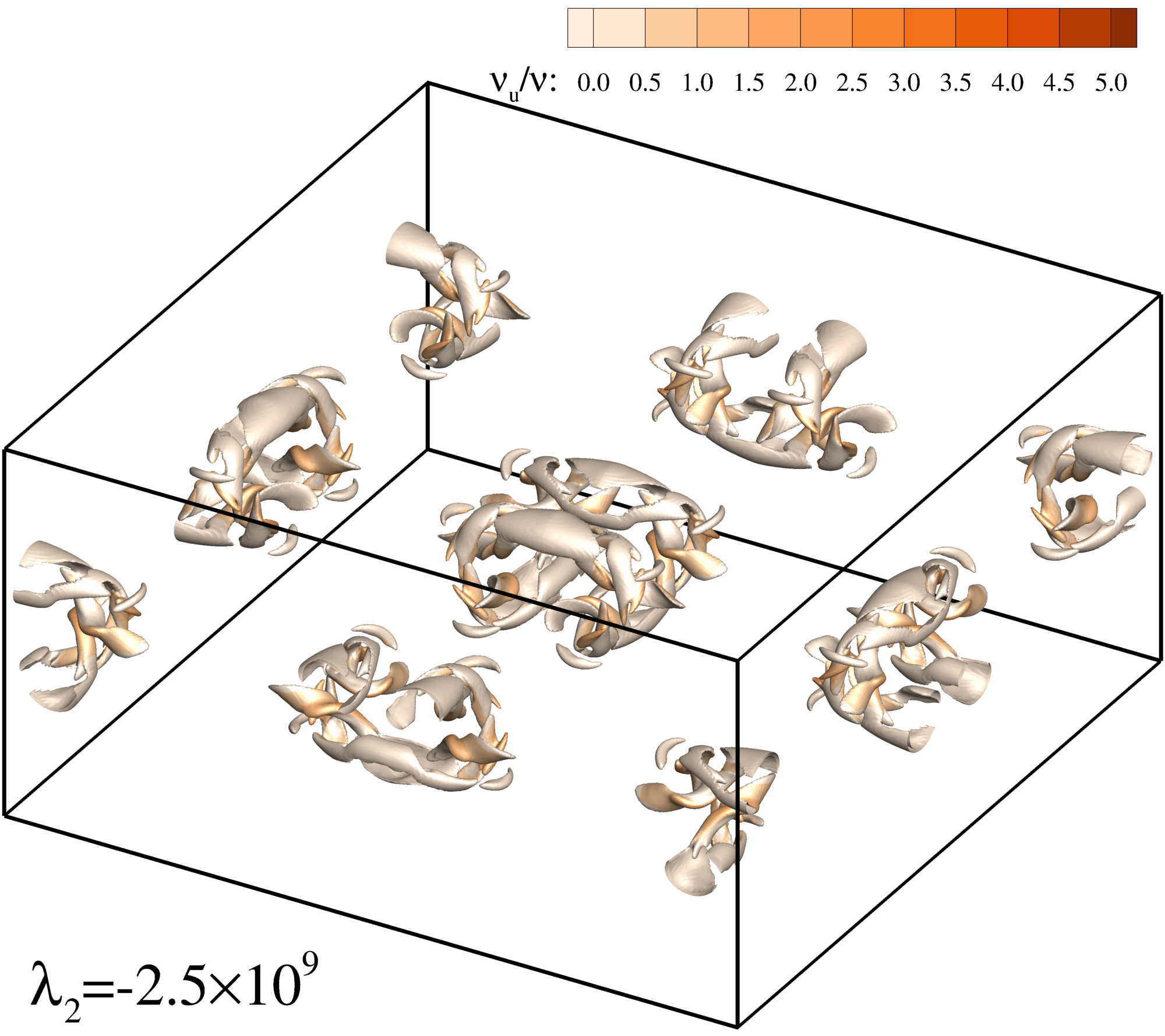}}
\\
\subfloat[$t=6.5$ at $f_k=0.35$.]{\label{fig:4.3.3_2c}
\includegraphics[scale=0.13,trim=0 0 0 0,clip]{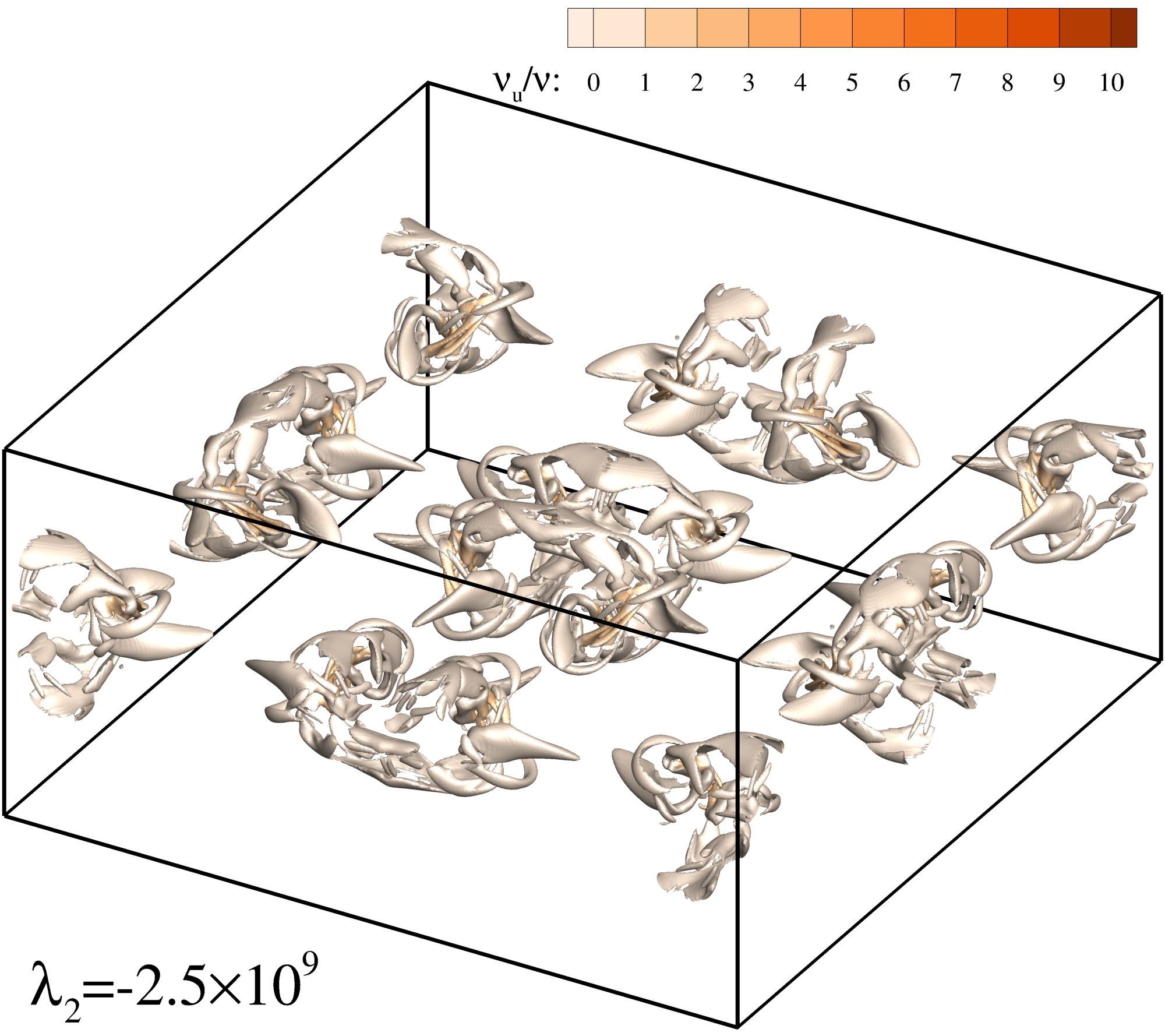}}
\hspace{5mm}
~
\subfloat[$t=6.5$ at $f_k=0.50$.]{\label{fig:4.3.3_2d}
\includegraphics[scale=0.13,trim=0 0 0 0,clip]{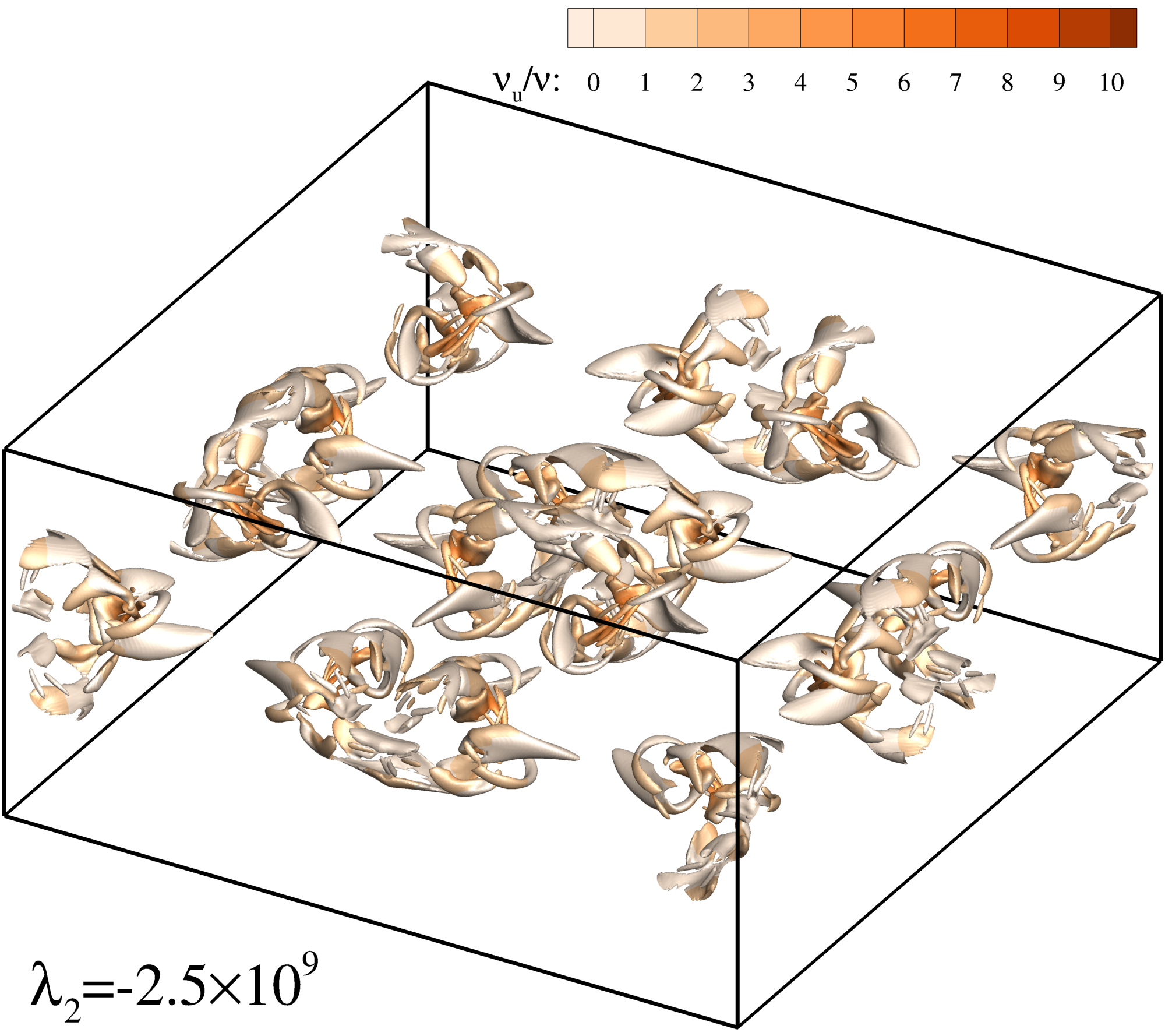}}
\\
\subfloat[$t=7.0$ at $f_k=0.35$.]{\label{fig:4.3.3_2e}
\includegraphics[scale=0.13,trim=0 0 0 0,clip]{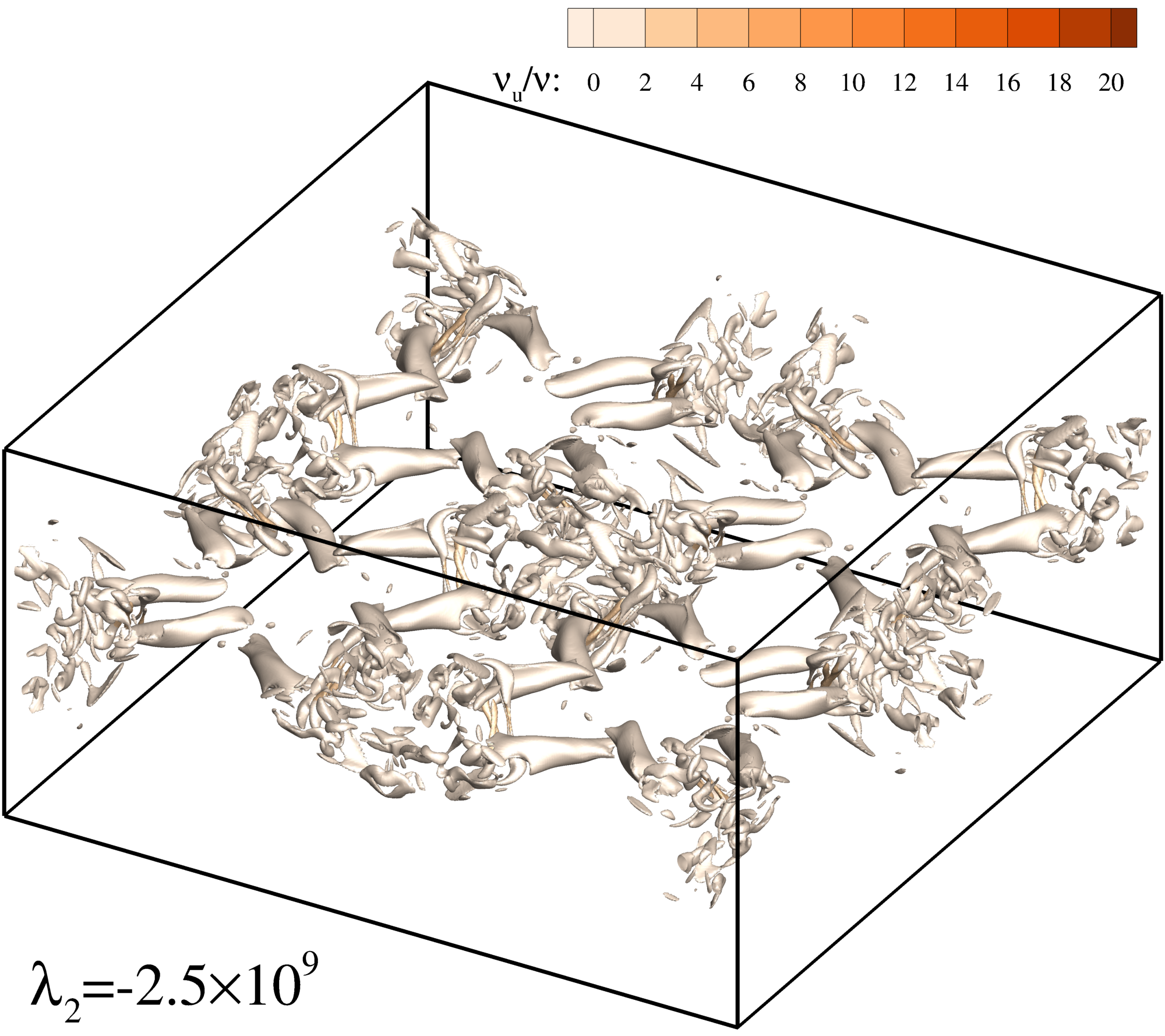}}
\hspace{5mm}
~
\subfloat[$t=7.0$ at $f_k=0.50$.]{\label{fig:4.3.3_2f}
\includegraphics[scale=0.13,trim=0 0 0 0,clip]{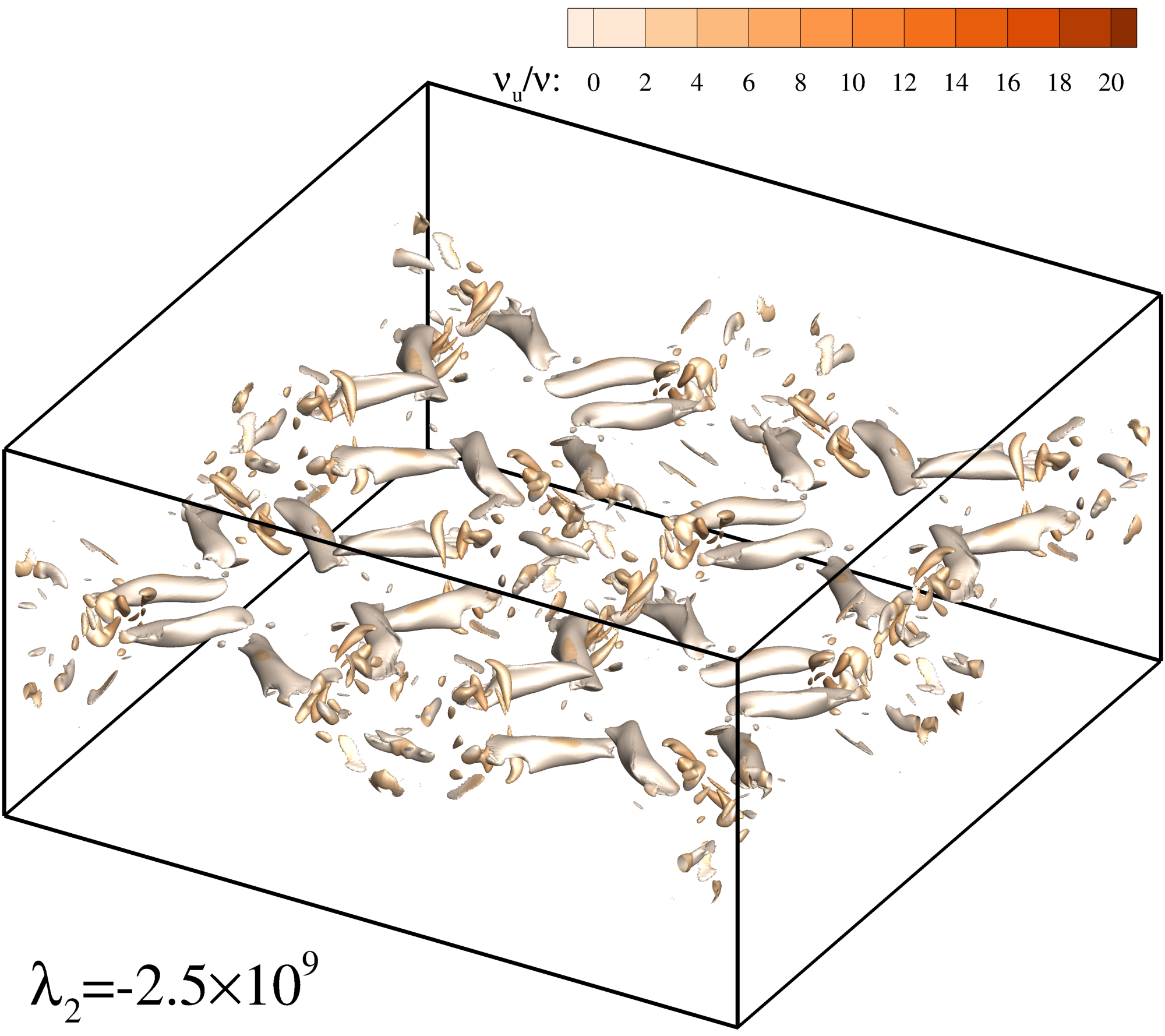}}
\caption{Temporal evolution of the vortical structures of the TGV flow for simulations at $f_k=0.35$ and $0.50$. Structures captured with $\lambda_2$-criterion ($\mathrm{s^{-2}}$) \cite{JEONG_JFM_1995}.}
\label{fig:4.3.3_2}
\end{figure*}

\begin{figure*}[t!]
\centering
\subfloat[$t=6.0$ at $f_k=0.75$.]{\label{fig:4.3.3_3a}
\includegraphics[scale=0.13,trim=0 0 0 0,clip]{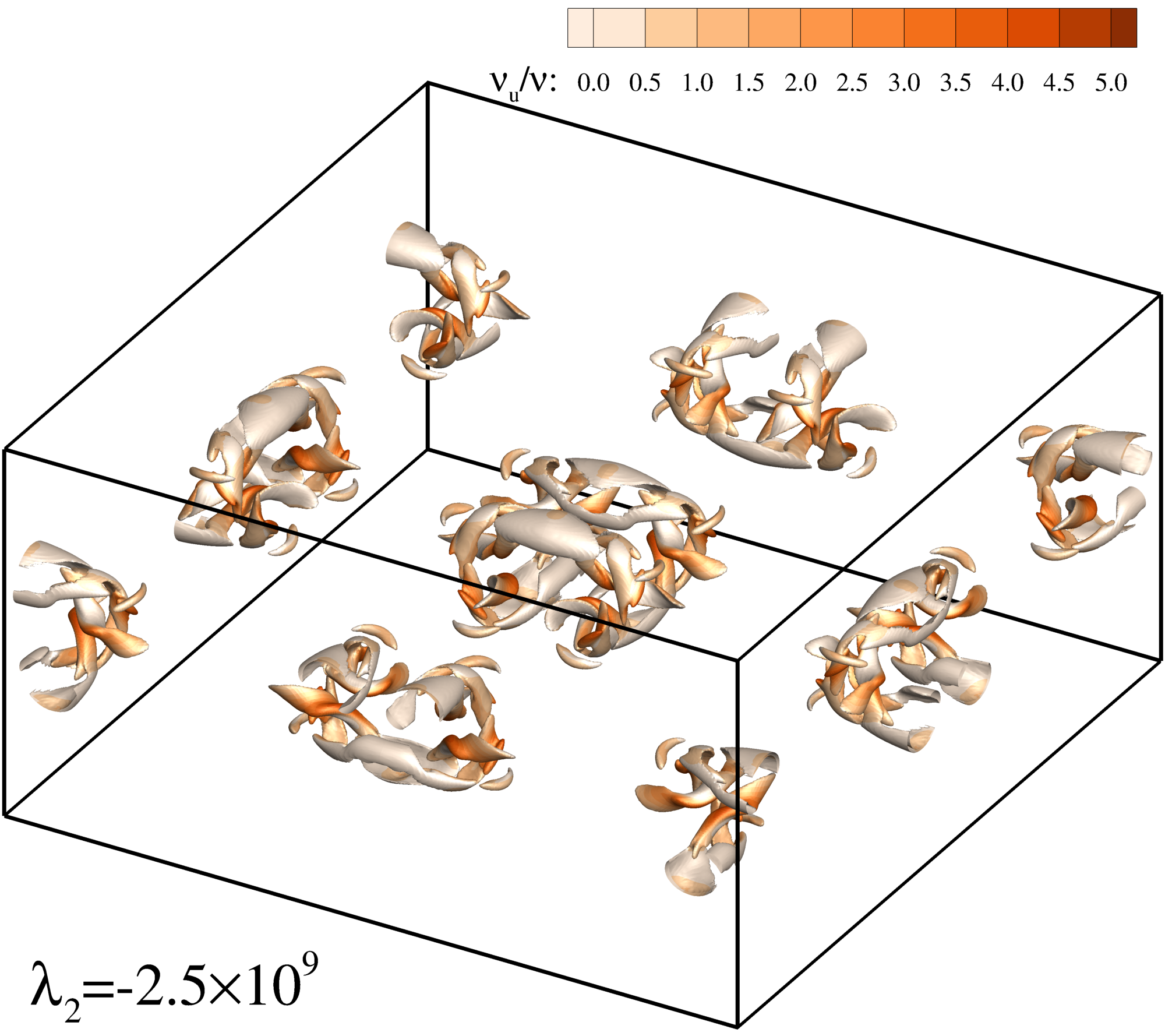}}
\hspace{5mm}
~
\subfloat[$t=6.0$ at $f_k=1.00$.]{\label{fig:4.3.3_3b}
\includegraphics[scale=0.13,trim=0 0 0 0,clip]{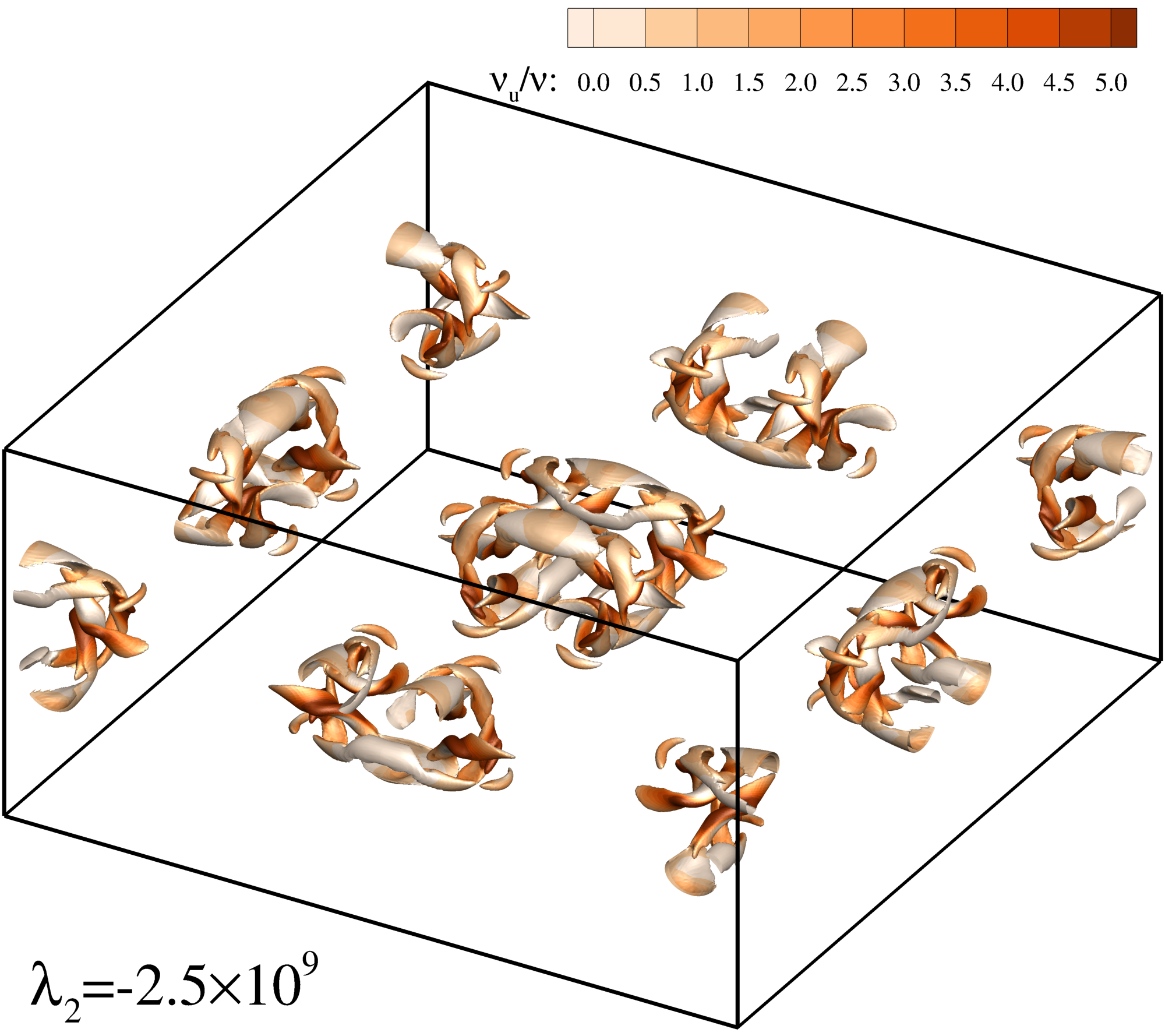}}
\\
\subfloat[$t=6.5$ at $f_k=0.75$.]{\label{fig:4.3.3_3c}
\includegraphics[scale=0.13,trim=0 0 0 0,clip]{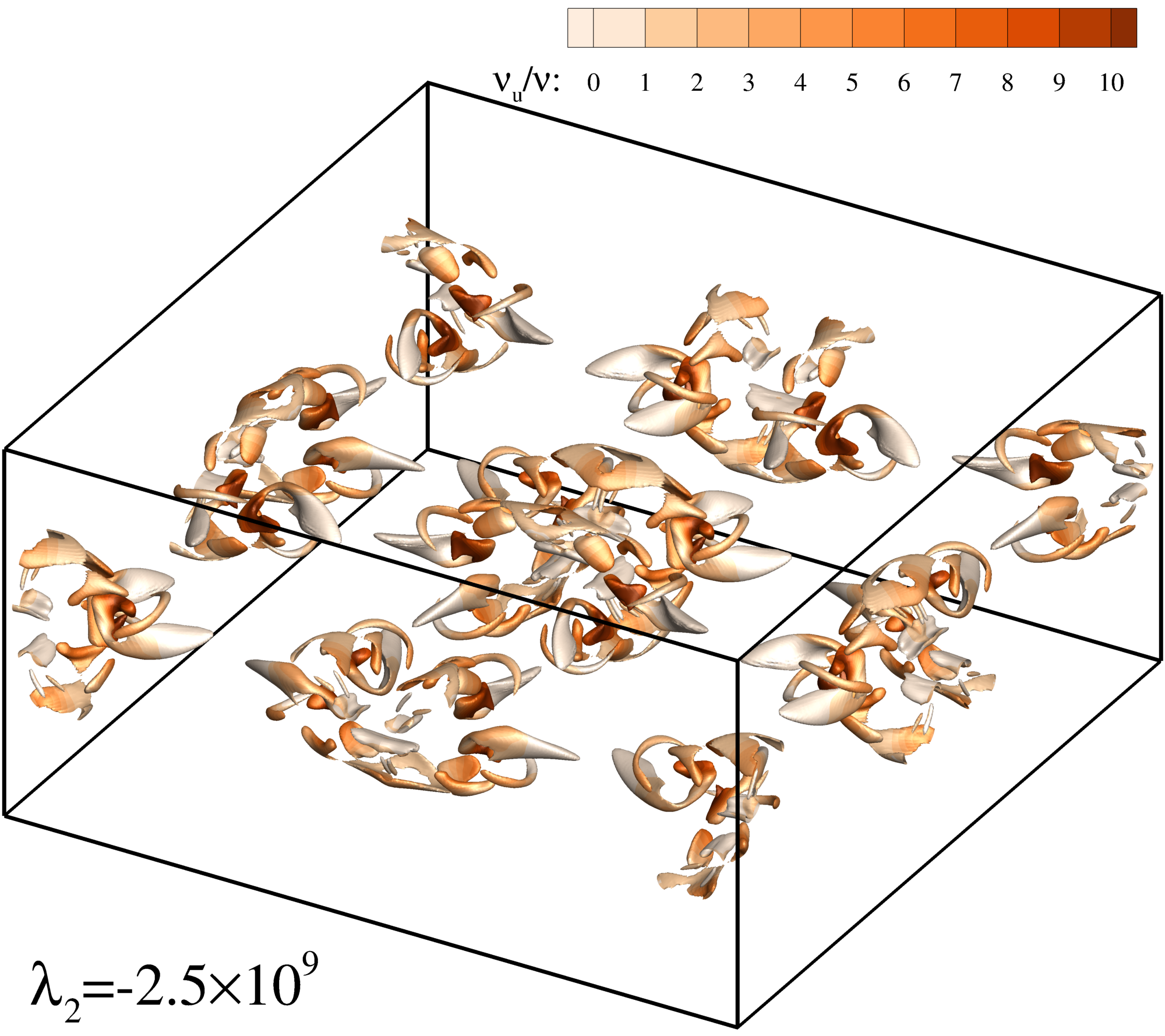}}
\hspace{5mm}
~
\subfloat[$t=6.5$ at $f_k=1.00$.]{\label{fig:4.3.3_3d}
\includegraphics[scale=0.13,trim=0 0 0 0,clip]{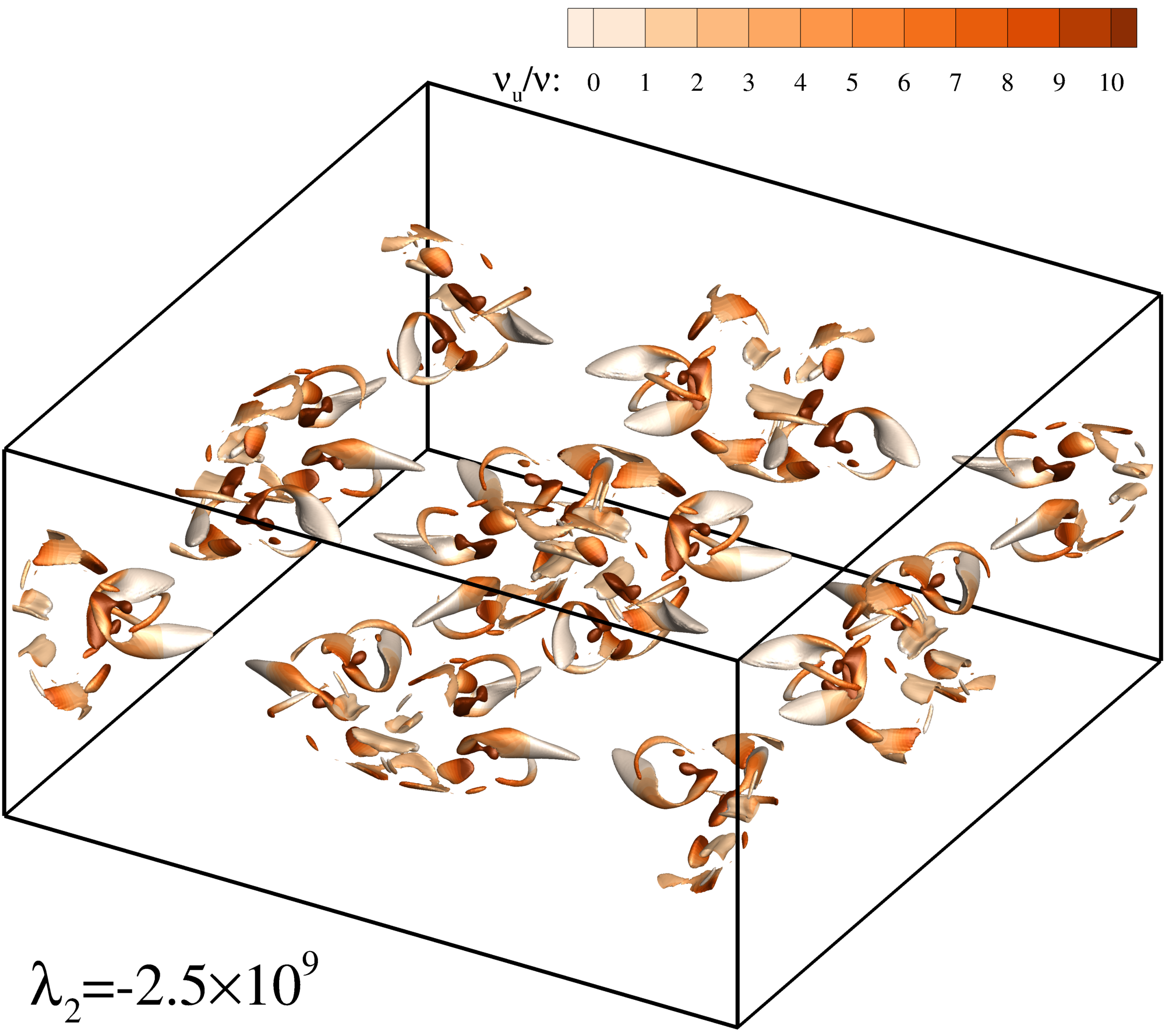}}
\\
\subfloat[$t=7.0$ at $f_k=0.75$.]{\label{fig:4.3.3_3e}
\includegraphics[scale=0.13,trim=0 0 0 0,clip]{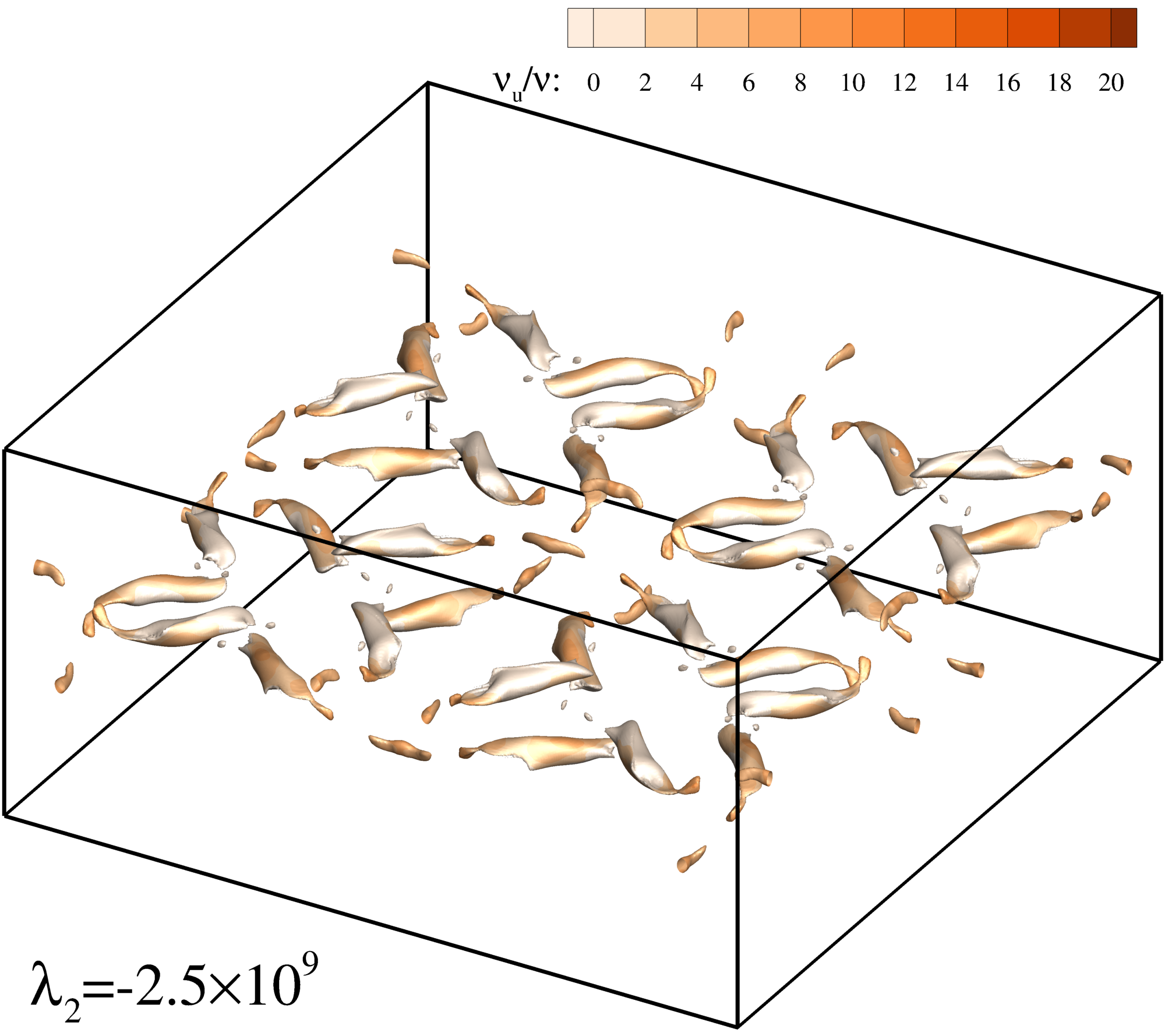}}
\hspace{5mm}
~
\subfloat[$t=7.0$ at $f_k=1.00$.]{\label{fig:4.3.3_3f}
\includegraphics[scale=0.13,trim=0 0 0 0,clip]{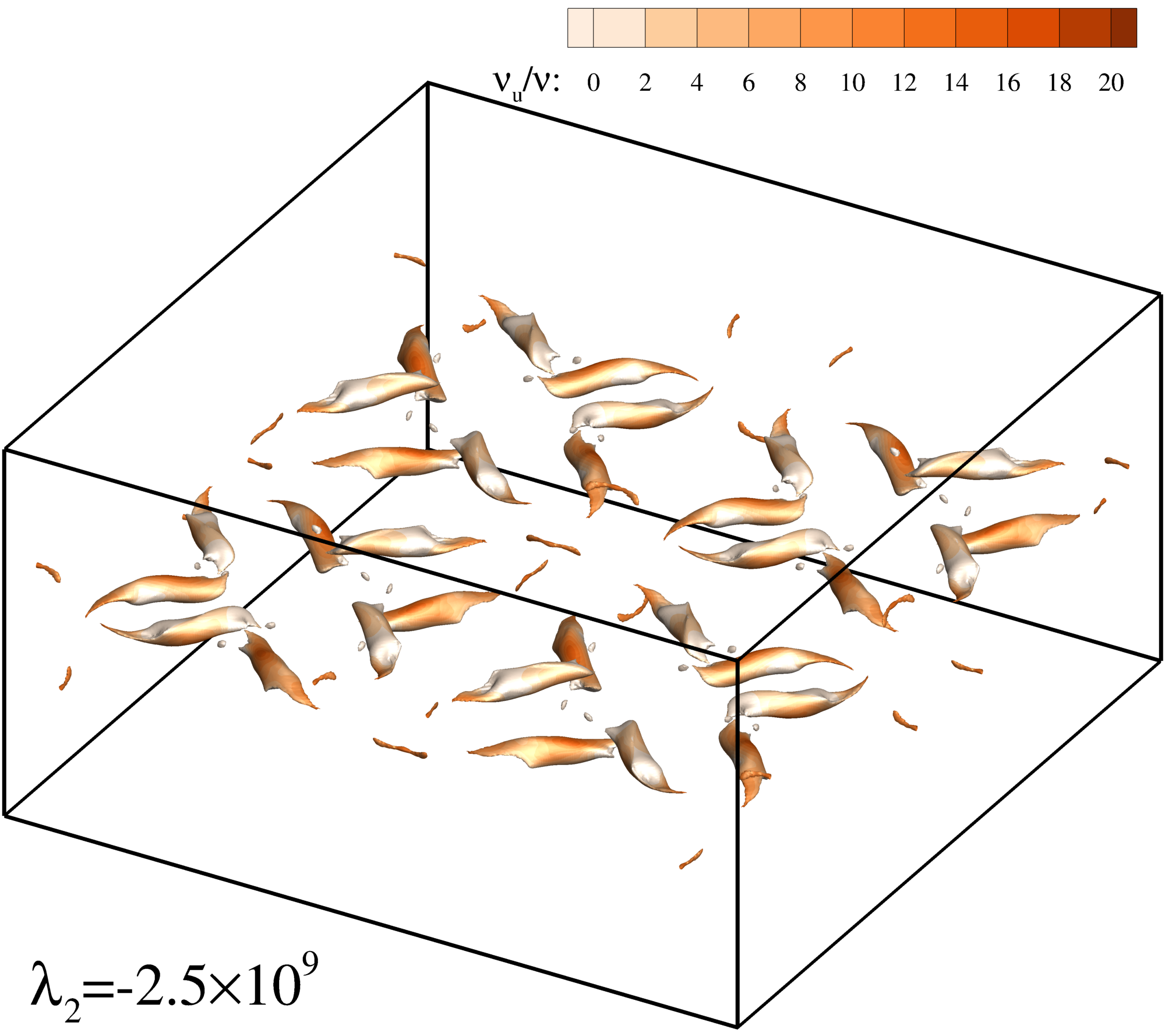}}
\caption{Temporal evolution of the vortical structures of the TGV flow for simulations at $f_k=0.75$ and $1.00$. Structures captured with $\lambda_2$-criterion ($\mathrm{s^{-2}}$) \cite{JEONG_JFM_1995}.}
\label{fig:4.3.3_3}
\end{figure*}

To investigate the origin of these discrepancies, figures \ref{fig:4.3.3_1}-\ref{fig:4.3.3_3} depict the main coherent structures observed at $t_c\leq t \leq 7.0$ and different values of $f_k$. These vortical structures are identified through the iso-surfaces of the $\lambda_2$-criterion \cite{JEONG_JFM_1995} ($\lambda_2=-2.5\times 10^9$) and colored by the magnitude of the turbulent-to-molecular kinematic viscosity ratio, $\nu_u/\nu$. The maximum values of $\nu_u/\nu$ obtained at each time instant and $f_k$ are presented in table \ref{tab:4.3.3_1}. Referring to the computations at $f_k\leq 0.25$, figure \ref{fig:4.3.3_1}, we observe that the coherent field predicted at these physical resolutions exhibits similar features. The plot shows that orthogonal pairs of counter-rotating vortices interact and reconnect, this leading to the onset of turbulence at $t=7.0$ (bursts of fine-scale turbulence). Also, table \ref{tab:4.3.3_1} indicates that $(\nu_u/\nu)_\mathrm{max}$ is small and does not exceed $0.25$ at $t=6.0$, $1.08$ at $t=6.5$, and $1.30$ at $t=7.0$. Figure \ref{fig:4.3.3_2} illustrates that coarsening the physical resolution to $f_k=0.50$ has a strong impact on the predicted coherent field. Whereas the simulation at $f_k=0.35$ can capture the vortical structures observed at $f_k\leq 0.25$ and exhibits mild levels of $(\nu_u/\nu)_\mathrm{max}$ not exceeding $2.9$, the computation at $f_k=0.50$ leads to significant morphological changes of the coherent field. At $t=7.0$, figure \ref{fig:4.3.3_2f}, part of the coherent field has been dissipated through the overprediction of turbulence and $\nu_u/\nu$. It is observed that $\nu_u$ can exceed its molecular counterpart by $1.7$  times at $t=6.0$, $6.3$ at $t=6.5$, and $6.6$ at $t=7.0$. As $f_k$ further coarsens, figure \ref{fig:4.3.3_3}, these phenomena become more pronounced. Taking the case of $f_k=1.00$, the corresponding figures show that the coherent field is strongly affected at $t=6.50$, and a significant fraction of the laminar vortical structures has been dissipated by the large values of $\nu_u/\nu$. Table \ref{tab:4.3.3_1} indicates that this quantity can reach $4.4$ at $t=6$.
\begin{figure*}[t!]
\centering
\subfloat[$f_k=0.00$.]{\label{fig4.3.4_1a}
\includegraphics[scale=0.08,trim=0 0 0 0,clip]{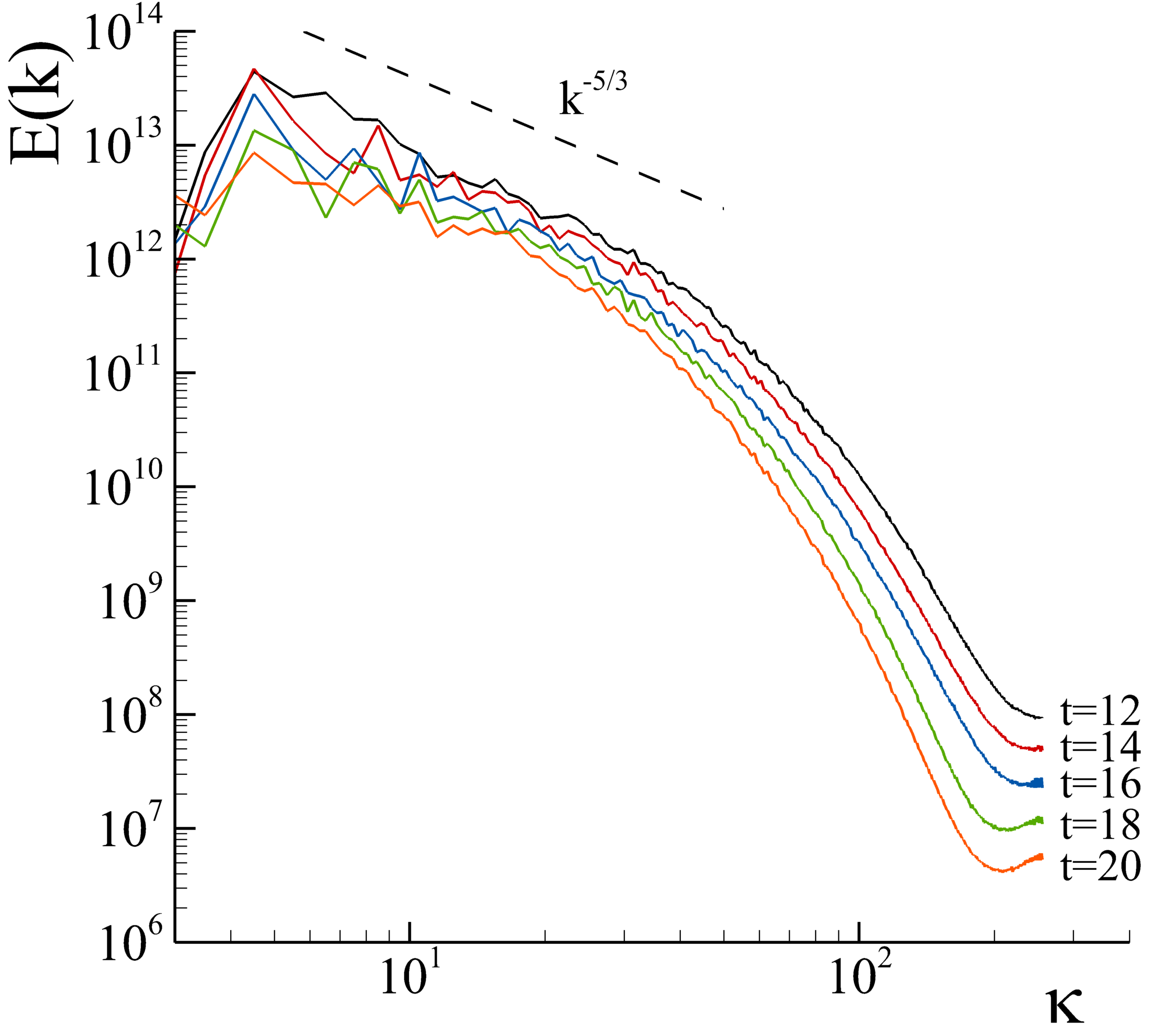}}
~
\subfloat[$f_k=0.25$.]{\label{fig4.3.4_1b}
\includegraphics[scale=0.08,trim=0 0 0 0,clip]{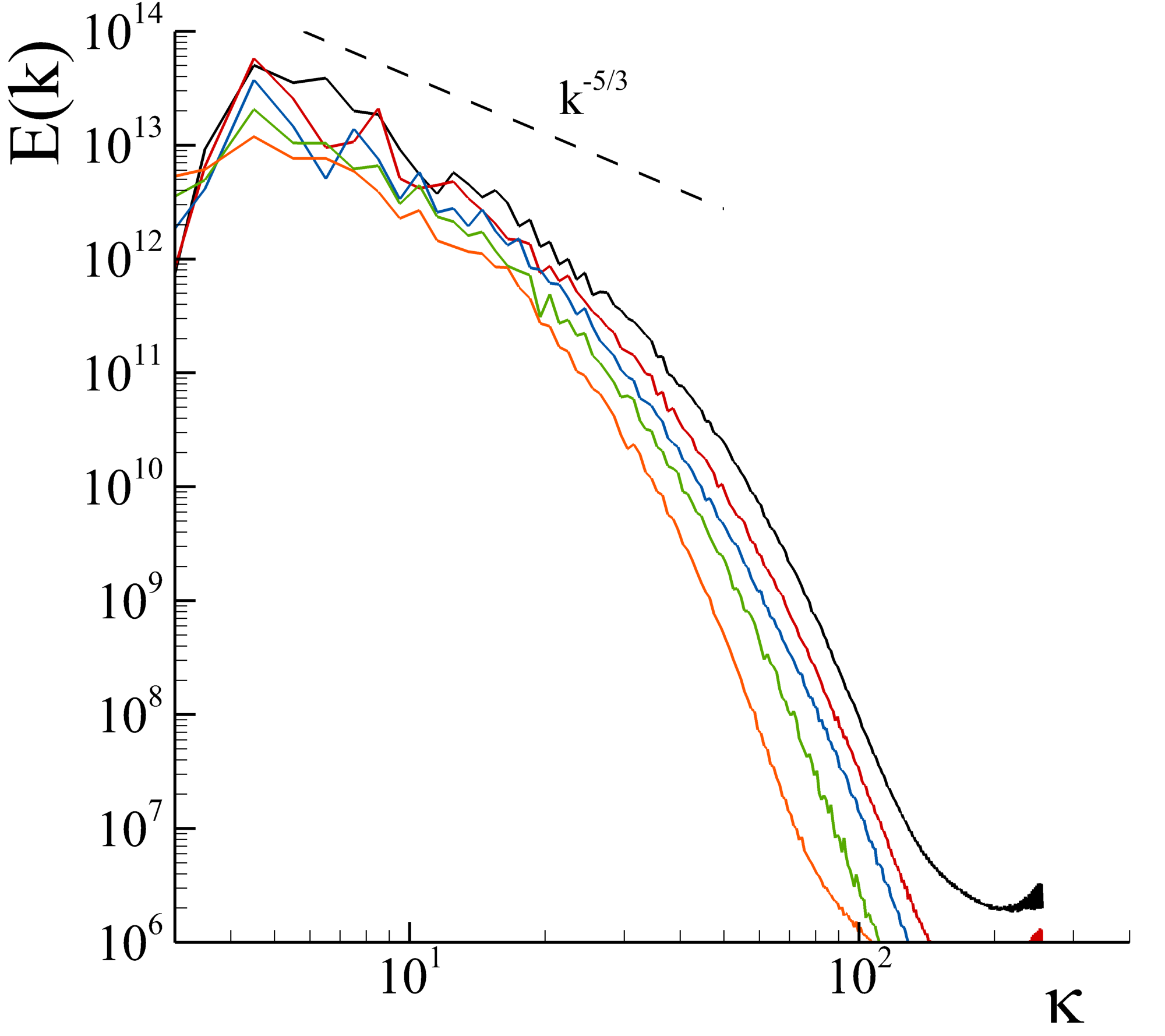}}
~
\subfloat[$f_k=0.35$.]{\label{fig4.3.4_1c}
\includegraphics[scale=0.08,trim=0 0 0 0,clip]{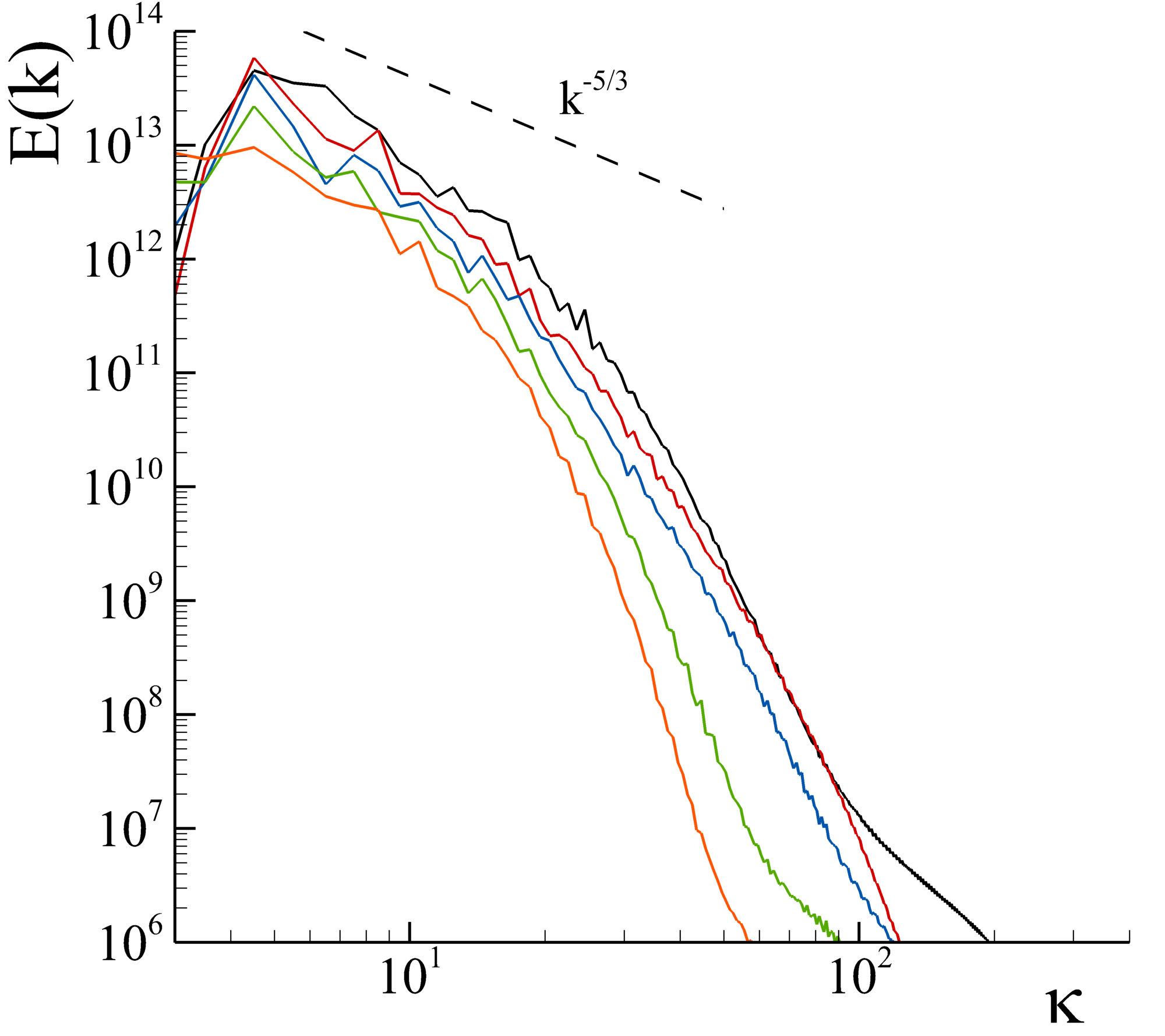}}
\\
\subfloat[$f_k=0.50$.]{\label{fig4.3.4_1d}
\includegraphics[scale=0.08,trim=0 0 0 0,clip]{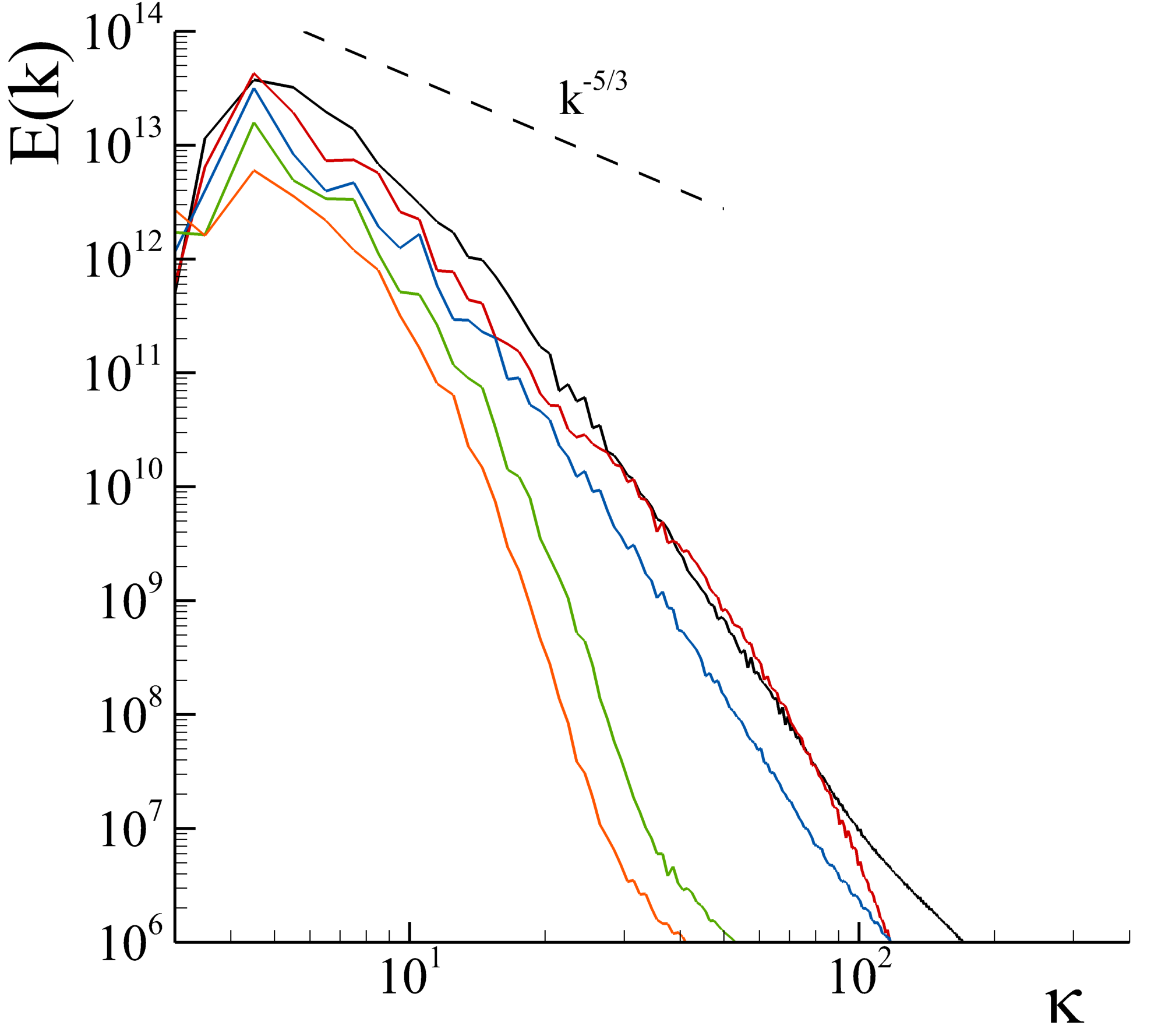}}
~
\subfloat[$f_k=0.75$.]{\label{fig4.3.4_1e}
\includegraphics[scale=0.08,trim=0 0 0 0,clip]{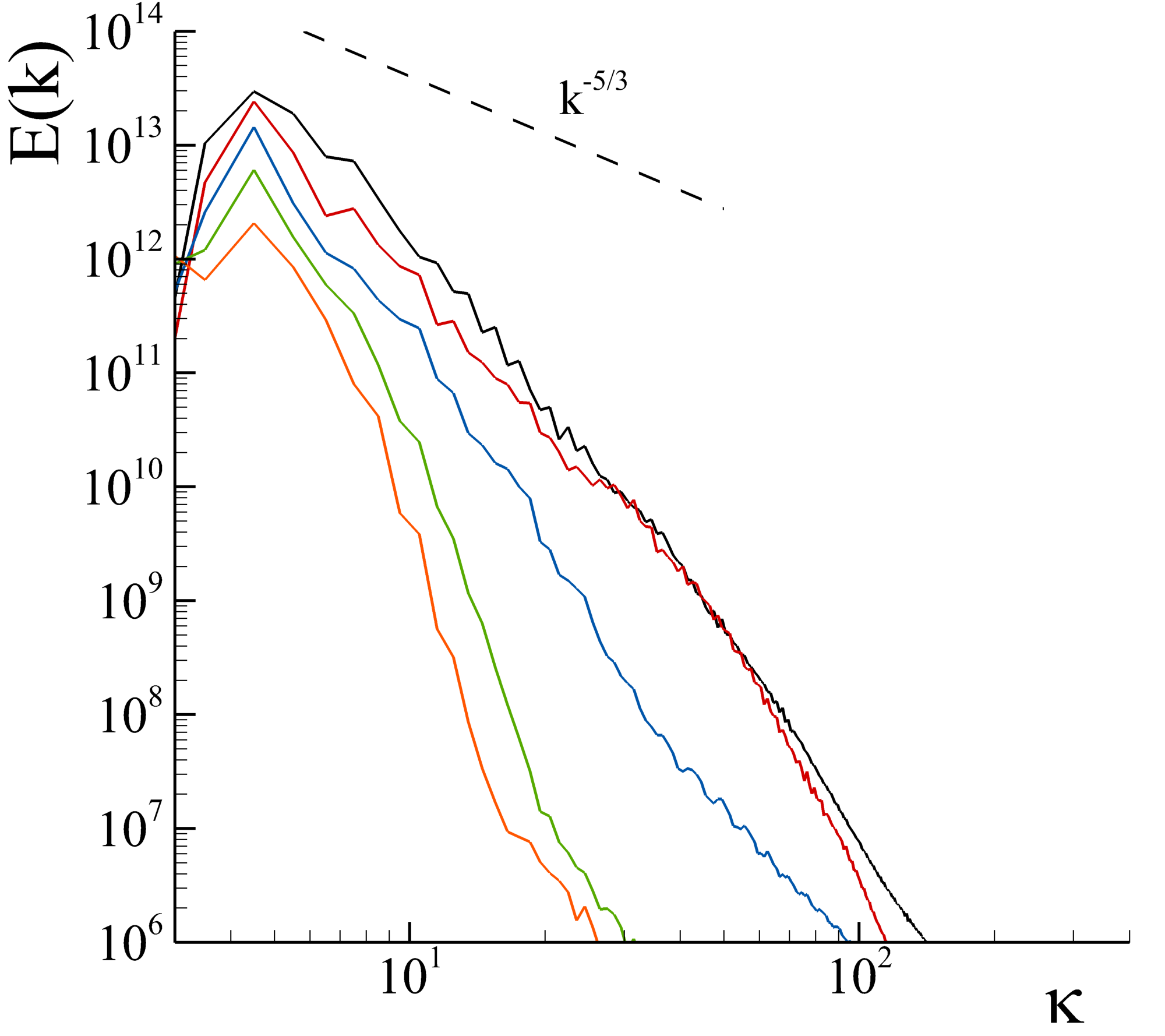}}
~
\subfloat[$f_k=1.00$.]{\label{fig4.3.4_1f}
\includegraphics[scale=0.08,trim=0 0 0 0,clip]{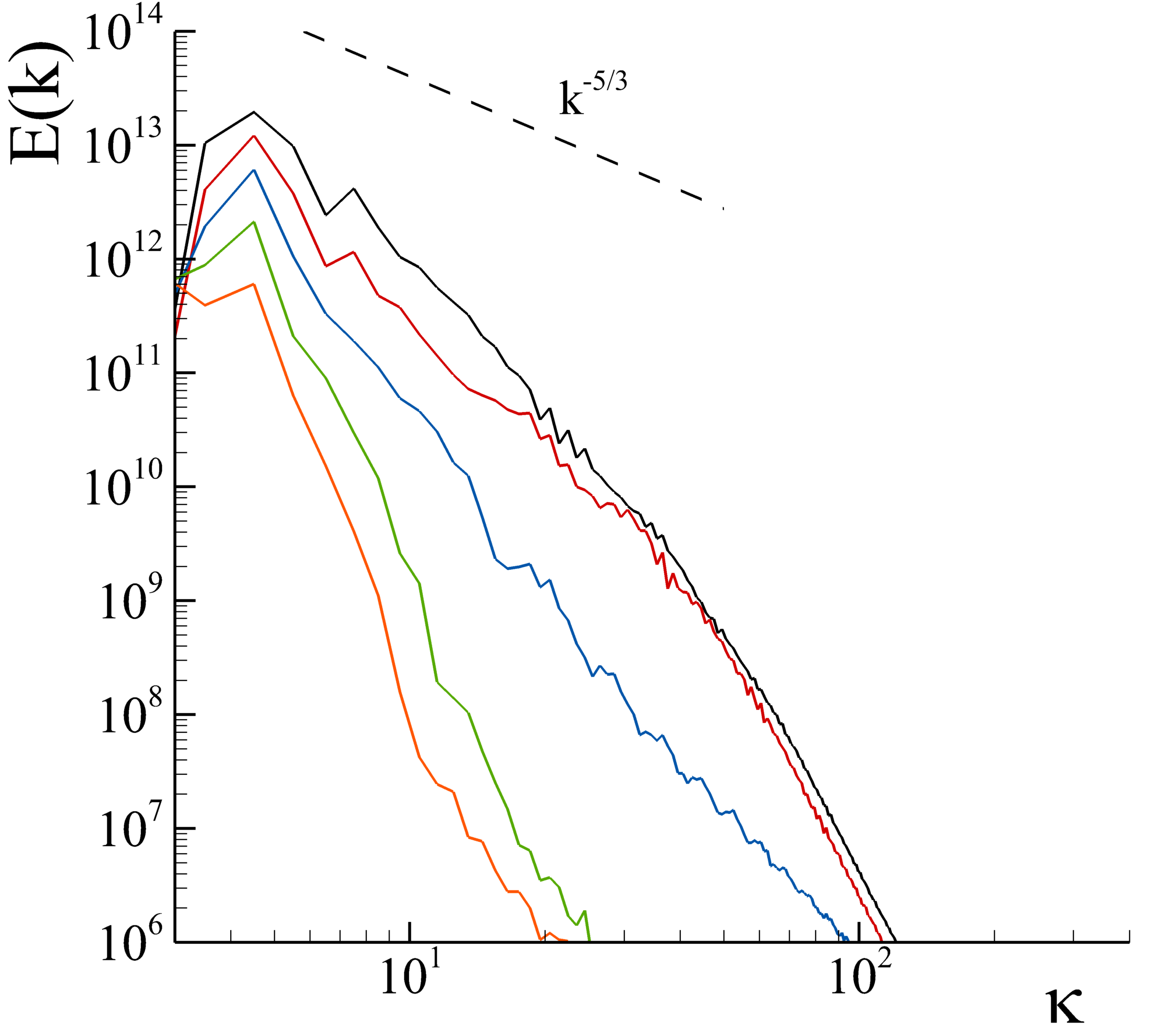}}
\caption{Temporal evolution of the turbulence kinetic energy spectrum, $E(k)$, for simulations using different values of $f_k$.}
\label{fig:4.3.4_1}
\end{figure*}

{\color{blue}
From a modeling perspective, it is important to emphasize that RANS simulations are not aimed to resolve the fine-scale turbulence observed in figures \ref{fig:4.3.3_1e}, \ref{fig:4.3.3_1f}, and \ref{fig:4.3.3_2e}. These are expected to be modeled by the closure. However, the results show that low-physical resolution simulations are also damping the largest and laminar coherent structures due to the magnitude of $\nu_t/\nu$. This results from the fact that one-point closures have not designed to represent the physics of such phenomena.

Figures \ref{fig:4.3.3_1}-\ref{fig:4.3.3_3} also demonstrate} that the accuracy of the simulations is determined by their ability to capture the vortical structures responsible for the onset of turbulence. Thus, the results reaffirm the findings of Pereira et al. \cite{PEREIRA_JCP_2018}. These indicate that efficient SRS calculations require the selection of the smallest physical resolution that enable the model to resolve only the flow scales not amenable to modeling. For the simulation of the TGV with PANS {\color{blue}BHR-LEVM}, this objective can be achieved by setting $f_k$ between $0.25$ and $0.35$.
%
%
\subsubsection{Spectral features}
\label{sec:4.3.4}

Next, we turn our attention to the effect of the physical resolution on the spectral features of the TGV flow. Figure \ref{fig:4.3.4_1} depicts the velocity spectra of the simulations at successively later time instants and different physical resolutions, $f_k$. The spectra obtained from high-physical resolution computations include an inertial range, in which the physical resolution determines the largest wavelength. As expected, it is observed that refining $f_k$ leads to longer inertial ranges but does not significantly affect the large, energy-containing scales; only the simulation at $f_k=0.00$ resolves the viscous scales (viscous sub-range). Considering the results of this study, this demonstrates that it is possible to obtain accurate simulations without resolving all/most turbulent scales. In contrast, the spectra obtained from low-physical resolution simulations are characterized by lower energy levels across the entire spectrum, including significant damping of the large-scale structures. Along with the reduced physical resolution, this result stems from the premature onset and overprediction of turbulence. For all values of $f_k$, it is possible to observe that the kinetic energy decreases with time. Yet, the decay rate increases with {\color{blue}$f_k\rightarrow 1.0$}. 

%
%
%
\subsection{Computational cost}
\label{sec:4.4}
%
\begin{table*}
\centering
\setlength\extrarowheight{3pt}
\caption{Computational cost (CPU.hours) of PANS simulations at different values of $f_k$ and grid resolutions. $^*$ indicates extrapolated value.}
\label{tab:4.4_1}    
\begin{tabular}{C{2.0cm}C{2.0cm}C{2.0cm}C{2.0cm}C{2.0cm}C{2.0cm}C{2.0cm}}
\hline 
$f_k$ & $0.00$ & $0.25$ & $0.35$ & $0.50$ & $0.75$ & $1.00$ \\[3pt]
\hline 
$N_c=1024^3$ 	& $5,782,595^*$& $-$		& $-$ 		& $-$ 		& $-$ 		& $-$ \\ [3pt]
$N_c=512^3$ 	& $322,284$ 	& $406,803$	& $405,435$ 	& $400,139$ 	& $393,947$ 	& $398,927$\\ [3pt]
$N_c=256^3$ 	& $17,962$ 	& $21,600$	& $21,408$ 	& $21,431$ 	& $21,216$ 	& $20,916$\\ [3pt]
$N_c=128^3$ 	& $1,204$ 		& $1,392$		& $1,375$ 		& $1,336$ 		& $1,321$ 		& $1,320$\\ [3pt] 
\hline
\end{tabular}
\end{table*}

The discussion of the results concludes with the evaluation of the simulations' computational intensity. Table \ref{tab:4.4_1} presents the cost in CPU.hours of the computations on different grid resolutions ($N_c$) and physical resolutions ($f_k$). Unfortunately, we cannot show the exact value for the cost of the simulations on $N_c=1024^3$. In this case, we present a very conservative estimate based on the cost of the computations on the grids with $256^3$ and $512^3$ cells.

As expected, table \ref{tab:4.4_1} shows that the computational cost of the simulations grows upon grid refinement. The results show that doubling the grid resolution increases the cost of the computations by more than one order of magnitude. Taking the case of $f_k=0.00$, the number of CPU.h grows $14.9$ times from simulations on $N_c=128^3$ to $256^3$, and $17.9$ times from $N_c=256^3$ to $512^3$. On the other hand, the comparison of the simulations' cost at different physical resolutions presents two distinct behaviors. First, computations resolving all flow scales ($f_k=0.00$) are computationally less intensive than the remaining. For the same grid, we obtain savings that can reach $20.8\%$. This reflects the cost of computing the closure model. Second, the cost of simulations at $f_k>0.00$ grows with the physical resolution. This is caused by the broader range of resolved scales that increases the steepness of the flow gradients (affecting iterative solvers' performance) and the implicit contribution of closure's discrete system of equations \cite{PEREIRA_PHD_2018,PEREIRA_OE2_2019,PEREIRA_ACME_2020}.

Although simulations at $f_k=0.00$ are computationally less intense than those at $f_k>0.00$ on the same grid, they require finer grid resolutions to obtain the same level of numerical uncertainty (a measure of the numerical error). The data show that computations at $f_k=0.00$ require a grid with a minimum of $N_c=1024^3$ cells to obtain similar magnitudes of numerical uncertainty as $f_k=0.25$ using $N_c=512^3$. This leads to a significant reduction of the simulations' cost from $5,782,595$ to $406,803$ CPU.h, which illustrates the potential of PANS method to predict the present transitional flow efficiently.
%
%
%
\section{Conclusions}
\label{sec:5}

A new PANS {\color{blue}model} is used to predict the transitional TGV flow at Re$=3000$. This is an archetypal problem widely utilized to assess the performance of turbulence models simulating transitional flows driven by vortex-stretching and reconnection mechanisms. Since these phenomena are present in numerous flows of variable-density (e.g., ocean and mixing problems), this work constitutes the first step toward extending PANS to this class of flows. 

The study starts by deriving the governing equations of the model. This includes the assessment of the parameters controlling the physical resolution of the model, $f_k$ and $f_\varepsilon$, and their dependence from the range of resolved scales. It is shown that the usual assumption of setting $f_\varepsilon=1.00$ is acceptable for values of $f_k$ of practical interest ($f_k\ge 0.20$). For this reason, all simulations are conducted using $f_\varepsilon=1.00$. In regard to $f_k$, we use different constant values of this parameter to evaluate the effect of the physical resolution on the simulations. This approach also prevents commutation errors, and enables robust verification and validations exercises since the governing equations of the model do not depend on the grid resolution.

The discussion of the results is preceded by a qualitative analysis of the flow dynamics and vortical structures using the simulation at $f_k=0.00$. It is inferred that the development of the TGV flow comprises two distinct periods: $i)$ in the first, the initially well-defined laminar vortices interact and deform, leading to the onset and development of turbulence through vortex-stretching and reconnection mechanisms. $ii)$ in the second, the flow exhibits high-intensity turbulence features and experiences a rapid decay of kinetic energy.

The results of PANS simulations demonstrate that the model can efficiently predict the TGV flow. Yet, the computations' accuracy is closely dependent on the physical resolution, and, most notably, on the ability to predict the first stage of the flow development $(t \leq 9)$. Whereas low-physical resolution ($f_k\ge 0.50$) simulations lead to poor representations of the flow dynamics, high-physical resolution ($f_k<0.50$) computations can accurately predict the TGV. The physical and modeling interpretation of the results indicates that the simulations' accuracy is determined by the ability to represent the vortex-reconnection mechanism driving the onset of turbulence. It is observed that low-physical resolution simulations overpredict turbulence, leading to the dissipation of these coherent structures and consequent poor representation of the flow physics. In contrast, high-physical resolution computations ($f_k <0.50$) can accurately simulate these phenomena and, consequently, the flow physics. Regarding the computations' cost, the numerical uncertainty of the results indicates that simulations at $f_k=0.00$ require grids two times finer to obtain values of numerical uncertainty similar to those obtained by simulations at $f_k>0$. Considering that all high-physical resolution computations can accurately predict the flow physics, the results illustrate that PANS ($0.25 \leq f_k \leq 0.35$) leads to cost savings exceeding one order of magnitude. These are expected to increase with Re.

In summary, this study reinforces the potential of the PANS method to calculate transitional flows of practical interest efficiently. {\color{blue}Also, it reiterates that the success of SRS methods depends upon the ability to identify and resolve the flow phenomena not amenable to accurately modeling \cite{PEREIRA_JCP_2018}. For the present flow, this requires resolving the vortex stretching and reconnection processes. The remaining flow scales can be accurately represented by the turbulence closure.}

%
%
%
\section*{Acknowledgments}

The authors would like to thank C. B. da Silva for sharing his DNS data sets and insightful discussions. {\color{blue}Also, we would like to thank the two reviewers for their suggestions that improved our paper.}

Los Alamos National Laboratory (LANL) is operated by TRIAD National Security, LLC for the US DOE NNSA. This research was funded by LANL Mix and Burn project under the DOE ASC, Physics and Engineering Models program. We express our gratitude to the ASC program and the High-Performance Computing (HPC) division for their dedicated and continued support of this work.
%
%
%
\bibliography{references}
\end{document}